\documentclass[prc,showpacs,floatfix]{revtex4}
\usepackage{bm}
\usepackage{psfig}
\usepackage{graphicx}
\usepackage{amsmath}
\usepackage{bbold}
\newcommand{\beq}{\begin{equation}}
\newcommand{\eeq}{\end{equation}}
\newcommand{\beqa}{\begin{eqnarray}}
\newcommand{\eeqa}{\end{eqnarray}}

\begin {document}
\title{
\hfill{\small {\bf MKPH-T-04-12}}\\
{\bf  General Survey of Polarization Observables in Deuteron 
Electrodisintegration
}}
\author{Hartmuth Arenh\"ovel$^{1}$, Winfried Leidemann$^{2}$,
and Edward L. Tomusiak$^{3}$}
\affiliation{                                                         
$^{1}$Institut f\"ur Kernphysik, Johannes Gutenberg-Universit\"at, D-55099 Mainz, Germany\\
$^{2}$Dipartimento di Fisica, Universit\`a di Trento, and Istituto Nazionale di Fisica Nucleare, Gruppo collegato di Trento, I-38050 Povo, Italy\\
$^{3}$Department of Physics and Astronomy, University of Victoria, Victoria, BC V8P 1A1, Canada}

\date{\today}

\begin{abstract}
\noindent
Polarization observables in inclusive and exclusive electrodisintegration of
the deuteron using a polarized beam and an oriented target are systematically 
surveyed using the standard nonrelativistic framework of nuclear 
theory but with leading order relativistic contributions included. The 
structure functions and the asymmetries corresponding to the various nucleon
polarization components are studied in a variety of kinematic regions 
with respect to their sensitivity to realistic $NN$-potential models, to 
subnuclear degrees of freedom in terms of meson exchange currents, 
isobar configurations and to 
relativistic effects in different kinematical regions, serving as a benchmark
for a test of present standard nuclear theory with effective degrees of 
freedom.
\end{abstract} 

\maketitle

\section{Introduction}\label{introduction}
Over the past decade we have made a systematic study of inclusive and 
exclusive deuteron electrodisintegration with special emphasis
on polarization observables \cite{LeT91,ArL92,ArL93,ArL95,ArL98,ArL00}. 
The main purpose of this study was to reveal to what extent the use of
polarized electrons, polarized targets and polarization analysis of the
outgoing nucleons will allow a more thorough and more detailed 
investigation of the dynamical features of the two-nucleon system than 
is possible without the use of polarization degrees of freedom (d.o.f.). 
Specifically, our interest was focused on the role of the $NN$-interaction 
model, of subnuclear degrees of freedom in terms of meson and isobar d.o.f. 
and in some cases on the role of relativistic effects. 

In the first paper of this investigation \cite{LeT91} we have considered 
the inclusive process, followed in \cite{ArL92} by the exclusive case, 
$\vec d(\vec e, e'N)N$, including beam and 
target polarization but without analysis of the outgoing 
nucleon polarization. In parallel, as an extension to 
previous work in photodisintegration \cite{Are88,ArS90}, we
have formally derived in \cite{ArL93} 
all possible polarization structure functions, in
total 648, and linear relations between them since only 324 can be linearly
independent considering the fact that each structure function is a
hermitean form of 18 independent complex $t$-matrix elements, provided 
parity conservation holds. Formal 
expressions for polarization observables, using a different representation 
scheme for the structure functions, have been given by {\sc Dmitrasinovic} and
{\sc Gross}~\cite{DmG89} where also the question of necessary and sufficient 
measurements for a complete determination 
of all transition amplitudes has been discussed in detail. 
In \cite{ArL95} we have continued our own 
study by looking at the polarization of one or both of the final state 
nucleons in the exclusive processes $d(\vec e,e'N)N$ and $d(\vec e,e'NN)$ with 
various combinations of beam and target polarizations. Although one has 324 
linearly independent observables, they are not independent in the more 
general sense of considering them as functions of the complex $t$-matrix 
elements. In view of the fact that for the 18 complex matrix elements one
phase can be arbitrarily chosen and thus all observables are functions of 
35 independent and real variables, it is obvious that the maximal number 
of independent observables is 35. Indeed, there exist quadratic relations 
among them, reducing the 324 linearly independent observables to the
required number. The remaining question then is, which one of the many 
possible subsets of 35 observables constitutes an independent set. 
This question has been investigated recently in~\cite{ArL98} for a two-body 
reaction of the type $a+b\rightarrow c+d$,
where we have derived a general criterion for the selection of a complete 
set of independent observables, which subsequently has been applied 
in~\cite{ArL00} to the electromagnetic deuteron break-up. In the latter work
we have also derived the relations between our structure functions and
the ones of~\cite{DmG89}. Most recently, we have obtained the general
multipole expansion of the structure functions in~\cite{ArL02}.

Besides this 
interest in the hadron dynamics, there is a second aspect which underlines 
the important role of the deuteron from a different perspective, 
namely in providing an effective neutron target because of its extremely 
weak binding. In fact, quasifree reactions on the deuteron are 
frequently used in order to investigate properties of the neutron. 
A prominent example is the current interest in the determination of the 
electric form factor of the neutron $G_{En}$ in quasifree electron 
scattering off the deuteron using longitudinally polarized electrons and 
either a vector polarized target or measuring the polarization of the 
outgoing neutron. Therefore, we also have studied in detail the sensitivity of 
polarization observables to $G_{En}$. 

The special and very fundamental role of the two-nucleon system 
for the investigation of the hadronic structure of nuclei, playing the same 
role as the hydrogen atom in atomic physics, is underlined first of all by 
the fact, that $NN$-scattering is of crucial importance for fitting realistic
$NN$-potential models. Secondly, the deuteron constitutes the simplest 
nucleus. It is very weakly bound and allows an exact theoretical 
treatment, at least in the nonrelativistic regime. Thus it should be clear 
that such an extensive survey is justified. Moreover, the electromagnetic 
probe allows a particularly clean and simple interpretation of the 
associated observables, because it possesses a well known but weak 
interaction so that in most cases lowest order approaches are sufficient. 

With the present work we want to give a concise and self-contained summary 
of this extensive study in order to provide the interested experimentalist 
and theorist a sort of handbook for the powerful tool of polarization 
observables in this fundamental process. We furthermore want to update 
our previous results with respect to (i) the recent high precision 
$NN$-potentials, (ii) consistent $\pi$- and $\rho$-like meson exchange 
currents, and (iii) complete and consistent inclusion of leading order 
relativistic contributions in a $v/c$-expansion. 
Thus we consider this work also as a 
benchmark for the status of present standard nuclear theory with effective 
degrees of freedom in terms of nucleon, meson and isobar d.o.f. 
To this end we will collect 
in the next section all material relevant for the formal aspects of this 
process including kinematic properties and relations, and definition of 
observables for the inclusive as well as for the exclusive reaction 
in terms of form factors and structure functions, respectively. In Section
\ref{completesets} we will discuss the experimental separation of structure 
functions and the question of how to select a complete set of observables.
Some calculational details will be presented in Section~\ref{ingredients} 
with respect to the hadronic interaction and the electromagnetic current. 
The latter comprises contributions from one-body and meson exchange currents 
(MEC) and isobar configurations (IC) as well as leading order relativistic 
contributions (RC). Then we will discuss in Section~\ref{examples} the 
influence of the above mentioned dynamical effects on various inclusive
and exclusive observables for different kinematic regions of energy and 
momentum transfers. Finally, we will close with a summary and an outlook. 
\section{General Formalism}\label{genform}

\subsection{Kinematics}
At first we will consider the kinematic properties of the disintegration 
process $e+d\rightarrow e'+n+p$. The kinematics is governed by the four 
momenta of the participating particles, i.e., the four-momenta $k_1$ and 
$k_2$ of incoming and scattered electron, respectively, and the four-momenta 
$p_d$, $p_p$ and $p_n$  of deuteron and outgoing 
proton and neutron, respectively. In view of on-shell conditions for the 
participating particles, four-momentum conservation, freedom in the 
choice of the incoming electron direction and in the choice of orientation of 
the scattering plane, one is left with five independent variables for 
this process. Different choices are possible and will be specified below. 

\begin{figure}
\includegraphics[scale=.5]{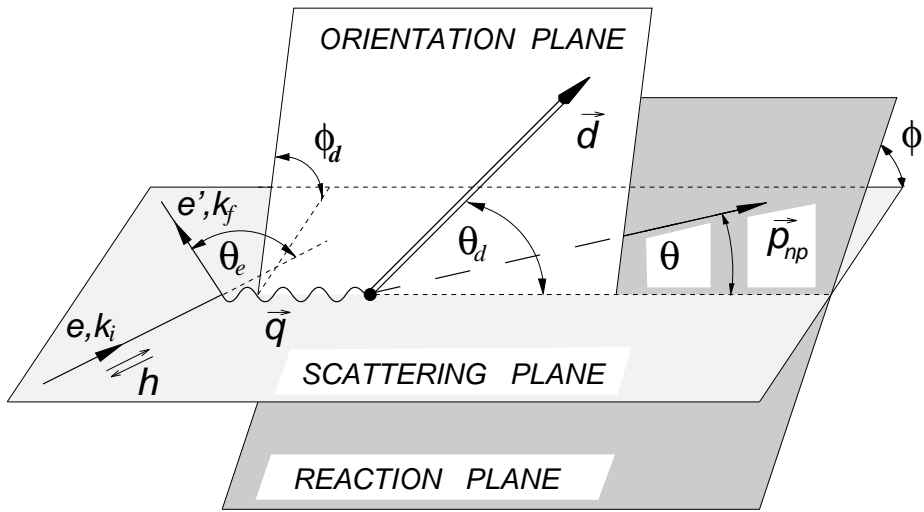}
\caption{Geometry of exclusive electron-deuteron scattering with
polarized electrons and an oriented deuteron target. 
The relative $np$-momentum, 
denoted by ${\vec p}_{np}$, is characterized by angles $\theta=\theta_{np}$ 
and $\phi=\phi_{np}$ where the deuteron orientation axis, denoted by 
$\vec d$, is specified by angles $\theta_d$ and $\phi_d$.\label{fig1}}
\end{figure}

The scattering geometry is illustrated in Fig.~\ref{fig1}. We distinguish 
three different planes which all intersect in one line as defined by the 
direction of the three-momentum transfer $\vec q = \vec k_1 - \vec k_2$. 
First, there is the 
scattering plane which is defined by the three-momenta of incoming and 
scattered electrons $\vec k_1$ and $\vec k_2$, 
respectively, then the reaction plane defined by the momentum 
transfer and the relative momentum 
$\vec p_{np}=(\vec p_{p}-\vec p_{n})/2$ of the two outgoing nucleons, 
and finally the orientation plane as defined again by the momentum transfer 
and the axis of orientation for a polarized deuteron. 

The principal frames of reference are associated with the scattering 
plane, namely the laboratory frame, and the c.m.\ frame of the final 
two nucleons, which is related to the former one by a boost along 
$\vec q$. The $z$-axis is chosen in both frames along $\vec q$ 
and the $y$-axis in the direction of $\vec k_1\times \vec k_2$, i.e., 
perpendicular to the scattering plane. Finally, the $x$-axis is defined by
$\vec e_x=\vec e_y \times \vec e_z$ in order to form a right-handed system. 
The laboratory frame is the natural 
choice for the experimental determination of observables, whereas the c.m.\ 
frame is very convenient for the theoretical calculation. Where necessary, 
we will indicate by a superscript ``$lab$'' or ``$c.m.$'' to which frame 
a given quantity refers. With respect to the c.m.\ frame, we will denote 
throughout this paper by $\theta$ and $\phi$ the spherical angles of the 
relative momentum $\vec p_{np}= (p^{\mathrm{c.m.}},\theta , \phi )$. 
Thus the spherical angles of proton and neutron momenta in this frame are 
$\theta^{\mathrm{c.m.}}_p=\theta$, $\phi^{\mathrm{c.m.}}_p=\phi$ and 
$\theta^{\mathrm{c.m.}}_n=\pi-\theta$, $\phi^{\mathrm{c.m.}}_n=\phi+\pi$. 
The final hadronic state is furthermore characterized by the excitation 
energy $E_{np}$ which is related to its invariant mass $W_{np}$ by
\begin{equation}
E_{np} = W_{np} -2\, M\,,\label{E_np}
\end{equation}
where $M$ denotes the average nucleon mass. Finally, $\theta_d$ and 
$\phi_d$ denote the lab frame spherical angles of the 
deuteron orientation axis in case a polarized deuteron target is employed. 

The relevant quantities in the lab frame are the four-momenta 
of the incoming and scattered electrons 
$k_1^{\mathrm{lab}}=(E_1^{\mathrm{lab}},\,\vec k_1^{\mathrm{lab}})$ and 
$k_2^{\mathrm{lab}}=(E_2^{\mathrm{lab}},\,\vec k_2^{\mathrm{lab}})$, 
respectively, the scattering 
angle $\theta_e^{\mathrm{lab}}$, and the proton and neutron three-momenta 
$\vec p_p^{\,lab}=(p_p^{\mathrm{lab}},\,\theta_p^{\mathrm{lab}},\,
\phi_p^{\mathrm{lab}})$ and 
$\vec p_n^{\,lab}=(p_n^{\mathrm{lab}},\,\theta_n^{\mathrm{lab}},\,
\phi_n^{\mathrm{lab}})$, respectively. 
For the five independent variables one convenient choice is 
$E_1^{\mathrm{lab}}$, 
$E_2^{\mathrm{lab}}$, $\theta_e^{\mathrm{lab}}$, 
$\theta_p^{\mathrm{lab}}$, and $\phi_p^{\mathrm{lab}}$. All 
other quantities are then determined by them. An alternative choice is 
$E_1^{\mathrm{lab}}$, $E_2^{\mathrm{lab}}$, $\theta_e^{\mathrm{lab}}$, 
and the spherical angles of the relative $np$-momentum in the c.m.\ system,
$\theta$, and $\phi$. Still another useful choice is the 
$np$-final state excitation energy $E_{np}$, the three momentum transfer 
$q^{\mathrm{c.m.}}$, again the angles $\theta$ and $\phi$, all with respect 
to the c.m.\ frame of the final $np$-state, and the lab electron scattering 
angle $\theta_e^{\mathrm{lab}}$. 
This choice is particularly useful for formal investigations of the 
structure functions, because the latter depend solely on  
$E_{np}$, $q^{\mathrm{c.m.}}$, and $\theta$, if calculated 
in the c.m.\ frame. 

We will now list the various relevant kinematic quantities in the lab and 
c.m.\ frames and their relation to the chosen independent variables. 
Throughout this work we will denote the square of a four-vector $x_{\nu}$ by 
$x_\nu^2=x_0^2-\vec x^{\,2}$ and use $x=|\vec x|$. For 
given electron momenta $\vec k_1^{\,\mathrm{lab}}$, 
$\vec k_2^{\,\mathrm{lab}}$ and 
scattering angle $\theta_e^{\mathrm{lab}}$, one has for the energy and 
momentum transfers in the lab frame
\beqa
\omega^{\mathrm{lab}}&=&E_{1}^{\mathrm{lab}}-E_{2}^{\mathrm{lab}}\,,\\
\vec q^{\,\, \mathrm{lab}}&=& \vec k_1^{\,\mathrm{lab}}
-\vec k_2^{\,\mathrm{lab}}\,,
\eeqa
and from this the invariant mass $W_{np}$ of the $np$-final state in terms 
of lab and c.m.\ frame quantities
\begin{subequations}
\beqa
W_{np}&=&\sqrt{(E_{np}^{\mathrm{lab}})^2-(q^{\mathrm{lab}})^2}\label{Wnp_lab}\\
&=& \sqrt{(M_d\,(M_d+2\,\omega^{\mathrm{lab}})+q_\nu^2)}\\
&=& \omega^{\mathrm{c.m.}}+E_d^{\mathrm{c.m.}}\,,\label{W_np_omega}
\eeqa
\end{subequations}
where
\begin{equation}
E_d^{\mathrm{c.m.}}=\sqrt{M_d^2+(q^{\mathrm{c.m.}})^2}\label{E_d_cm}
\end{equation}
denotes the deuteron c.m.\ energy, and
\begin{equation}
E_{np}^{\mathrm{lab}}=\omega^{\mathrm{lab}}+M_d
\end{equation}
the lab energy of the final hadronic state. At the photon point 
$\omega^{\mathrm{lab}}=q^{\mathrm{c.m.}}$ one finds according to 
(\ref{W_np_omega}) and (\ref{E_d_cm}) the relation
\beqa
q^{\mathrm{c.m.}}&=&\frac{W_{np}^2-M_d^2}{2\,W_{np}}\nonumber\\
&=&(E_{np}+\varepsilon_d)\Big(1-
\frac{E_{np}+\varepsilon_d}{2(E_{np}+2\,M)}\Big)\,,\label{photonline}
\eeqa
where $\varepsilon_d$ denotes the deuteron binding energy.

The boost parameter $\gamma$, which governs the transformation from the lab to 
the c.m.\ frame and vice versa, is given by 
\begin{subequations}
\beqa
\gamma&=&\frac{E_{np}^{\mathrm{lab}}}{W_{np}}\\
&=&\frac{E_d^{\mathrm{c.m.}}}{M_d}\,.
\eeqa
\end{subequations}
Energy and momentum transfers in the lab and c.m.\ frames are related to 
each other by
\begin{subequations}
\beqa
\omega^{\mathrm{c.m.}}&=&\frac{1}{W_{np}}
(M_d\,\omega^{\mathrm{lab}}+q_\nu^2)\,,\\
q^{\mathrm{c.m.}}&=&\frac{M_d}{W_{np}}q^{\mathrm{lab}}\,.\label{qcm_lab}
\eeqa
\end{subequations}
Similarly, one has for the relative $np$-momentum
\begin{subequations}
\beqa
p_{np}^{\mathrm{lab}}&=&p^{\mathrm{c.m.}}\,\sqrt{1+
\Big(\frac{q^{\mathrm{c.m.}}}{M_d}\Big)^2
\cos^2\theta}\,,\\
\cos \theta_{np}^{\mathrm{lab}}&=& \frac{E_d^{\mathrm{c.m.}}}
{\sqrt{M_d^2+(q^{\mathrm{c.m.}})^2\,\cos^2\theta}}
\,\cos \theta \,,\\
p^{\mathrm{c.m.}}&=&p_{np}^{\mathrm{lab}}\,\sqrt{1+
\Big(\frac{q^{\mathrm{lab}}}{W_{np}}\Big)^2\,
\cos^2\theta_{np}^{\mathrm{lab}}}\,,\\
\cos \theta&=& \frac{E_{np}^{\mathrm{lab}}}{
\sqrt{W_{np}^2+(q^{\mathrm{lab}})^2\,\cos^2\theta_{np}^{\mathrm{lab}}}}
\,\cos \theta_{np}^{\mathrm{lab}} \,.
\eeqa
\end{subequations}
In the c.m.\ frame, the nucleon energies are 
(neglecting the small proton-neutron mass difference)
\begin{equation}
E_p^{\mathrm{c.m.}}=E_n^{\mathrm{c.m.}}=E^{\mathrm{c.m.}}=
\frac{W_{np}}{2}=\sqrt{M^2+(p^{\mathrm{c.m.}})^2}\,,
\end{equation} 
and their three-momenta 
are given by the relative $np$-momentum
\begin{equation}
\vec p_p^{\,c.m.}= (p^{\mathrm{c.m.}},\theta,\phi)\,,\qquad
\vec p_n^{\,c.m.}= (p^{\mathrm{c.m.}},\pi-\theta,\phi+\pi)\,.
\end{equation}
The same quantities in the lab frame may be expressed in terms of 
the c.m.\ variables by
\begin{subequations}
\beqa
E_{p/n}^{\mathrm{lab}}&=&\frac{E_d^{c.m}}{M_d}\,\Big(\frac{W_{np}}{2}\pm
\frac{q^{\mathrm{c.m.}}\,p^{\mathrm{c.m.}}\,\cos\theta}{E_d^{c.m}}\Big)\,,\\
(p_{p/n}^{\mathrm{lab}})^2&=&\Big(\frac{E_d^{c.m}\,p^{\mathrm{c.m.}}}{M_d}
\Big)^2
+\frac{1}{M_d^2}\Big((q^{\mathrm{c.m.}})^2\Big[M^2+(p^{\mathrm{c.m.}})^2\,
\cos^2\theta\Big]
\pm E_d^{c.m}\,W_{np}\,q^{\mathrm{c.m.}}\,p^{\mathrm{c.m.}}\,\cos\theta\Big)\,.
\eeqa
\end{subequations}

The Jacobian for the transformation  
$\Omega_{np}^{\mathrm{c.m.}}\rightarrow\Omega_i^{\mathrm{lab}}$ 
($i\in \{n,p\}$) is given by
\begin{equation}
\frac{\partial\Omega_{np}^{\mathrm{c.m.}}}{\partial\Omega_i^{\mathrm{lab}}}
=\frac{1}{\gamma}\,\Big(\frac{\beta_i^{\mathrm{lab}}\,\gamma_i^{\mathrm{lab}}}
{\beta^{\mathrm{c.m.}}\,\gamma^{\mathrm{c.m.}}}\Big)^3
\Big(1+\frac{\beta}{\beta^{\mathrm{c.m.}}}\cos \theta_i^{\mathrm{c.m.}}\Big)
^{-1}\,,
\end{equation}
where the angle $\theta_i^{\mathrm{c.m.}}$ or $\theta_i^{\mathrm{lab}}$ denotes
the angle of the momentum of the particle ``$i$'' in the indicated frames of 
reference, and $\beta=\sqrt{\gamma^2-1}/\gamma$. Furthermore,
\begin{equation}
\gamma^{\mathrm{c.m.}}=\frac{E^{\mathrm{c.m.}}}{M}\,,
\end{equation}
is the boost parameter that takes particle ``$i$'' from its rest system to 
the c.m.\ frame. It is the same for both particles. Similarly, for the boost 
from the particle rest frame to the lab one has
\begin{equation}
\gamma_i^{\mathrm{lab}}=\gamma\,\gamma^{\mathrm{c.m.}}\,
(1+\beta\,\beta^{\mathrm{c.m.}}\,\cos \theta_i^{\mathrm{c.m.}})\,.
\end{equation}
Note the relations $\theta_p^{\mathrm{c.m.}}=\theta$, and 
$\theta_n^{\mathrm{c.m.}}=\pi-\theta$ (see Fig.~\ref{fig1}). 
Furthermore, the particle 
lab angle is given by
\begin{equation}
\theta_i^{\mathrm{lab}}=\arcsin \Big[
\frac{\beta^{\mathrm{c.m.}}\gamma^{\mathrm{c.m.}}}
{\beta_i^{\mathrm{lab}}\,\gamma_i^{\mathrm{lab}}}\,
\sin \theta_i^{\mathrm{c.m.}}\Big]\,.
\end{equation}

For the description of the polarization components of the outgoing particle, 
one associates with each particle ``$i$'' a frame of reference according 
to the Madison convention, for which the $z$-axis is taken along the 
particle momentum, i.e., in the reaction plane, the $y$-axis along 
$\vec q\times \vec p_i$, i.e., perpendicular to the reaction plane, and the 
$x$-axis is then 
determined by the requirement to form a right-handed system. 
Often the polarization components are evaluated in the c.m.\ system whereas
the experimental measurement is done in the lab frame. Then it is necessary 
to convert these observables to the laboratory system. 
Applying nonrelativistic kinematics, the spin eigenstates in either system 
are simply related by a rotation, 
$\theta_i^{\mathrm{c.m.}}-\theta_i^{\mathrm{lab}}$ about the $y$-axis. 
However, it is well known (cf.~refs.~\cite{Hal68,MaS70,Gie85}, 
for example) that for relativistic kinematics there is 
a correction such that the actual angle of rotation, the Thomas-Wigner 
angle $\theta_i^W$, is given by
\begin{equation}
\theta_i^W=\arcsin \Big[
\frac{1+\gamma}{\gamma^{\mathrm{c.m.}}+\gamma_i^{\mathrm{lab}}}\,
\sin (\theta_i^{\mathrm{c.m.}}-\theta_i^{\mathrm{lab}})\Big]\,.
\end{equation}
One readily observes that for nonrelativistic boosts the Wigner angle becomes 
simply the $\theta_i^{\mathrm{c.m.}}-\theta_i^{\mathrm{lab}}$. 
Since the rotation is about the $y$-axis, the $y$-components of the 
polarization of the outgoing nucleons undergo no change while the $x$- and 
$z$-components mix according to
\begin{equation}
P_k^{\mathrm{lab}}(i)=R^W_{kl}(i)\,P_l^{\mathrm{c.m.}}\,,\quad i\in \{p,n\}\,,
\end{equation}
where
\begin{equation}
R^W(i)=\left(
\begin{matrix}
\cos \theta_i^W & 0 & \sin \theta_i^W\cr
        0 & 1 & 0 \cr
       -\sin \theta_i^W & 0 & \cos \theta_i^W \cr
\end{matrix}\right)\,.
\end{equation}
Similarly, double polarization observables transform as 
\begin{equation}
P_{kl}^{\mathrm{lab}}=R^W_{kk'}(p)\,R^W_{ll'}(n)\,P_{k'l'}^{\mathrm{c.m.}}
\,.
\end{equation}


\subsection{Definition of observables}

The most general form of an observable in deuteron electrodisintegration is 
\beqa
{\cal O}(\Omega_X)&=&P_X\,S_0\nonumber\\
&=&
tr({\cal T}^\dagger\Omega_X {\cal T}\rho_i)\,,\label{obs1gen}
\eeqa
where 
\begin{equation}
S_0=\frac{d^3\sigma_0}{dk_2^{\mathrm{lab}}d\Omega_e^{\mathrm{lab}}d
\Omega_{np}^{\mathrm{c.m.}}}\label{diffcrossS0}
\end{equation}
denotes the unpolarized cross section. $\Omega_X$ is an operator in the final 
two-nucleon spin space with $P_X$ as corresponding polarization observable.
Its specific form depends on the analysis of the hadronic final state, i.e.,
whether or not polarization components of one or both outgoing nucleons 
are measured, and is defined below. Polarization analysis of the scattered 
electron is not considered here. 
${\cal T}$ denotes the reaction matrix, and $\rho_i$ the density 
matrix for the spin degrees of the initial system.
The trace refers to all initial state spin degrees of freedom 
comprising incoming electron and target deuteron.

In the one-photon-exchange approximation the reaction matrix ${\cal T}$ 
separates into a leptonic and a hadronic part, and 
one obtains from (\ref{obs1gen}) the well-known expression
\beqa
{\cal O}(\Omega_X)&=&3\,c(k_1^{\mathrm{lab}},\,k_2^{\mathrm{lab}})\, 
tr(T^\dagger\Omega_X T\rho_i)\,,\label{obs1}
\eeqa
where the hadronic part is represented by the $T$-matrix which is 
related to the current matrix 
element between the initial deuteron state and the final $np$-scattering state.
The electron kinematics refers to the lab frame while the $T$-matrix and 
all quantities of the final $np$-state refer according to our choice 
to the final state c.m.\ system. 
In (\ref{obs1}) the initial state density matrix $\rho_i$ refers now to the 
spin degrees of the exchanged virtual photon and the deuteron, i.e.\ 
the virtual photon polarizations $\lambda$ $(=0,\,\pm 1)$ and the deuteron 
spin projections $\lambda_d$ with respect to a chosen quantization axis, here 
parallel to $\vec q$. Furthermore, the kinematic factor in (\ref{obs1}) is 
\begin{equation}
c(k_1^{\mathrm{lab}},\,k_2^{\mathrm{lab}}) = 
\frac{\alpha}{6 \pi^2} \frac{k^{\mathrm{lab}}_2}{k_1^{\mathrm{lab}} 
Q^4}\,,
\end{equation}
with $\alpha$ denoting the fine structure constant and 
$Q^2=-q_{\nu} ^2$ the four-momentum transfer squared $(q = k_1 - k_2)$. 
This factor is related to the Mott cross section $\sigma_{\mathrm{Mott}}$ by
\begin{equation}
c(k_1^{\mathrm{lab}},\,k_2^{\mathrm{lab}}) =\frac{1}{6\pi^2\alpha}
\,\frac{{\rm tan}^2(\theta_e^{\mathrm{lab}}/2)}{Q^2}\,
\sigma_{\mathrm{Mott}}\,,\quad \mathrm{where}\quad 
\sigma_{\mathrm{Mott}}=\frac{\alpha^2}{4\,(E_{1}^{\mathrm{lab}})^2}
\frac{{\rm cos}^2(\theta_e^{\mathrm{lab}}/2)}
{{\rm sin}^4(\theta_e^{\mathrm{lab}}/2)}\,.\label{mott}
\end{equation}

The explicit form of the unpolarized cross section in terms 
of structure functions is given below in (\ref{S_0xsection}).
The spin degrees of the final state may be taken as $s$, the total 
spin of the $np$-final state, and $m_s$ its projection on the relative 
$np$-momentum $\vec p_{np}$ in the final $np$-c.m.\ system. Another 
convenient choice are the helicities $\lambda_p$ and $\lambda_n$ of proton 
and neutron, respectively. 

Then the $T$-matrix of (\ref{obs1}) between the initial deuteron state 
$|\lambda_d\rangle$ and the final $np$-scattering state $|m_1 m_2\rangle$, 
both in non-covariant normalization, is given by
\begin{subequations}\label{Tmatrix}
\beqa
T_{m_1 m_2 \lambda \lambda_d}(\theta,\phi)&=&
-\pi\sqrt{2\alpha \,p_{np}E^{\mathrm{c.m.}}E_d^{\mathrm{c.m.}}/M_d}\,
\langle m_1 m_2|J_ \lambda(\vec q\,)|\lambda_d\rangle\label{TmatrixJ}\\
&=&e^{i(\lambda + \lambda_d)\phi}t_{m_1 m_2 \lambda \lambda_d}(\theta)\,,
\label{redtmatrix}
\eeqa
\end{subequations}
where $\lambda=0,\pm 1$, and the spherical angles of the relative 
momentum $\vec p_{np}$ of the final neutron-proton state in the c.m.\ 
system are denoted by $(\theta,\phi)$ as already defined above. 
Here, $J_0(\vec q\,)$ denotes the {\sc Fourier} 
component of the charge density operator and $J_{\pm 1}(\vec q\,)$ 
the {\sc Fourier} components of the transverse current density operator. Furthermore,
$(m_1,m_2)$ stands for the spin quantum numbers of the final two-nucleon
state, either in the standard (coupled) representation $(s,m_s)$ of 
the total spin $s$ of the outgoing nucleons and its projection $m_s$ 
on the relative momentum, or 
in the helicity (uncoupled) basis $(\lambda_p,\lambda_n)$. 
The transformation from one representation to the other is 
simply given by a Clebsch-Gordan coefficient
\begin{equation}
t_{\lambda_p \lambda_n \lambda \lambda_d}=\sum_{s m_s} (-)^{m_s}\hat s
 \left(
\begin{matrix}
\frac{1}{2} & \frac{1}{2} & s \cr
    \lambda_p  & \lambda_n & -m_s \cr\end{matrix}
 \right) t_{s m_s \lambda \lambda_d}\,.
\end{equation}
In Eq.~(\ref{TmatrixJ}) noncovariant state normalization has been assumed 
and the hadronic c.m.\ motion has been eliminated already. Thus initial 
and final hadronic states refer to the relative two-body motion in the 
hadronic rest frame. Eq.~(\ref{redtmatrix}) defines the reduced 
$t$-matrix. If parity is conserved, it obeys the symmetry relation 
\begin{equation}
t_{s -m_s -\lambda -\lambda_d}=(-)^{1+s+ m_s+ \lambda+ \lambda_d}\,
t_{s m_s \lambda \lambda_d}\,,
\label{sym1}
\end{equation}
for the standard representation, and
\begin{equation}
t_{-\lambda_p -\lambda_n -\lambda -\lambda_d}=(-)^{\lambda_p+\lambda_n+ \lambda+\lambda_d}t_{\lambda_p \lambda_n \lambda \lambda_d}\,,
\label{sym2}
\end{equation}
for the helicity representation. This relation reduces the number of 
independent $t$-matrix elements to 18, six for the longitudinal 
($\lambda=0$) and twelve for the transverse ($\lambda=\pm 1$) matrix elements. 

The initial state density matrix $\rho_i$ in (\ref{obs1}) is a direct product 
of the density matrices $\rho^\gamma$ of the virtual photon and $\rho^d$ of 
the deuteron
\begin{equation}
\rho_i=\rho^\gamma\otimes\rho^d \,.
\end{equation}
For the evaluation of $\rho^\gamma$ of the virtual photon, we allow the
incoming electrons to be partially longitudinally polarized of degree $h$.
This restriction does not mean a loss of generality because, as has been 
shown in \cite{ArL93}, one obtains already for this case the maximal 
number of linearly independent observables. The virtual photon density
matrix can be split into an unpolarized and a polarized part
\begin{equation}
\rho^\gamma_{\lambda \lambda'}=
\rho^0_{\lambda \lambda'}+h\rho'_{\lambda \lambda'}\,,
\end{equation}
where $\rho^0$ and $\rho'$ can be expanded in terms of independent components 
$\rho_\alpha$ and $\rho^{\prime}_\alpha$ $(\alpha\in\{L,\,T,\,LT,\,TT\})$ 
according to the various combinations of longitudinal and transverse 
polarization
\begin{subequations}
\beqa
\rho^0_{\lambda \lambda'}&=&\sum_{\alpha=L,\,T,\,LT,\,TT}
\delta^\alpha_{\lambda \lambda'}\rho_\alpha\,,\\
\rho'_{\lambda \lambda'}&=&\sum_{\alpha=L,\,T,\,LT,\,TT}
\delta^{\prime\,\alpha}_{\lambda \lambda'}\rho'_\alpha\,,
\eeqa
\end{subequations}
with
\begin{equation}
\begin{array}{ll}
\delta^L_{\lambda \lambda'}=\delta_{\lambda \lambda'}\delta_{\lambda 0}\,,
&\delta^{LT}_{\lambda \lambda'}=\lambda'\delta_{\lambda 0}+
\lambda\delta_{\lambda' 0}\,,\cr
&\cr
\delta^T_{\lambda \lambda'}=\delta_{\lambda \lambda'}|\lambda|\,,
&\delta^{TT}_{\lambda \lambda'}=\delta_{\lambda,\, -\lambda'}|\lambda|
\,,\cr
&\cr
\delta^{\prime\,L}_{\lambda \lambda'}=0\,,
&\delta^{\prime\,LT}_{\lambda \lambda'}=|\lambda'|\delta_{\lambda 0}+
|\lambda|\delta_{\lambda' 0}\,,\cr
&\cr
\delta^{\prime\,T}_{\lambda \lambda'}=\delta_{\lambda \lambda'}
\lambda\,,
&\delta^{\prime\,TT}_{\lambda \lambda'}=0.\cr
\end{array}
\end{equation}
They obey the symmetries 
\begin{subequations}
\beqa
\delta^{\alpha}_{\lambda' \lambda}&=&\delta^{\alpha}_{\lambda \lambda'}
=(-)^{\lambda+\lambda'}\delta^{\alpha}_{-\lambda' -\lambda}\;,\\
\delta^{\prime\,\alpha}_{\lambda' \lambda}&=&\delta^{\prime\,\alpha}
_{\lambda \lambda'}
=(-)^{1+\lambda+\lambda'}\delta^{\prime\,\alpha}_{-\lambda' -\lambda}\,.
\eeqa
\end{subequations}

The independent components $\rho_\alpha$ and $\rho^{\prime}_\alpha$ 
are given by the well-known expressions (note $Q^2=-q_\nu^2>0$)
\begin{eqnarray}
\begin{array}{ll}
 \rho_L=\rho_{00}^0=\beta^2 Q^2\frac{\xi^2}{2\eta} 
\,,\quad& \rho_T=\rho_{11}^0
  =\frac{1}{2}Q^2\,\Big(1+\frac{\xi}{2 \eta} \Big) \,,\cr
&\cr
 \rho_{LT}=\rho_{01}^0=\beta Q^2 \frac{\xi}{\eta}\,
 \sqrt{\frac{\eta+ \xi}{8}}
\, ,& \rho_{TT}=\rho_{-11}^0=-Q^2\frac{\xi}{4 \eta} \,,\cr
&\cr
 \rho_{LT}^{\prime}=\rho_{01}^{\prime}=
 \frac{1}{2}\,\beta\frac{Q^2}{\sqrt{2\eta}}\,\xi \,,\quad&
 \rho_T^{\prime}=\rho_{11}^{\prime}=
  \frac{1}{2}Q^2\, \sqrt{\frac{\eta+\xi}{\eta}} \, ,\cr
\end{array}
\end{eqnarray}
with
\begin{equation}
\beta = \frac{ q^{\,lab}}{q^{\,c}},\,\,\,\,\,
\xi = \frac{Q^2}{({q}^{\,lab})^{\,2}},\,\,\,\,\,
\eta = {\rm tan}^2(\frac{\theta_e^{\mathrm{lab}}}{2})\;,\label{betaxieta}
\end{equation}
where $\beta$ expresses the boost from the lab system to the frame 
in which the 
hadronic current is evaluated and $\vec q^{\,c}$ denotes the momentum transfer 
in this frame. If, as is the case here, one calculates the observables 
in the final $np$-c.m. system, one has $\vec q^{\,c}=\vec q^{\,c.m.}$. 
We further note the simple relation to the often used parametrization of 
the virtual photon density matrix in terms of the quantities 
$v_{\alpha^{(\prime)}}$ of Ref.~\cite{DoS86} (for $\beta=1$)
\begin{eqnarray}
\rho_\alpha^{(\prime)} &=& \frac{Q^2}{2\eta}\, v_{\alpha^{(\prime)}}\,,
\end{eqnarray}
where $\alpha \in\{ L,\, T,\, LT,\, TT\}$.

Furthermore, the deuteron density matrix $\rho^d$ can be expressed in terms of 
irreducible spin operators $\tau^{[I]}$ with respect to the 
deuteron spin space 
\begin{equation}
\rho_{\lambda_d\, {\lambda_d}'}^d=\frac{1}{3}
\sum_{I\,M}(-)^M\hat{I}\,\langle 1\lambda_d|\tau^{[I]}_M |1\lambda_d' \rangle
P^d_{I-M}\,,
\end{equation}
where $P^d_{00}=1$, and $P^d_{1-M}$ and $P^d_{2-M}$ describe vector 
and tensor polarization components of the deuteron, respectively. 
We use throughout the notation $\hat I=\sqrt{2I+1}$. 
The spin operators are defined by their reduced matrix elements
\begin{equation}
\langle 1||\tau^{[I]}||1 \rangle = \sqrt{3}\,\hat I
\quad\mbox{for}\quad I=0,1,2\,.
\end{equation}
From now on we will assume that the deuteron density matrix is diagonal 
with respect to an orientation axis $\vec d$ having spherical angles 
$(\theta_d,\phi_d)$ with respect to the coordinate system associated with 
the scattering plane in the lab frame (see Fig.~\ref{fig1}). Then
one has with respect to $\vec d$ as quantization axis
\begin{equation}
\rho_{m\,m'}^d=p_m\,\delta_{m\,m'}\,,
\end{equation}
where $p_m$ denotes the probability for finding a deuteron spin projection $m$ 
on the orientation axis. With respect to this axis one has 
$P^d_{I\,M}(\vec d\,)=P^d_I\,\delta_{M,0}$, where the orientation 
parameters $P_I^d$ are related to the $p_m$ by
\beqa
P_I^d&=&\sqrt{3}\,\hat{I}\sum_{m}(-)^{1-m}
\left( 
\begin{matrix}
1&1&I \cr m &-m & 0 \cr
\end{matrix} \right)p_m\nonumber\\
&=& \delta_{I,0} + \sqrt{\frac{3}{2}}(p_1-p_{-1})\,\delta_{I,1} 
+\frac{1}{\sqrt{2}}\,(1-3\,p_0)\,\delta_{I,2}\,.
\eeqa
The polarization components in the chosen lab frame are obtained 
from the $P^d_I$ by a rotation
\beq
P^d_{IM}(\vec z\,)=P^d_Ie^{iM\phi_d}d^I_{M0}(\theta_d)\,,
\eeq
where $d^j_{m m'}$ denotes a small rotation matrix~\cite{Ros57}.
Thus the deuteron density matrix becomes finally
\begin{equation}
\rho_{\lambda_d\, {\lambda_d}'}^d=\frac{1}{\sqrt{3}}(-)^{1-\lambda_d}
\sum_{I\,M}\hat{I}
\left( 
\begin{matrix}
1&1&I \cr \lambda_d'&-\lambda_d&M \cr
\end{matrix} \right) P_I^d
e^{-iM\phi_d}d^I_{M0}(\theta_d)\,. \label{rhod}
\end{equation}
This means, 
the deuteron target is characterized by four parameters, namely the 
vector and tensor polarizations $P_1^d$ and $P_2^d$, respectively,
and by the orientation angles $\theta_d$ and $\phi_d$.  Note that the 
deuteron density matrix undergoes no change in the 
transformation from the lab to the c.m. system, since the boost to the c.m.\ 
system is collinear with the deuteron quantization axis~\cite{Rob74}.

Now we turn to the definition of the operator $\Omega_X$ characterizing
the various observables. 
One has 16 independent observables according to all combinations of the four 
operators $(\mathbb 1_2,\,\vec \sigma)$ in the spin space of each 
of the two nucleons. In detail, if no 
polarization analysis of the outgoing nucleons is performed, one has  
\beqa
\Omega_{1}=\Omega_{00}&=& \sigma_0(p)\otimes \sigma_0(n)\,,
\eeqa
where we have defined $\sigma_0=\mathbb 1_2$. 
If the polarization component $x_i$ of the outgoing proton or neutron, 
respectively, is measured, the corresponding operator is
\beqa
\Omega_{i0}&=&\sigma_{i}(p)\otimes\sigma_0(n)\quad \mbox{or}\quad
\Omega_{0i}=\sigma_0(p)\otimes\sigma_{i}(n) \,.
\eeqa
Finally, the combined measurement of the polarization components $x_i(p)$ 
and $x_j(n)$ of both final particles is represented by 
\beqa
\Omega_{ij}&=& \sigma_{i}(p)\otimes\sigma_{j}(n)\,.
\eeqa
Thus each observable $X$ is represented by a pair 
$X=(\alpha'\alpha)$ with $\alpha',\,\alpha=0,\dots,3$ and related 
to the operator 
$\Omega_{\alpha'\alpha}=\sigma_{\alpha'}(p)\otimes \sigma_{\alpha}(n)$. 
Since the $T$-matrix is calculated in the $np$-c.m.\ system, 
the spin operators of both particles refer to the same reference frame 
with $z$-axis parallel to $\vec p_{np}$ and $y$-axis along 
$\vec q \times \vec p_{np}$, i.e., perpendicular to the reaction plane. Thus 
the polarization components of the proton are chosen according to the Madison 
convention while for the neutron the $y$- and $z$-components 
of $\vec P$ have to be reversed in order to comply with this convention. 
The resulting observables are listed in Table~\ref{tab1} and are divided 
into two sets, called $A$ and $B$, according to their behaviour under a parity
transformation \cite{ArS90}. 

\begin{table}[h]
\caption{Notation for the cartesian components of the spin observables and 
their division into sets $A$ and $B$.}
\begin{ruledtabular}
\begin{tabular}{cccccccccc}
observable & 1 & $x_p$ & $y_p$ & $z_p$ & $x_n$ & $y_n$ & $z_n$ &  &  \\ 
set & $A$ & $B$ & $A$ & $B$ &  $B$ & $A$ & $B$ &  &  \\ 
\colrule
observable & $x_px_n$ & $x_py_n$ & $x_pz_n$ & $y_px_n$ & $y_py_n$ & 
$y_pz_n$ & $z_px_n$ & $z_py_n$ & $z_pz_n$ \\
set & $A$ & $B$ & $A$ & $B$ & $A$ & $B$ & $A$ & $B$ & $A$ \\
\end{tabular}
\end{ruledtabular}
\label{tab1}
\end{table}

For real photons, the photon density matrix contains transverse components
only, and thus, in order to obtain the corresponding observables in 
photodisintegration, one has to make in (\ref{obs1}) the replacements 
\beqa
\begin{array}{ccc}
\rho_L \rightarrow 0,\,&\rho_{LT} \rightarrow 0,\,&
\rho'_{LT} \rightarrow 0,\,\cr
c(k_1^{\mathrm{lab}},k^{\mathrm{lab}}_2)\,\rho_T \rightarrow 1/6,\,&
h\,c(k_1^{\mathrm{lab}},k^{\mathrm{lab}}_2)\,\rho'_T \rightarrow 
-P^\gamma_c/6,\,& 
\,c(k_1^{\mathrm{lab}},k^{\mathrm{lab}}_2)\,\rho_{TT} \rightarrow 
P^\gamma_l/6\,,
\end{array}
\eeqa
where $P^\gamma_l$ and $P^\gamma_c$ denote the degree of linear and circular
photon polarization, respectively, and $P^\gamma_l< 0$ means linear 
polarization along the $x$-axis while along the $y$-axis for $P^\gamma_l> 0$.
Furthermore, $P^\gamma_c> 0$ or $P^\gamma_c< 0$ describe right or left 
handed circular polarization, respectively.


\subsection{Structure functions}

For each observable $X$ a set of structure functions is defined
as quadratic hermitean forms of the $t$-matrix elements by 
\begin{subequations}
\label{strucfunall}
\begin{eqnarray}
f_{L}^{IM}(X)&=&\frac{2}{1+\delta_{M0}}\Re e\left(i^{\bar \delta^X_I}
{\cal U}^{00 I M}_{X}\right)\,,\label{strucfunL}\\
f_{T}^{IM}(X)&=&\frac{4}{1+\delta_{M0}}\Re e\left(i^{\bar \delta^X_I}
{\cal U}^{11 I M}_{X}\right)\,,\label{strucfunT}\\
f_{LT}^{IM\pm}(X)&=&\frac{4}{1+\delta_{M0}}\Re e\left[i^{\bar \delta^X_I}
\left({\cal U}^{01 I M}_{X}\pm(-)^{I+M+\delta_{X,\,B}}
{\cal U}^{01 I -M}_{X}\right)\right]\,,\label{strucfunLT}\\
f_{TT}^{IM\pm}(X)&=&\frac{2}{1+\delta_{M0}}\Re e\left[i^{\bar \delta^X_I}
\left({\cal U}^{-11 I M}_{X}\pm(-)^{I+M+\delta_{X,\,B}}
{\cal U}^{-11 I -M}_{X}\right)\right]\,,\label{strucfunTT}\\
f_{T}^{\prime\, IM}(X)&=&\frac{4}{1+\delta_{M0}}\Re e\left(i^{1+\bar \delta^X_I}
{\cal U}^{11 I M}_{X}\right)\,,\label{strucfunTs}\\
f_{LT}^{\prime\, IM\pm}(X)&=&\frac{4}{1+\delta_{M0}}\Re e\left[i^{1+\bar \delta^X_I}
\left({\cal U}^{01 I M}_{X}\pm(-)^{I+M+\delta_{X,\,B}}
{\cal U}^{01 I -M}_{X}\right)\right]\,.\label{strucfunLTs}
\end{eqnarray}
\end{subequations}
Here $\bar \delta_I^X$ is defined by
\begin{equation}
\bar \delta_I^X= (\delta_{X,B}-\delta_{I1})^2, \mbox{ with }
\delta_{X,B}:=\left\{
\begin{matrix}
0 & \mbox{for}\; X\in A\cr 1 & 
\mbox{for}\; X\in B\cr\end{matrix} \right\},
\end{equation}
distinguishing the two sets of observables $A$ and $B$. In the 
foregoing expressions, the ${\cal U}$'s are given as bilinear hermitean 
forms in the reduced $t$-matrix elements, i.e., for $X=(\alpha'\alpha)$
\begin{equation}
{\cal U}_{\alpha'\,\alpha}^{\lambda' \lambda I M}= 
\sum_{m_1'm_2'\lambda_d' m_1 m_2 \lambda_d}
t^*_{m_1'm_2'\lambda'\lambda_d'}\langle m_1'm_2'|
\sigma_{\alpha'}(p)\sigma_{\alpha}(n)|m_1 m_2\rangle
t_{m_1 m_2 \lambda \lambda_d}\langle \lambda_d|\tau^{[I]}_M|\lambda_d'\rangle
\,.\label{ulamcart}
\end{equation}
Although the ${\cal U}$'s are independent of the chosen representation 
for the matrix elements, their explicit form in terms of the $t$-matrix 
elements depends certainly on the representation for the initial and final 
spin states. We have already mentioned that two conventions are in 
common use, the helicity representation
with spin quantum numbers $(\lambda_p,\, \lambda_n)$ and the standard one with 
$(s,\, m_s)$. 
A third representation called hybrid basis, where the quantization axis 
is chosen perpendicular to the reaction plane, was introduced in \cite{DmG89}. 
Explicit expressions are 
listed in the Appendix~\ref{Uexplicit} for the $(s,m_s)$-representation. 
More general representations, which are obtained by arbitrary rotations 
of the quantization axes of initial and final spin states, are considered 
in~\cite{ArL00}. However one should keep in mind that the observables 
and thus the structure functions are independent of the representation 
because they are defined as traces over the spin degrees of freedom 
(see (\ref{obs1})).

Note that 
$f_\alpha^{00-}(X)$, $f_\alpha^{20-}(X)$ and 
$f_\alpha^{10+}(X)$ vanish identically for $X\in A$ and correspondingly 
$f_\alpha^{00+}(X)$, $f_\alpha^{20+}(X)$ and 
$f_\alpha^{10-}(X)$ for $X\in B$.
For this reason we often use the notation $f_\alpha(X)$, $f_\alpha^{10}(X)$ and
$f_\alpha^{20}(X)$ instead of $f_\alpha^{00\pm}(X)$, $f_\alpha^{10\mp}(X)$
and $f_\alpha^{20\pm}(X)$, respectively.

The structure functions $f^{(\prime)\,IM(\pm)}_\alpha(X)$ 
($\alpha\in\{L,\,T,\,LT,\,TT\}$) (primed and unprimed structure functions 
$f^{\prime \,IM(\pm)}_{\alpha}(X)$ and $f^{IM(\pm)}_{\alpha}(X)$ 
are here referred to collectively as $f^{(\prime )IM(\pm)}_{\alpha}(X)$) 
contain the complete information on the
dynamical properties of the $np$-system available in deuteron 
electrodisintegration. They are functions of the $np$-angle $\theta$,
the relative $np$-energy $E_{np}$, and the
three-momentum transfer squared $( q ^{\mathrm{c.m.}})^{2}$, 
all in the c.m.\ system.

In terms of these structure functions a general observable in $d(e,e'N)N$ 
and $d(e,e'np)$ is given by
\beqa
{\cal O}(\Omega_X)
=c(k_1^{\mathrm{lab}},\,k_2^{\mathrm{lab}})\,
 \sum _{I=0}^2 P_I^d \sum _{M=0}^I
 &\Bigl\{  &\Big(\rho _L f_L^{IM}(X) + \rho_T f_T^{IM}(X) + 
\rho_{LT} {f}_{LT}^{IM+}(X) \cos \phi \nonumber\\
&&+  \rho _{TT} {f}_{TT}^{IM+}(X) \cos2 \phi
\Big)\cos (M\tilde{\phi}-\bar\delta_{I}^{X} \frac{\pi}{2}) \nonumber\\
&&- \Big(\rho_{LT} {f}_{LT}^{IM-}(X) \sin \phi 
+ \rho _{TT} {f}_{TT}^{IM-}(X) \sin2 \phi\Big) 
\sin (M\tilde{\phi}-\bar\delta_{I}^{X} \frac{\pi}{2})\nonumber\\
&&+  h \Big[ \Big(\rho'_T f_T^{\prime IM}(X) 
+ \rho '_{LT} {f}_{LT}^{\prime IM-}(X) \cos \phi \Big) 
\sin (M\tilde{\phi}-\bar\delta_{I}^{X} \frac{\pi}{2}) \nonumber\\
&&+  \rho '_{LT} {f}_{LT}^{\prime IM+}(X) \sin \phi 
\cos (M\tilde{\phi}-\bar\delta_{I}^{X} \frac{\pi}{2})\Big] 
\Big\} d_{M0}^I(\theta_d)\,,\label{obsfin}
\eeqa
where we have introduced $\tilde \phi = \phi - \phi_d$. In particular, one 
obtains for $X=1$ and $P^d_I=\delta_{I,0}$ the unpolarized cross section as
\begin{equation}
S_0= c(k_1^{\mathrm{lab}},\,k_2^{\mathrm{lab}})\,
(\rho _L f_L + \rho_T f_T + \rho_{LT} {f}_{LT} \cos \phi
+\rho _{TT} {f}_{TT} \cos2 \phi)\,,\label{S_0xsection}
\end{equation}
using as a shorthand $f_{\alpha}=f_{\alpha}^{00+}(1)$. One should remember 
that the nucleon angles and polarization components 
refer to the c.m.\ frame. The transformation to the lab frame is  
described in the previous subsection. 
 
\begin{table}[h]
\caption{Listing of ($IM$)- and ($I M \pm$)-values of nonvanishing structure 
functions $f^{(\prime)\,IM}_{\alpha}(X)$ and 
$f^{(\prime)\,IM \pm}_{\alpha}(X)$, respectively.}
\begin{ruledtabular}
\begin{tabular}{cccccccc}
\multicolumn{2}{c}{$f_L^{IM}(X),\,f_T^{IM}(X)$} &
\multicolumn{2}{c}{$f^{\prime IM}_T(X)$} &
\multicolumn{4}{c}{$f_{LT}^{IM\pm}(X),\,f_{TT}^{IM\pm}(X),
\,f_{LT}^{\prime IM\pm}(X)$} \\ \colrule
 $X\in A$ & $X\in B$ & $X\in A$ & $X\in B$ & \multicolumn{2}{c}{$X\in A$} & \multicolumn{2}{c}{$X\in B$} 
\\ \colrule
$00$ & $10$ & $10$ & $00$ & $00+$ & $10-$ & $10+$ & $00-$ \\
$11$ & $11$ & $11$ & $11$ & $11+$ & $11-$ & $11+$ & $11-$ \\
$20$ & $21$ & $21$ & $20$ & $20+$ & $21-$ & $21+$ & $20-$ \\
$21$ & $22$ & $22$ & $21$ & $21+$ & $22-$ & $22+$ & $21-$ \\
$22$ &      &      & $22$ & $22+$ &       &       & $22-$ \\ 
\end{tabular}
\end{ruledtabular}
\label{tab4}
\end{table}
\begin{table}[h]
\caption{Number of nonvanishing structure functions 
$f^{(\prime)\,IM}_\alpha(X)$ and 
$f^{(\prime)\,IM \pm}_{\alpha}(X)$ for an observable $X\in A$ or $X\in B$.}
\begin{ruledtabular}
\begin{tabular}{cccccccc}
set & $L$ & $T$ & $T'$ & $LT$ & $LT'$ & $TT$ & total \\ \colrule
$A$ & 5 & 5 & 4 & 9 & 9 & 9 & 41 \\
$B$ & 4 & 4 & 5 & 9 & 9 & 9 & 40 \\ 
\end{tabular}
\end{ruledtabular}
\label{tab5}
\end{table}

The possible $(I,\,M)$-values are listed in Table~\ref{tab4} and the total 
number of structure functions for each observable $X$ and each $\alpha$ 
are listed in Table~\ref{tab5}. As mentioned in the introduction, 
one finds altogether a total number of 648 observables, each of which is 
a hermitean form of the $t$-matrix elements. However, since the 
$t$-matrix has only $n=18$ independent complex amplitudes, only $n^2=324$ 
linearly independent hermitean forms can exist. Indeed, one finds 
$n^2$ linear relations 
between the observables which are presented in the next section. The 
remaining structure functions are linearly independent so that 
indeed the maximal information can be obtained by using longitudinally 
polarized electrons alone. Transverse polarization is not necessary.

On the other hand, since each reaction matrix element is in general a complex 
number, but one overall phase is undetermined, a set of $2n-1$ properly 
chosen observables should suffice to determine completely all 
matrix elements. This seeming contradiction is resolved by the 
observation, that the linearly independent observables are 
not completely independent of each other in a more general sense. In fact, 
any bilinear form $t_{j'}^\ast t_j$ can be expressed as a linear form
in the observables (see~\cite{ArL98} and also Sec.~\ref{linrel}), and 
for these bilinear forms one can find 
exactly $(n-1)^2$ quadratic relations (see Appendix~\ref{quadrel}), thus 
reducing the total number of independent observables just to the required 
number $2n-1$. Consequently, one can determine all matrix elements from
$2n-1$ properly chosen observables. However, one should keep in mind that 
the solution is in general not unique but contains discrete ambiguities. 
This is discussed in Sect.~\ref{completesets}.

To close this section, we will give for the transverse structure functions 
($\alpha\in\{T,\,T',\,TT\}$) the correspondence to the observables in 
photodisintegration derived in \cite{ArS90}. The formal definition of 
observables is completely analogous except for the fact that in 
photodisintegration only transverse current components contribute. 
Taking into account the slightly different definition of the $T$-matrix 
(compare $T$ of (\ref{TmatrixJ}) with the definition of $T^\gamma$ 
in~\cite{Par64,ArS91}), i.e.
\begin{equation}
T_{m_1 m_2 \lambda \lambda_d}(\theta,\phi)=\sqrt{\frac{W_{np}\,q^{c.m.}}{M_d}}
T_{m_1 m_2 \lambda \lambda_d}^{\gamma}(\theta,\phi)\,,\label{tmatrixratio}
\end{equation}
one has the following relations at the photon point 
with respect to the general 
form of an observable in photodisintgration as given in (14) and (15) 
of \cite{ArS90}
\begin{subequations}\label{equivobs}
\beqa
f_T^{IM}(X)&=&(-)^{\bar\delta_I^X}\,\frac{6\,W_{np}\,q^{c.m.}}{M_d}\,P^{0,\,IM}_X
\,\frac{d\sigma}{d\Omega_{np}}
\,,\label{equiv1}\\
f_T^{\prime\,IM}(X)&=
&-\frac{6\,W_{np}\,q^{c.m.}}{M_d}\,P^{c,\,IM}_X\,\frac{d\sigma}{d\Omega_{np}}
\,,\\
f_{TT}^{IM\pm}(X)&=&\mp(-)^{\bar\delta_I^X}\,\frac{W_{np}\,q^{c.m.}}{M_d}\,
\frac{6}{1+\delta_{M0}}\,
\left(P^{l,\,IM}_X\pm(-)^{I+M+\delta_{X,B}}P^{l,\,I-M}_X\right)
\frac{d\sigma}{d\Omega_{np}}
\,.\label{equiv4}
\eeqa
\end{subequations}


\subsection{Linear relations between structure functions}\label{linrel}

As is shown in detail in \cite{ArL93}, the derivation of linear relations 
among observables is based on the inversion of (\ref{ulamcart}) 
expressing any bilinear form $t_{s'm_s'\lambda' m'}^*t_{s m_s\lambda m}$ 
as a linear superposition of observables. This inversion 
can be done analytically (see~\cite{ArL93}). In general one obtains 
two types of relations among the structure functions of an observable 
$X$ and those of another observable $X'(X)$, uniquely related to $X$. 
Explicitly, one finds as the first type of equations 
\begin{subequations}
\beqa
g^{00}(X)&=&\frac{1}{3}p(X)\Big(\bar g^{00}(X')-\sqrt{2}\bar g^{20}(X')
-\sqrt{3}\bar g^{22}(X')\Big)\,,\\
g^{11}(X)&=&p(X)\bar g^{11}(X')\,,\\
g^{20}(X)&=&\frac{1}{3}p(X)\Big(-\sqrt{2}\bar g^{00}(X')+2\bar g^{20}(X')
-\sqrt{\frac{3}{2}}\bar g^{22}(X')\Big)\,,\\
g^{21}(X)&=&p(X)\bar g^{21}(X')\,,\\
g^{22}(X)&=&-\frac{2}{\sqrt{3}}p(X)\Big(\bar g^{00}(X')+\frac{1}{\sqrt{2}}
\bar g^{20}(X')\Big)\,.
\eeqa\label{grelation}
\end{subequations}
The second type of equations reads 
\begin{subequations}
\beqa
h^{10}(X)&=&-\frac{1}{\sqrt{2}}p(X)\bar h^{22}(X')\,,\\
h^{11}(X)&=&p(X)\bar h^{21}(X')\,,\\
h^{21}(X)&=&-p(X)\bar h^{11}(X')\,,\\
h^{22}(X)&=&\sqrt{2}p(X)\bar h^{10}(X'))\,, 
\eeqa\label{hrelation}
\end{subequations}
where $X'(X)$ and $p(X)$ are listed in Table~\ref{tab6} for 8 observables. 
For the remaining other 8 observables, not listed in Table~\ref{tab6}, one 
obtains $X'(X)$ from Table~\ref{tab6} with the help of the relation
\begin{equation}
X'(X'(X))=X\,,\label{xprime}
\end{equation}
and $p(X)$ from the relation
\begin{equation}
p(X'(X))=(-)^{\delta_{X,B}}\,p(X)\,.\label{pxprime}
\end{equation}
Table~\ref{tab7} shows which of these two types of 
relations holds for a specific 
structure function, depending on whether $X$ belongs to an observable of 
set $A$ or $B$. Which structure functions are related to each other in 
these relations, i.e., $ g^{IM} (X)$ to $\bar g^{I'M'}(X')$ and 
$ h^{IM} (X)$ to $\bar h^{I'M'} (X')$, is also listed in Table~\ref{tab7}. 

\begin{table}
\caption{Definition of $X'(X)$ and $p(X)$.}
\begin{ruledtabular}
\begin{tabular}{ccccccccc}
$X$ & $1$ & $x_px_n$ & $x_pz_n$ & $y_p$ & $x_p$ & $x_n$ & $z_p$ & $z_n$ \\ 
\colrule
$X'(X)$ & $y_py_n$ & $z_pz_n$ & $z_px_n$ & $y_n$ & $z_py_n$ & $y_pz_n$ & 
$x_py_n$ & $y_px_n$ \\ 
$p(X)$ & 1 & $-1$ & 1 & 1 & $-1$ & $-1$ & 1 & 1 \\ \colrule
set & $A$ & $A$ & $A$ & $A$ & $B$ & $B$ & $B$ & $B$ \\ 
\end{tabular}
\end{ruledtabular}
\label{tab6}
\end{table}
\begin{table}
\caption{Listing of structure functions $g^{IM}(X)$ and $\bar g^{I'M'}(X')$ 
for observables $X,X'$ which fulfill the relations 
(\protect{\ref{grelation}}), and of structure functions $h^{IM}(X)$ and 
$\bar h^{I'M'}(X')$ for observables which fulfill the relations 
(\protect{\ref{hrelation}}). The associated observable $X'(X)$ is either 
listed in Table~\protect{\ref{tab6}} or
can be obtained using (\protect\ref{xprime}).}
\begin{ruledtabular}
\begin{tabular}{cccccc}
& \multicolumn{5}{c}{$X \in A\, [B]$} \\ \colrule
$g\,[\bar h]^{IM}(X)$ & $f^{IM}_L$ & $f^{IM+}_{LT}$ & $f^{\prime \,IM+}_{LT}$ 
& $f^{IM}_T$ & $f^{IM+}_{TT}$ \\ 
$\bar g\,[\bar h]^{I'M'}(X')$ & $f^{I'M'}_L$ & $f^{I'M'+}_{LT}$ & 
$f^{\prime \,I'M'+}_{LT}$ & $-f^{I'M'+}_{TT}$ & $-f^{I'M'}_{T}$ \\ 
\colrule
\colrule
& \multicolumn{5}{c}{$X \in B\, [A]$} \\ \colrule
$g\,[h]^{IM}(X)$ & $f^{IM-}_{LT}$ & $f^{\prime \,IM-}_{LT}$ 
& $f^{\prime\,IM}_T$ & $f^{IM-}_{TT}$ & \\ 
$\bar g\,[\bar h]^{I'M'}(X')$ & $f^{\prime\,I'M'-}_{LT}$ & 
$-f^{I'M'-}_{LT}$ & $f^{I'M'-}_{TT}$ & 
$-f^{\prime\,I'M'}_{T}$ & \\ 
\end{tabular}
\end{ruledtabular}
\label{tab7}
\end{table}

At the end of this section, 
we will give two examples of how to find the proper relation for a 
given structure function of an observable $X$. As first example we choose 
the $y$-component of the neutron polarization, i.e.\  $X=y_n$ 
belonging to set $A$. According to Table~\ref{tab6} its structure 
functions are related to the ones of the $y$-component of the proton 
polarization. With the help of (\ref{xprime}) and (\ref{pxprime}) 
one finds $X'(y_n)=y_p$ and $p(y_n)=1$ and in view of Table~\ref{tab7}
the relations (\ref{grelation}) apply, e.g.\
\beqa
f_L^{00}(y_n)&=&\frac{1}{3}\Big(f_L^{00}(y_p)-\sqrt{2} f_L^{20}(y_p)
-\sqrt{3}f_L^{22}(y_p)\Big)\,,\\
f_{T}^{00}(y_n)&=&-\frac{1}{3}\Big(f_{TT}^{00}(y_p)-\sqrt{2} f_{TT}^{20}(y_p)
-\sqrt{3}f_{TT}^{22}(y_p)\Big)\,.
\eeqa
For the second example we choose $X=z_p\,y_n$, 
belonging to set $B$, and the structure function $f_L^{10}(z_p\,y_n)$. 
From Table~\ref{tab6} with the help of (\ref{xprime}) and (\ref{pxprime}) 
one finds $X'(z_p\,y_n)=x_p$ and $p(z_p\,y_n)=1$. Furthermore, according to 
Table~\ref{tab7} the relation (\ref{hrelation}) applies, resulting, for
example, in
\begin{equation}
f_L^{10}(z_p\,y_n)= -\frac{1}{\sqrt{2}}\,f_L^{22}(x_p)\,.
\end{equation}

 
\subsection{Multipole decomposition}

A convenient parametrization of the angular dependence of observables and 
structure functions is provided by an expansion in terms of the small rotation
matrices $d^j_{m'm}$~~\cite{Are88,Kaw58,CaM82,RaD89}. Explicit expressions
for deuteron electrodisintegration have been derived recently 
in~\cite{ArL02}. They facilitate the analysis of the contributions of the 
various charge, electric, and magnetic transition multipole moments to the 
different structure functions. This expansion is based on the multipole 
expansion of the $t$-matrix. We take the outgoing $np$-state in the form of 
the Blatt-Biedenharn convention~\cite{BlB52}
\begin{equation}
|\vec p\, s\,m_s\rangle^{(-)}=\sum_{\mu j m_j l}\hat l\,(l0sm_s|jm_s)\,
e^{-i\delta^j_\mu}\,U^j_{ls\mu}\,D^j_{m_j m_s}(R)\,|\mu j m_j\rangle\,,
\end{equation}
where $D^j_{m_j m_s}(R)$ denote the rotation matrices in the
convention of Rose~\cite{Ros57} and 
$\mu=1,\dots,4$ numbers the four possible partial waves for a
given total angular momentum $j>0$. For $j=0$ one has only two partial waves. 
The phase shifts are denoted by $\delta^j_\mu$, and the matrix $U^j_{ls\mu}$ 
is determined by the mixing parameters $\epsilon_j$ as listed in 
Table~\ref{tabU}. 

\begin{table}
\caption{Listing of the matrix $U^j_{ls\mu}$.}
\begin{ruledtabular}
\begin{tabular}{cccccc}
$l$ & $s$ & $\mu=1$ & 2 & 3 & 4 \\
\colrule
$j-1$ & 1 & $\cos{\epsilon_j}$ & 0 & $-\sin{\epsilon_j}$ & 0\\
$j$ & 0 & 0 & 1 & 0 & 0\\
$j+1$ & 1 & $\sin{\epsilon_j}$ & 0 & $\cos{\epsilon_j}$ & 0 \\
$j$ & 1 & 0 & 0 & 0 & 1\\
\end{tabular}
\end{ruledtabular}
\label{tabU}
\end{table}

Furthermore, $R=(0,-\theta,-\phi)$ rotates the
chosen quantization axis into the direction of the relative
$np$-momentum $\vec p$. The partial waves 
\begin{equation}
|\mu j m_j\rangle = \sum_{l's'}U^j_{l's'\mu}\,|\mu (l's') j m_j\rangle
\end{equation}
are solutions of a system of coupled equations of $NN$-scattering. 
In this convention, the $t$-matrix reads 
\beqa
t_{sm_s \lambda \lambda_d}(\theta)
&=& (-)^\lambda \sqrt{1 + \delta_{\lambda 0} } 
\,\sum_{L l j m_j \mu}\frac{\hat l}{\hat \jmath} (1 \lambda_d L \lambda |j m_j)
(l 0 s m_s | j m_s) {\cal O}^{L\lambda} (\mu j l s)\, d^j_{m_j m_s}
(\theta)\,, \label{tmatrixmult}
\eeqa
with
\begin{equation}
{\cal O}^{L\lambda} (\mu j l s) = \sqrt {4\pi}\,e^{i\delta_\mu^j} U^j_{l s
,\mu}N^L_\lambda(\mu j)\,,\label{Olmj}
\end{equation}
and
\begin{equation}
N^L_\lambda(\mu j)= \delta_{|\lambda| 1}
\Big(E^L(\mu j) + \lambda M^L(\mu j)\Big)
+ \delta_{\lambda 0} C^L(\mu j)\,,
\end{equation}
where $E^L(\mu j)$, $M^L(\mu j)$ and $C^L(\mu j)$ denote the reduced 
electric, magnetic and charge multipole matrix elements, respectively, 
between the deuteron state and a final 
state partial wave $|\mu j\rangle$ in the Blatt-Biedenharn parametrization.
Parity conservation implies the selection rules
\begin{subequations}
\beqa
(C/E)^L(\mu j)&=& 0\quad \mbox{for}\quad (-)^{L+j+\mu}=-1\,,\\
M^L(\mu j)&=& 0\quad \mbox{for}\quad (-)^{L+j+\mu}=1\,,
\eeqa
\end{subequations}
which leads to the relation
\begin{equation}
{\cal O}^{L-\lambda} (\mu j l s) = (-)^{L+l}
{\cal O}^{L\lambda} (\mu j l s) \,. \label{symO}
\end{equation}
As is shown in detail in~\cite{ArL02}, one obtains for a structure 
function the general multipole expansion according to the expressions 
in (\ref{strucfunall}) from the one of 
${\cal U}_{\alpha'\,\alpha}^{\lambda' \lambda I M}$ in (\ref{ulamcart})
which reads 
\begin{equation}
{\cal U}_{\alpha'\,\alpha}^{\lambda' \lambda I M}= 
\sum_{K,\,\kappa\in \kappa_X}
{\cal U}_{\alpha'\,\alpha}^{\lambda' \lambda I M,\,K\kappa}\,
d^K_{\lambda' -\lambda-M,\kappa}(\theta)\,,\label{Umultipole}
\end{equation}
where $d^j_{m'm}(\theta)$ denotes the small $d$-function of the rotation 
matrices. 

\begin{table}
\caption{Listing of the sets $\kappa_X$ determining the summation values 
$\kappa$ in the multipole decomposition (\ref{fmultipole}) of a structure 
function for an observable $X=(\alpha'\alpha)$.}
\begin{ruledtabular}
\begin{tabular}{c|cccc|cccccccc|cccc}
$\alpha'$ & 0&3&0&3 & 1&0&2&0&3&1&3&2 & 1&2&1&2 \\
$\alpha$  & 0&0&3&3 & 0&1&0&2&1&3&2&3 & 1&1&2&2 \\
\hline
$\kappa_X$ & \multicolumn{4}{c|}{$\{0\}$} & \multicolumn{8}{c|}{$\{-1,\,1\}$} 
& \multicolumn{4}{c}{$\{-2,\,0,\,2\}$} \\
\end{tabular}
\end{ruledtabular}
\label{kappa_X}
\end{table}

The sets $\kappa_X$ of 
the possible $\kappa$-values are listed in Table~\ref{kappa_X} and the 
coefficients are given by
\begin{equation}
{\cal U}_{\alpha'\,\alpha}^{\lambda' \lambda I M,\,K\kappa}=
4\pi\,i^{\delta_{(\alpha',\alpha)}^{(2)}}\,\sum_{L' L \mu' j' \mu j} 
{\cal C}^{\lambda' \lambda I M K}(L' j' L j)\,
\widetilde{{\cal D}}_{\alpha'\alpha}^{K\kappa}(\mu' j' \mu j)\,
\widetilde N^{L'\ast}_{\lambda'}(\mu' j')\,\widetilde N^L_\lambda(\mu j)\,,
\label{Umultcoeff}
\end{equation}
where we have defined 
\begin{subequations}
\beqa
\delta_{(\alpha',\alpha)}^{(k)}&=& \delta_{\alpha',k}+\delta_{\alpha,k}\,,
\eeqa
and
\beqa
{\cal C}^{\lambda' \lambda I M,K}(L' j' L j)&=&
(-)^{\lambda'+L}\,2\,
\sqrt{3\,(1 + \delta_{ \lambda'0})(1 + \delta_{ \lambda0})}\,
\hat \jmath' \,\hat \jmath\,\hat I\, {\hat K}^2\,
\nonumber\\& &
\sum_{J}{\hat J}^2 \,
\left(
\begin{matrix}
J&I&K \cr \lambda -\lambda'&M& \lambda'- \lambda-M\cr
\end{matrix}\right)
\left(\begin{matrix}
L'&L& J \cr \lambda' &-\lambda&\lambda-\lambda' \cr\end{matrix}\right) 
\left\{ 
\begin{matrix}
j'&j&K \cr L'&L&J \cr 1&1&I \cr\end{matrix} \right\}\,,\label{CJ}\\
\widetilde{{\cal D}}_{\alpha'\alpha}^{K\kappa}(\mu' j' \mu j)&=&
(-i)^{\delta_{(\alpha',\alpha)}^{(2)}}
\sum_{l' s' l s}{{\cal D}}_{\alpha'\alpha}^{K\kappa}(j' l' s'j l s)\,
U^{j'}_{l' s',\mu'}\,U^j_{l s,\mu}\,.\label{tildeD}
\eeqa
with
\beqa
{\cal D}_{\alpha'\alpha}^{K\kappa}(j' l' s'j l s)
&=&
(-)^{l+s'+s}
\, \hat l' \,\hat l\, \hat s' \,\hat s  
\sum_{\tau'\nu'\tau\nu}(-)^{\tau'+\tau}\,\hat \tau'\,\hat \tau\,
s_{\alpha'}^{\tau'\nu'}s_{\alpha}^{\tau\nu}\,
\Bigg[\sum_{S}{\hat S}^2\,
\left(\begin{matrix}
\tau' & \tau & S \cr \nu' & \nu & -\kappa \cr\end{matrix}\right)
\left\{ \begin{matrix}
\frac{1}{2}&\frac{1}{2}&\tau'\cr 
\frac{1}{2}&\frac{1}{2}&\tau \cr s'&s &S \cr\end{matrix} \right\}
\nonumber\\
& &\Big[\sum_{K'}{\hat K}^{\prime\,2}\,
\left(\begin{matrix}
 S & K & K' \cr \kappa & -\kappa & 0 \cr\end{matrix}\right)
\left( \begin{matrix}
K' & l & l'\cr 0 & 0 & 0\cr\end{matrix} \right)
\left\{ \begin{matrix}
S&K&K' \cr s&j&l \cr s'&j'&l' \cr\end{matrix} \right\}\Big]\Bigg]\,.
\eeqa
\end{subequations}
The definition of $s_{\alpha}^{\tau\nu}$ is given in (\ref{s_alpha}) of 
Appendix~\ref{Uexplicit}. Furthermore, in~(\ref{Umultcoeff}) we have 
incorporated the phase shift for convenience into the quantity
$\widetilde N^L_\lambda(\mu j)=e^{i\delta_\mu^j}\,N^L_\lambda(\mu j)$. 

\begin{table}
\caption{Listing of the values of $\beta(\alpha)$ in the multipole 
decomposition (\ref{fmultipole}).}
\begin{ruledtabular}
\begin{tabular}{cccc}
$\alpha$ & $L/T$ & $LT$ & $TT$ \\
\hline
$\beta(\alpha)$ & 0 & 1 & 2 \\
\end{tabular}
\end{ruledtabular}
\label{beta_alpha}
\end{table}

Then the general multipole decomposition reads 
\begin{equation}
f^{(\prime)\,IM(\pm)}_\alpha(X) = \sum_{K,\,\kappa\in \kappa_X}
f^{(\prime)\,IM(\pm),\,K\,\kappa}_\alpha(X)\,
d^K_{-M-\beta(\alpha),\kappa}(\theta)\,,\label{fmultipole}
\end{equation}
where $\beta(\alpha)$ is listed in 
Table~\ref{beta_alpha} and the coefficients 
$f^{(\prime)\,IM(\pm),\,K\,\kappa}_\alpha(X)$ are obtained via 
(\ref{strucfunall}) from the foregoing multipole expansion.
In detail one has for the longitudinal and transverse structure 
functions of an observable $X$ 
\begin{subequations}
\label{multstrucfundiag}
\beqa
f_{L}^{IM,\,K\kappa}(X)&=&
\sum_{L' \mu' j' L \mu j} 
\widetilde {\cal C}_{L}^{\,I M,K}(L' j' L j)\,
\widetilde{{\cal D}}_{\alpha'\alpha}^{K\kappa}(\mu' j' \mu j)\,
\Re e\Big(i^{\bar \delta^X_I+\delta_{(\alpha',\alpha)}^{(2)}}
\widetilde C^{L'\ast}(\mu' j')\,\widetilde C^L(\mu j)\Big)\,,
\label{smulttrucfunL}\\
f_{T}^{IM,\,K\kappa}(X)&=&
\sum_{L' \mu' j' L \mu j} 
\widetilde {\cal C}_{T}^{\,I M,K}(L' j' L j)\,
\widetilde{{\cal D}}_{\alpha'\alpha}^{K\kappa}(\mu' j' \mu j)\,
\Re e\Big(i^{\bar \delta^X_I+\delta_{(\alpha',\alpha)}^{(2)}}
\widetilde N^{L'\ast}_{1}(\mu' j')\,\widetilde N^L_1(\mu j)
\Big)\,,\label{smulttrucfunT}\\
f_{T}^{\prime\,IM,\,K\kappa}(X)&=&
-\sum_{L' \mu' j' L \mu j} 
\widetilde {\cal C}_{T}^{\,I M,K}(L' j' L j)\,
\widetilde{{\cal D}}_{\alpha'\alpha}^{K\kappa}(\mu' j' \mu j)\,
\Im m\Big(i^{\bar \delta^X_I+\delta_{(\alpha',\alpha)}^{(2)}}
\widetilde N^{L'\ast}_{1}(\mu' j')\,\widetilde N^L_1(\mu j)
\Big)\,,\label{smulttrucfunTprime}
\eeqa
\end{subequations}
and for the interference ones, distinguishing observables of type $A$
\begin{subequations}
\label{multstrucfunintA}
\beqa
f_{TT}^{IM\pm,\,K\kappa}(X)&=&
\sum_{L' \mu' j' L \mu j} 
\widetilde {\cal C}_{TT}^{\,I M\pm,K}(L' j' L j)\,
\widetilde{{\cal D}}_{\alpha'\alpha}^{K\kappa}(\mu' j' \mu j)\,
\Re e\Big(i^{\bar \delta^X_I+\delta_{(\alpha',\alpha)}^{(2)}}
\widetilde N^{L'\ast}_{-1}(\mu' j')\,\widetilde N^L_1(\mu j)
\Big)\,,\label{multstrucfunTTA}\\
f_{LT}^{IM\pm,\,K\kappa}(X)&=&
\sum_{L' \mu' j' L \mu j} 
\widetilde {\cal C}_{LT}^{\,I M\pm,K}(L' j' L j)\,
\widetilde{{\cal D}}_{\alpha'\alpha}^{K\kappa}(\mu' j' \mu j)\,
\Re e\Big(i^{\bar \delta^X_I+\delta_{(\alpha',\alpha)}^{(2)}}
\widetilde C^{L'\ast}(\mu' j')\,\widetilde N^L_1(\mu j)\Big)
\,,\label{multstrucfunLTA}\\
f_{LT}^{\prime\, IM\pm,\,K\kappa}(X)&=&
-\sum_{L' \mu' j' L \mu j} 
\widetilde {\cal C}_{LT}^{\,I M\pm,K}(L' j' L j)\,
\widetilde{{\cal D}}_{\alpha'\alpha}^{K\kappa}(\mu' j' \mu j)\,
\Im m\Big(i^{\bar \delta^X_I+\delta_{(\alpha',\alpha)}^{(2)}}
\widetilde C^{L'\ast}(\mu' j')\,\widetilde N^L_1(\mu j)\Big)
\,,\label{multstrucfunLTsA}
\end{eqnarray}
\end{subequations}
and observables of type $B$
\begin{subequations}
\label{multstrucfunintB}
\beqa
f_{TT}^{IM\pm,\,K\kappa}(X)&=&
\sum_{L' \mu' j' L \mu j} 
\widetilde {\cal C}_{TT}^{\,I M\mp,K}(L' j' L j)\,
\widetilde{{\cal D}}_{\alpha'\alpha}^{K\kappa}(\mu' j' \mu j)\,
\Re e\Big(i^{\bar \delta^X_I+\delta_{(\alpha',\alpha)}^{(2)}}
\widetilde N^{L'\ast}_{-1}(\mu' j')\,\widetilde N^L_1(\mu j)
\Big)\,,\label{multstrucfunTTB}\\
f_{LT}^{IM\pm,\,K\kappa}(X)&=&
\sum_{L' \mu' j' L \mu j} 
\widetilde {\cal C}_{LT}^{\,I M\mp,K}(L' j' L j)\,
\widetilde{{\cal D}}_{\alpha'\alpha}^{K\kappa}(\mu' j' \mu j)\,
\Re e\Big(i^{\bar \delta^X_I+\delta_{(\alpha',\alpha)}^{(2)}}
\widetilde C^{L'\ast}(\mu' j')\,\widetilde N^L_1(\mu j)\Big)
\,,\label{multstrucfunLTB}\\
f_{LT}^{\prime\, IM\pm,\,K\kappa}(X)&=&
-\sum_{L' \mu' j' L \mu j} 
\widetilde {\cal C}_{LT}^{\,I M\mp,K}(L' j' L j)\,
\widetilde{{\cal D}}_{\alpha'\alpha}^{K\kappa}(\mu' j' \mu j)\,
\Im m\Big(i^{\bar \delta^X_I+\delta_{(\alpha',\alpha)}^{(2)}}
\widetilde C^{L'\ast}(\mu' j')\,\widetilde N^L_1(\mu j)\Big)
\,.\label{multstrucfunLTsB}
\end{eqnarray}
\end{subequations}
Here the coefficients $\widetilde {\cal C}_{\alpha}$ are defined by
\begin{subequations}
\label{LT-TT}
\beqa
\widetilde {\cal C}_{L}^{\,IM,K}(L'j'Lj)&=&\frac{8\,\pi}{1+\delta_{M0}}\,
{\cal C}^{00 I M,K}(L' j' L j)\,,\\
\widetilde {\cal C}_{T}^{\,IM,K}(L'j'Lj)&=&\frac{16\,\pi}{1+\delta_{M0}}\,
{\cal C}^{11 I M,K}(L' j' L j)\,,\\
\widetilde {\cal C}_{LT}^{\,IM\pm,K}(L'j'Lj)&=&\frac{16\,\pi}{1+\delta_{M0}}\,
\Big({\cal C}^{01 I M,K}(L' j' L j)\pm(-)^{I+M}
{\cal C}^{01 I -M, K}(L' j' L j)\Big)\,,\\
\widetilde {\cal C}_{TT}^{\,IM\pm,K}(L'j'Lj)&=&\frac{8\,\pi}{1+\delta_{M0}}\,
\Big({\cal C}^{-11 I M,K}(L' j' L j)\pm(-)^{I+M}
{\cal C}^{-11 I -M, K}(L' j' L j)\Big)\,.
\eeqa
\end{subequations}

More detailed expressions for the coefficients of the structure 
functions of the differential cross section are listed in the 
Appendix~\ref{formfactors}. Explicit results for the coefficients 
$\widetilde {\cal C}$ and $\widetilde {\cal D}$ 
for a maximal multipolarity $L_{max}=3$ may be found in~\cite{ArL02}.


\subsection{Inclusive process and form factors}

The inclusive cross section is obtained by integration over the solid angle
$\Omega _{np} ^{\mathrm{c.m.}}=(\theta,\,\phi)$ yielding 
\begin{eqnarray}
\frac{d \sigma}{dk _2 ^{\mathrm{lab}} d \Omega _e ^{\mathrm{lab}}}&=&
6\,c(k_1^{\mathrm{lab}},\,k_2^{\mathrm{lab}})\, \bigl\{\rho _L F_L + 
\rho _T  F _T - P_1^d \rho_{LT} F_{LT}^{1-1} \sin \phi _d d_{10}^1
(\theta_d) \nonumber\\
& &{} + P _2 ^d \bigl[( \rho _L F _L ^{20} + \rho _T F _T ^{20}) d_
{00} ^{2} (\theta _d)- \rho _{LT} F _{LT} ^{2-1} \cos \phi _d d_{10} ^{2}
(\theta _d)
+ \rho _{TT} F _{TT} ^{2-2} \cos 2 \phi _d d _{20} ^{2}
(\theta _d) \bigr] \nonumber\\
& &{} + h P_1 ^d \bigl[ -\rho ' _T F^{\prime 10}_T d _{00}
^{1} (\theta _d) + \rho ' _{LT} F^{\prime 1-1}_{LT} \cos \phi _d d _
{10} ^{1} (\theta _d) \bigr]
- h P_2^d \rho '_{LT} F^{\prime 2-1}_{LT} 
\sin \phi _d d^2_{10}(\theta_d) \bigr\} \nonumber\\
&\equiv&  \sigma (h, P^d_1 , P^d_2)\,. \label{incl_cross} 
\end{eqnarray}
It is governed by a set of inclusive form 
factors $F^{(\prime)I-M}_\alpha$ ($M\ge 0$) as given by
\begin{eqnarray}
F_{\alpha} ^{(\prime)I-M}& = & (-)^{I+M}\,(1+\delta_{M0})\,
\frac{\pi}{6} \int d (\cos \theta)
(f_{\alpha} ^{(\prime )IM+}-f_{\alpha} ^{(\prime )IM-})\,,
\end{eqnarray}
for $\alpha\in\{L,\,T,\,LT,\,TT\}$. This equation corresponds to Eqs.~(13) 
and (14) of \cite{LeT91} except for the fact that the primed form factors 
$F_T^{\prime\,10}$ and 
$F_{LT}^{\prime\,1-1}$ differ in sign from the ones given in \cite{LeT91} 
due to a redefinition of the primed structure functions incorporating a 
phase factor $(-)^I$ (see the remark in \cite{ArL95} before Eq.\ (9)). 
Altogether, the inclusive cross section depends on
ten form factors: $F_L,$ $F_T$, $F^{1-1}_{LT}$, $F_L ^{20}$, $F_T
^{20}$, $F_{LT} ^{2-1}$, $F_{TT} ^{2-2}$, $F^{\prime\, 10}_T$,
$F^{\prime\, 1-1}_{LT},$
and $F^{\prime\, 2-1}_{LT},$ of which $F_{LT}^{1-1}$ and
$F^{\prime\, 2-1}_{LT}$ vanish below pion threshold due to time reversal 
invariance. 

The multipole decomposition of the inclusive form factors are given by the 
($K=0$)-coefficients of the multipole expansion of the structure functions 
of the differential cross section $(X=1=(0,0))$, as listed in 
(\ref{fmultipole}), i.e.
\begin{equation}
F_{\alpha} ^{(\prime)I-M} =  (-)^{I+M}\,(1+\delta_{M0})\,
\frac{\pi}{3} \,(f_{\alpha} ^{(\prime )IM+,\,0}
-f_{\alpha} ^{(\prime )IM-,\,0})\,.
\end{equation}
Explicit expressions are listed in Appendix~\ref{formfactors}.
 
At the photon point
one can relate the purely transverse form factors to the total
photoabsorption cross section $\sigma _{tot}$ of deuteron
photodisintegration for unpolarized photons and
deuterons and to the corresponding beam and target asymmetries of the
total cross section as defined in~\cite{ArS91}. Taking into account the
relations (\ref{equivobs}) one obtains respectively
\begin{equation}
\sigma _{tot} = \frac{M_d}{{W_{np}\,q^{c.m.}}} F_T\,,\quad
\tau ^0 _{20} = \frac{F^{20} _T}{F_T}\,\quad
\tau ^c _{10} = \frac{F^{\prime\, 10} _T}{ F_T}\,,\quad
\tau ^l _{22} = \frac{F^{2-2} _{TT}}{F_T}\,,
\end{equation}
where $W_{np}$ and $q^{c.m.}$ denote the invariant mass of the $np$ 
system and the photon c.m.\ momentum, respectively.



\section{Separation of structure functions and complete sets}
\label{completesets}

\subsection{Experimental separation of structure functions}

The experimental separation of structure functions has been discussed 
in detail in~\cite{ArL92,ArL95}. It is based on the general definition of 
asymmetries of a polarization observable with respect to the beam and target 
polarization parameters $h$, $P_1^d$, and $P_2^d$, respectively. 
To this end one writes a general polarization 
observable ${\cal O}(\Omega_X)$ as given in (\ref{obs1}) and (\ref{obsfin}) 
in the form
\begin{equation}
P_X=A_0(X)+P_1^d\,A_d^V(X)+P_2^d\,A_d^T(X)
     +h\,\Big[A_e(X)+P_1^d\,A_{ed}^V(X)+P_2^d\,A_{ed}^T(X)\Big]
\,,\label{asy}
\end{equation}
defining implicitly the various asymmetries, i.e.\ $A_e(X)$ for beam 
polarization, $A_d^V(X)$ and $A_d^T(X)$ for vector and tensor target 
polarization, respectively, and $A_{ed}^V(X)$ and $A_{ed}^T(X)$ for 
the corresponding beam-target asymmetries. 
Their explicit form can be read from (\ref{obsfin}) 
\begin{subequations}\label{asyall}
\begin{eqnarray}
A_0(X)=&\frac{c(k_1^{\mathrm{lab}},\,k_2^{\mathrm{lab}})}{ S_0}&
 \Big[ \Big(\rho _L f_L^{00}(X) + \rho_T f_T^{00}(X) + 
\rho_{LT} {f}_{LT}^{00+}(X) \cos \phi
+\rho _{TT} {f}_{TT}^{00+}(X) \cos2 \phi
\Big)\delta_{X,A}\,\nonumber\\
& &\;\;+\Big(\rho_{LT} {f}_{LT}^{00-}(X) \sin \phi 
+ \rho _{TT} {f}_{TT}^{00-}(X) \sin2 \phi\Big)\delta_{X,B}\Big]\,,
\label{asy0}\\
A_d^V(X)=&
\frac{c(k_1^{\mathrm{lab}},\,k_2^{\mathrm{lab}})}{ S_0}&\sum _{M=0}^1
\Big[ \Big(\rho _L f_L^{1M}(X) + \rho_T f_T^{1M}(X) + 
\rho_{LT} {f}_{LT}^{1M+}(X) \cos \phi\nonumber\\
&&+ \rho _{TT} {f}_{TT}^{1M+}(X) \cos2 \phi
\Big)\cos (M\tilde{\phi}-\delta_{X,A} \frac{\pi}{2}) \nonumber\\
& &-\Big(\rho_{LT} {f}_{LT}^{1M-}(X) \sin \phi 
+ \rho _{TT} {f}_{TT}^{1M-}(X) \sin2 \phi\Big) 
\sin (M\tilde{\phi}-\delta_{X,A} \frac{\pi}{2})\Big] d_{M0}^1(\theta_d)\,,
\label{asydv}\\
A_d^T(X)=&
\frac{c(k_1^{\mathrm{lab}},\,k_2^{\mathrm{lab}})}{ S_0}&\sum _{M=0}^2
\Big[ \Big(\rho _L f_L^{2M}(X) + \rho_T f_T^{2M}(X) + 
\rho_{LT} {f}_{LT}^{2M+}(X) \cos \phi \nonumber\\
&&+ \rho _{TT} {f}_{TT}^{2M+}(X) \cos2 \phi
\Big)\cos (M\tilde{\phi}-\delta_{X,B} \frac{\pi}{2}) \nonumber\\
& &-\Big(\rho_{LT} {f}_{LT}^{2M-}(X) \sin \phi 
+ \rho _{TT} {f}_{TT}^{2M-}(X) \sin2 \phi\Big) 
\sin (M\tilde{\phi}-\delta_{X,B} \frac{\pi}{2})\Big] d_{M0}^2(\theta_d)\,,
\label{asydt}\\
A_e(X)=&\frac{c(k_1^{\mathrm{lab}},\,k_2^{\mathrm{lab}})}{ S_0}&
 \Big[ -\Big(\rho'_T f_T^{\prime\, 00}(X) 
+ \rho '_{LT} {f}_{LT}^{\prime \,00-}(X) \cos \phi \Big)\delta_{X,B}
+ \rho '_{LT} {f}_{LT}^{\prime \,00+}(X) \sin \phi\, 
\delta_{X,A} \Big] \,,\label{asye}\\
A_{ed}^V(X)=&
\frac{c(k_1^{\mathrm{lab}},\,k_2^{\mathrm{lab}})}{ S_0}&\sum _{M=0}^1
 \Big[ \Big(\rho'_T f_T^{\prime\, 1M}(X) 
+ \rho '_{LT} {f}_{LT}^{\prime\, 1M-}(X) \cos \phi \Big) 
\sin (M\tilde{\phi}-\delta_{X,A} \frac{\pi}{2}) \nonumber\\
& & + \rho '_{LT} {f}_{LT}^{\prime\, 1M+}(X) \sin \phi 
\cos (M\tilde{\phi}-\delta_{X,A} \frac{\pi}{2})\Big] 
 d_{M0}^1(\theta_d)\,,\label{asyedv}\\
A_{ed}^T(X)=&
\frac{c(k_1^{\mathrm{lab}},\,k_2^{\mathrm{lab}})}{ S_0}&\sum _{M=0}^2
 \Big[ \Big(\rho'_T f_T^{\prime\, 2M}(X) 
+ \rho '_{LT} {f}_{LT}^{\prime\, 2M-}(X) \cos \phi \Big) 
\sin (M\tilde{\phi}-\delta_{X,B} \frac{\pi}{2}) \nonumber\\
& &\qquad + \rho '_{LT} {f}_{LT}^{\prime\, 2M+}(X) \sin \phi 
\cos (M\tilde{\phi}-\delta_{X,B} \frac{\pi}{2})\Big] 
 d_{M0}^2(\theta_d)\,,\label{asyedt}
\end{eqnarray}
\end{subequations}
where the unpolarized differential cross section $S_0$ is defined in 
(\ref{S_0xsection}). For simplicity, we will also call $A_0(X)$ an 
asymmetry although it is not one in the strict sense. The nonvanishing 
structure functions contributing to an asymmetry 
of a given observable are listed in Table~\ref{tab4}.
For the differential cross section ($X=1=(00)$) we remind the reader
that one has with respect to the notation in \cite{ArL92,ArL88}
\begin{equation}
A_0(1)=1,\quad A_d^{V/T}(1)=A_d^{V/T},\quad A_e(1)=A_e,\quad A_{ed}^{V/T}(1)=
A_{ed}^{V/T}\,.
\end{equation}

For the simplest case, namely in the absence of beam and target polarization, 
the four structure functions $f_\alpha(X)$ can be separated choosing 
first different $\phi$-angles, yielding $f_{LT}$, $f_{TT}$ and a linear 
superposition of $f_L$ and $f_T$ and subsequently a {\sc Rosenbluth} separation 
for disentangling $f_L$ and $f_T$. 
In the general case, by a proper variation of the longitudinal 
electron polarization $h$ and the deuteron vector 
and tensor polarization parameters $P_1^d$ and $P_2^d$, respectively,
one can first separate the various beam, target and beam-target 
asymmetries as listed in (\ref{asyall}).
These asymmetries are functions of the deuteron orientation angles
$\theta_d$ and $\phi_d$, viz. $\tilde{\phi}=\phi-\phi_d$, and the azimuthal or 
out-of-plane angle $\phi$. One can now utilize these variables for the 
further separation of the different structure functions. 

This is achieved by observing that the general functional form of 
an asymmetry is
\begin{equation}
A^I(\phi,\tilde \phi,\theta_d)=\sum_{M=0}^I \alpha_{IM}(\phi, 
\tilde{\phi)}d_{M0}^I(\theta_d)\,,\quad (I=0,1,2)\,,
\end{equation}
where
\begin{equation}\label{alphaIM}
\alpha_{IM}(\phi, \tilde{\phi})=c_{IM}(\phi) \cos M\tilde{\phi} +
s_{IM}(\phi) \sin M\tilde{\phi}\;,
\end{equation}
and the $\phi$-dependent functions $c_{IM}(\phi)$ and $s_{IM}(\phi)$ 
have either the form
\begin{subequations}\label{csIM}
\begin{equation}
a_0 + a_1 \cos \phi + a_2 \cos 2\phi \label{(34)}
\end{equation}
or
\begin{equation}
b_1 \sin \phi + b_2 \sin 2\phi\;.
\end{equation}
\end{subequations}
For a given $I$ the $M$-components $\alpha_{IM}(\phi, \tilde{\phi})$ 
of the asymmetry 
$A^I(\phi,\tilde \phi,\theta_d)$ can be separated 
by a proper choice of $\theta_d$ exploiting the properties of the small
$d^I_{M0}$-functions. For $I=1$ (vector asymmetries), taking 
$\theta_d=0$ or $\pi /2$, i.e.\ $d^1_{M0}(0)=\delta_{M0}$ or 
$d^1_{M0}(\pi /2)=M/\sqrt{2}$, yields  $\alpha_{10}$ or $\alpha_{11}$, 
respectively, and for the tensor asymmetries ($I=2$) one may first choose 
$\theta_d=0$ yielding with $d^2_{M0}(0)=\delta_{M0}$ directly 
$\alpha_{20}$. The latter being determined, then setting $\theta_d={\pi /4}$ 
and $\pi /2$, one can obtain the remaining two terms $\alpha_{21}$ and 
$\alpha_{22}$. For the separation of $\alpha_{21}$ and $\alpha_{22}$ one 
can also choose $\theta_d=\theta_d^0=\mbox{arcos}\,(1/\sqrt{3})$ together with 
$\tilde{\phi}$ and $\tilde{\phi}+\pi$. Then the sum and difference of the 
corresponding asymmetries result in $\alpha_{21}$ and $\alpha_{22}$, 
respectively.

In the next step, in order to separate the two contributions $c_{IM}$ 
and $s_{IM}$ in (\ref{alphaIM}), one can take first $\tilde{\phi}=0$ 
giving $c_{IM}$ 
and then $\tilde{\phi}={\pi / 2M}$  for $M\neq 0$ which yields 
directly $s_{IM}$. 
The remaining separation of the coefficients $a_n$ or $b_n$ in 
(\ref{csIM}) is then achieved by appropriate 
choices of $\phi$. In a few cases the constant term $a_0$ in (\ref{(34)}) 
will contain two structure functions in the combination 
$\rho_L f_L^{IM}(X)+\rho_T f_T^{IM}(X)$. In this case one needs a {\sc Rosenbluth} 
separation in addition. 

A different task than the complete separation of all structure functions 
is to find an optimal way for separating a specific structure
function $f_{\alpha}^{IM(\pm)}(X)$ ($I > 0$) or 
$f_{\alpha}^{\prime\,IM(\pm)}(X)$. In other words, the question is: 
what is the minimal 
number of measurements necessary for the separation of a specific structure 
function? This has been discussed in~\cite{ArL95} and is described in 
detail in Appendix~\ref{sepstrucfun}.


\subsection{Complete sets of observables}
We have already mentioned the fact that in deuteron 
electrodisintegration the total 
number of independent complex $t$-matrix elements is 18, while for 
photodisintegration the number is 12. Since one phase remains arbitrary 
this means that this process is determined by 35 independent observables, 
whereas in the corresponding photoreaction one needs 23. 
The question is how to choose from the much larger set of 324 
(or 144 in photodisintegration) linearly independent observables an 
appropriate set of 35 (or 23). In \cite{ArL98} we had 
derived a general criterion which allows one to decide uniquely whether for 
a reaction with $n$ independent $t$-matrix elements a set of $2n-1$ 
observables, taken from the set of $n^2$ linearly independent observables, 
constitutes a complete set. Subsequently this criterion has been applied 
to deuteron 
electro- and photodisintegration in~\cite{ArL00}. A brief review of 
the main results of \cite{ArL98} and \cite{ArL00} is appropriate. 


\subsubsection{General criterion~\cite{ArL98}}
Any observable in a reaction with $n$ independent, complex matrix 
elements can be represented by a $n\times n$ hermitean form $f^\alpha$ 
in the complex $n$-dimensional variable $z=(z_1,\dots,z_n)$  
\begin{eqnarray}
f^\alpha(z)&=&\frac{1}{2}\sum_{j'j} z_{j'}^* F_{j'j}^\alpha z_j\,,\label{genf}
\end{eqnarray}
where hermiticity requires 
\begin{eqnarray}
(F_{j'j}^{\alpha})^*=F_{jj'}^\alpha\,,
\end{eqnarray}
and $z$ comprises all independent reaction matrix elements labeled by 
$j$. 

For the application of our criterion, derived in \cite{ArL98}, 
one first has to rewrite the 
hermitean form in (\ref{genf}) into a real quadratic form by introducing
\begin{eqnarray}
z&=&x+i y\,,\\
F^\alpha&=&A^\alpha+i\, B^\alpha \,,
\end{eqnarray}
where $A^\alpha$ and $B^\alpha$ are real matrices, and $A^\alpha$ is 
symmetric whereas $B^\alpha$ is antisymmetric.
Considering further the fact that one overall phase is arbitrary, one may 
choose $y_{j_0}=0$ for an arbitrary index $j_0$ and then one finds for the 
given observable
\begin{eqnarray}
f^\alpha(x+iy)&=&\frac{1}{2}\Big[\sum_{j'j} x_{j'} A_{j'j}^\alpha x_j
+\sum_{\tilde j'\tilde j} y_{j'} A_{j'j}^\alpha y_j 
+2\sum_{\tilde j'j} y_{j'} B_{j'j}^\alpha x_j\Big]
\,,
\end{eqnarray}
where the tilde over a summation index indicates that the index $j_0$ has to  
be left out. Introducing now a $(m=2n-1)$-dimensional real vector $u$ by 
\begin{eqnarray}
u=( x_1, \dots , x_n , y_1, \dots , y_{j_0-1}, y_{j_0+1}, \dots , y_n)\,,
\end{eqnarray}
one can represent the $n\times n$ hermitean form by a $m\times m$
real quadratic form
\begin{eqnarray}
\widetilde f^\alpha(u)&=&\frac{1}{2}\sum_{l'l=1}^{m} u_{l'} 
                     \widetilde F_{l'l}^\alpha u_l\,,
\end{eqnarray}
where the $m\times m$-matrix $\widetilde F^\alpha$ is given by 
\begin{eqnarray}
\widetilde F^\alpha=\left(\begin{array}{cc} A^\alpha & 
(\widetilde B^\alpha)^T \\
\widetilde B^\alpha &  {\widehat A}^\alpha \end{array}\right)\,.
\end{eqnarray}
Here $\widetilde B^\alpha$ is obtained from $B^\alpha$ by canceling the 
$j_0$-th row, and ${\widehat A}^\alpha$ from $A^\alpha$ by canceling 
the $j_0$-th row and column. Thus $\widetilde B^\alpha$ is a 
$(n-1)\times n$-matrix and ${\widehat A}^\alpha$ a $(n-1)\times(n-1)$-matrix. 

Now, for checking the completeness of a chosen set of $2n-1$ observables 
one has to construct the $m\times m$ corresponding matrices 
$\widetilde F^\alpha$, and then one builds from their columns for all possible
sets $\{k_1,\dots,k_m;\, k_\alpha \in \{1,\dots,m\}\}$ the matrices  
\begin{eqnarray}
\widetilde W(k_1,\dots,k_m)=\left(\begin{array}{ccc} 
\widetilde F^1_{1k_1} & \cdots &\widetilde  F^m_{1k_m} \\ 
\vdots & &\vdots \\ \widetilde F^1_{mk_1} & \cdots & \widetilde F^m_{mk_m} 
\end{array}\right)
\,.
\end{eqnarray}
Note that the $k_\alpha$ need not be different. If at least one of the 
determinants of $\widetilde W(k_1,\dots,k_m)$ is 
nonvanishing then one has a complete set. 

\subsubsection{Complete sets for photo- and 
electrodisintegration~\protect\cite{ArL00}}\label{complsets}
In order to apply our criterion, one has to construct the matrices 
$\widetilde F$ which represent the structure functions as 
hermitean forms 
in the reaction matrix elements as
\begin{equation}
f_\alpha^{(\prime)\, I M\pm}(X)= 
t^\dagger\,\widetilde F^{(\prime)\, I M \pm,\,\alpha}\, t\,,\label{Fmatrix}
\end{equation}
where $t$ is a vector comprising all reduced $t$-matrix elements in a 
certain labeling. 
It is convenient to arrange the labeling of the $t$-matrix elements in 
such a way that the longitudinal ones belong to $j=1,\dots,6$ and the 
transverse ones to $j=7,\dots,18$. Thus the general structure of these 
matrices then is  
\begin{equation}
\widetilde F^{(\prime)\, I M \pm,\,\alpha} = 
   \left(\begin{array}{cc}
   A^{(\prime)\, I M \pm,\,\alpha} & C^{(\prime)\, I M \pm,\,\alpha}\\
   (C^{(\prime)\, I M \pm,\,\alpha})^\dagger & B^{(\prime)\, I M \pm,\,\alpha}
   \end{array}\right),\,
\end{equation}
where $A^{(\prime)\, I M \pm,\,\alpha}$ is a $(6\times 6)$-matrix, 
$C^{(\prime)\, I M \pm,\,\alpha}$ a $(6\times 12)$-matrix, and 
$B^{(\prime)\, I M \pm,\,\alpha}$ a $(12\times 12)$-matrix. In particular
one has 
\begin{subequations}
\beqa
\widetilde F^{I M \pm,\,L} &=& 
   \left(\begin{array}{cc}
   A^{I M \pm,\,L }& 0 \\
   0 & 0
   \end{array}\right),\\
\widetilde F^{(\prime)\,I M \pm,\,T/TT} &=& 
   \left(\begin{array}{cc}
   0 & 0 \\
   0 & B^{(\prime)\,I M \pm,\,T/TT}
   \end{array}\right),\\
\widetilde F^{(\prime)\,I M \pm,\,LT} &=& 
   \left(\begin{array}{cc}
   0 & C^{(\prime)\,I M \pm,\,LT} \\
   (C^{(\prime)\,I M \pm,\,LT})^\dagger & 0
   \end{array}\right).
\eeqa
\end{subequations}
The explicit forms of these 
matrices are obtained from the matrix representation of
\begin{equation}
{\cal U}^{\lambda' \lambda I M}_{X}=
  \sum_{j'j} t^*_{j'}\widetilde C^{I M \lambda' \lambda}_{j'j}
  (X) t_j\,.
\end{equation}
Comparison with (\ref{ulamcart}) gives for the matrix elements
\begin{equation}
\widetilde C^{I M \lambda' \lambda}_{j'j}(X) = \langle m_1'm_2'|
\sigma_{\alpha'}(p)\sigma_{\alpha}(n)|m_1 m_2\rangle
\langle \lambda_d|\tau^{[I]}_M|\lambda_d'\rangle\,,\label{CIMX}
\end{equation}
where the labeling is to be understood as $j^{(\prime)}=(m^{(\prime)}_1,
m^{(\prime)}_2, \lambda^{(\prime)}, \lambda^{(\prime)}_d)$. 
Detailed expressions of the $\widetilde C^{I M \lambda' \lambda}(X)$'s 
for several representations are easily obtained 
from the expressions listed in Appendix~\ref{MatRep} (see also \cite{ArL00}). 

The structure of these matrices is such that the longitudinal ($L$) and the 
transverse ($T,\,TT$) observables are decoupled filling separated $6\times 6$-
and $12\times 12$-submatrices, respectively, whereas the $LT$-type
observables are represented by $18\times 18$-matrices. These features offer
various kinds of strategies for selecting complete sets. 
\begin{itemize}
\item[(i)]
One may independently select complete sets of observables for the 
longitudinal and transverse cases, i.e., a set of 11 longitudinal and 23 
transverse structure functions for a check of completeness.
With respect to the latter, one has in view of the linear relations between 
the $T$- and the $TT^+$-type and between the $T'$- and $TT^-$-type 
observables, different choices, taking either $T$- and $T'$-type or 
$TT^{\pm}$-type observables or even mixing different types of observables. 
The missing relative phase between the longitudinal and transverse 
$t$-matrix elements can then be provided by any one of the $LT$-observables. 
The advantage of this approach is that in this way one automatically obtains 
complete sets of observables for the case of photodisintegration as well, 
namely from the transverse ones. 
\item[(ii)]
Again one may start with a selection of 11 longitudinal structure functions. 
But then instead of choosing transverse observables, one may directly choose
24 linearly independent $LT$-type observables which then constitute a simple 
system of linear equations for the missing transverse matrix elements, 
because the longitudinal ones are then known from the first step. 
\item[(iii)]
Complementary to case (ii) one may start with a selection of 23 transverse
observables taking one of the alternatives listed in (i). Then a proper set of 
12 $LT$-type observables provides a set of linear equations from which
the missing longitudinal $t$-matrix elements can be obtained. 
\item[(iv)]
Another alternative would be a selection of 
35 structure functions of $LT$-type. However, in this case the completeness 
check would be much more involved due to the considerably higher dimension 
of the determinants to be checked. 
\end{itemize}
Which of these strategies is most advantageous will depend on
the experimental conditions. Often $L$- and $T$-type structure functions 
are easier to determine in an experiment although the required 
{\sc Rosenbluth} 
separation introduces some unwanted complication. In view of the fact that 
the strategies (i) through (iii) require the determination of either $L$- 
or $T$-type observables or both, we have considered in \cite{ArL98,ArL00} 
exclusively the question of complete sets for longitudinal and 
transverse structure functions. 

The choice of complete sets of longitudinal structure functions, 
containing eleven structure functions, from
a set of linearly independent ones has been discussed in detail 
in~\cite{ArL98}. It turned out, 
that there is only a very weak restriction on the choice of 
possible complete longitudinal 
sets. In fact one may select from the chosen set of 36 linearly 
independent observables any subset of eleven structure functions, 
which does not contain more than eight of the type $X^{10}$ 
and $X^{22}$. This has been discussed explicitly for the linearly 
independent set $X\in \{1^{IM},xx^{IM},xz^{IM},y_1^{IM},x_1^{IM},
x_2^{IM},z_1^{IM},z_2^{IM}\}$, and possible complete sets are listed
in Tables 3 and 4 of~\cite{ArL98}. 
The case of the transverse observables for which a complete set 
contains 23 structure functions, has been discussed in~\cite{ArL00}. 
Again with respect to the general question of a choice of a 
complete set of structure functions, we found the general statement, 
that one may pick from the chosen set of 144 linearly independent 
ones any subset of 23 structure 
functions with the only restriction, that not more than 16 should 
be of the type $X^{(\prime)\,10}$ and $X^{(\prime)\,22}$. 

In Ref.~\cite{ArL98} we furthermore simulated an experimental study for the 
determination of the longitudinal $t$-matrix elements in the helicity basis 
from a given set of ``measured'' observables whose numerical values 
were taken from a calculation. Various complete sets were
selected and the arising system of 11 nonlinear equations for the
$t$-matrix elements was solved numerically. Since the solutions were not
unique we had to calculate additional observables, called 
``check observables'', taking as input the obtained 
solutions for the $t$-matrix elements, and compared them to their ``measured''
values. For the arbitrarily chosen kinematics (internal excitation energy
$E_{np}=100$ MeV, momentum transfer 
$(q^{\mathrm{c.m.}})^2=5$ fm$^{-2}$, various $np$-angles 
$\theta$) we found that one of the considered complete
sets was particularly suitable (first set of Table 6 in Ref.~\cite{ArL98}). 
In this case only one additional check
observable ($f^{10}_L(x_2)$) was sufficient to determine the correct solution. 
In \cite{ArL00} this simulation was extended to a somewhat more 
realistic experimental situation using the same kinematics 
again and taking the same specific set but allowing for errors in the 
measured observables. The simulation of an experimental situation  
showed that one can get quite 
reliable results for the $t$-matrix elements even if experimental errors 
are taken into account. The results can be greatly improved if 
additional check observables are considered.
For the case studied in \cite{ArL00}, it was sufficient to consider two 
such observables. 
If on the other hand one uses no check observables at all, 
one gets rather unreliable results since other types of solutions of
the nonlinear system of equations are mixed in. 
In fact, performing such a simulation without any check observable leads 
to large errors in the resulting $t$-matrix elements 
(average error more than 100 \%) and also to strong 
average deviations of the mean values from the true values of the $t$-matrix
elements (about 50 \%).  

A similar study has been performed in \cite{ArL00} for the determination 
of the
transverse $t$-matrix elements from observables, but without introducing 
experimental errors. Although the transverse case is much more complicated 
than the longitudinal one due to the higher dimensionality (12 instead of 
6 complex $t$-matrix elements), it was found that in principle the method 
works also for the transverse case.

\subsection{Analytic expressions of the $t$-matrix elements in terms of 
observables}\label{bilin}

One can also derive an analytic solution of the reaction matrix 
elements in terms of observables, because one can 
express all bilinear forms $t_{j'}^*t_j$ as linear 
forms in the structure functions $f^{(\prime) I M\pm}_\alpha (X)$, i.e.,
\begin{equation}
t_{j'}^*t_j=T_{j'j}[f^{(\prime) I M\pm}_\alpha (X)]\,\label{ttobs}=
\sum_{\alpha I M, sig=\pm}\Big(T_{j'j}^{\alpha I M sig}f^{I M sig}_\alpha (X)
+T_{j'j}^{\prime \alpha I M sig}f^{\prime I Msig}_\alpha (X)\Big)\,,
\end{equation}
where the explicitly appearing
structure functions $f^{(\prime) I M\pm}_\alpha (X)$ constitute a
complete, linearly independent set. Here the square bracket of
$T_{j'j}[f]$ indicates the functional dependence on the structure
functions. This relation has also been used in the derivation of the
linear relations between observables (see
Sect.~\ref{linrel}). Explicit expressions for the coefficients 
$T_{j'j}^{(\prime) \alpha I M sig}$ are derived in various
representations in~\cite{ArL00}. To give an example, we list here two
cases, a diagonal and an interference term, of the longitudinal
$t$-matrix elements in a rotated helicity basis, where the final
helicity states are rotated into the $y$-axis, 
\begin{subequations}
\beqa
t^\ast_{\frac{1}{2} \frac{1}{2} 0 0}t_{\frac{1}{2} \frac{1}{2} 0 
  0} &=& \frac{1}{6}\Big(f_{L}^{0 0} - {\sqrt{2}}\,f_{L}^{2 0} + 
f_{L}^{0 0}(y_1) - 
   {\sqrt{2}}\,f_{L}^{2 0}(y_1)\Big) \,,\\
t^\ast_{-\frac{1}{2} -\frac{1}{2} 0 0}t_{\frac{1}{2} \frac{1}{2}
   0 0} &=& \frac{1}{6}\Big(-f_{L}^{0 0}(zz) + {\sqrt{2}}\,f_{L}^{2 0}(zz)+
  i\,\Big( f_{L}^{0 0}(xz) - {\sqrt{2}}\,f_{L}^{2 0}(xz) \Big) \Big) \,.
\eeqa
\end{subequations}
For a complete listing and further details 
we refer to the Appendix E of~\cite{ArL00}.

The linear relations in (\ref{ttobs}) can be exploited in various ways. 
One possibility is to choose a specific matrix element, say $t_{j_0}$, as real
and positive. Then all other matrix elements $t_j$ with $j\neq j_0$ are 
uniquely determined relative to $t_{j_0}$ and are given as linear forms of
appropriate structure functions \cite{ArS90}
\begin{equation}
t_j = \frac{1}{t_{j_0}} T_{j_0j}[f^{(\prime)IM\pm}_\alpha(X)]\,.\label{t_j}
\end{equation}
Finally, for the determination of the missing matrix element $t_{j_0}$ one
has to choose only one additional structure function, say 
\begin{equation}
f_0=\sum_{j'j}t_{j'}^* \tilde F_{j'j}t_j\,,\label{f_0}
\end{equation}
yielding 
\begin{equation}
t_{j_0}=\frac{1}{\sqrt{f_0}}\sqrt{\sum_{j'j}T_{j'j_0}[f^{(\prime)IM\pm}_\alpha(X)]
\tilde F_{j'j} T_{j_0 j}[f^{(\prime)IM\pm}_\alpha(X)]}\,.\label{tj0}
\end{equation}
Thus (\ref{t_j}) in conjunction with (\ref{tj0}) 
constitutes a nonlinear functional in the structure functions
$f^{(\prime) I M\pm}_\alpha (X)$.
However, proceeding in this way, one needs in general a much larger number of 
observables for the complete determination of the $t$-matrix 
than the required minimal number of $2n-1$ of a complete set of a 
$n$-dimensional $t$-matrix. 

\begin{figure}
\includegraphics[scale=.5]{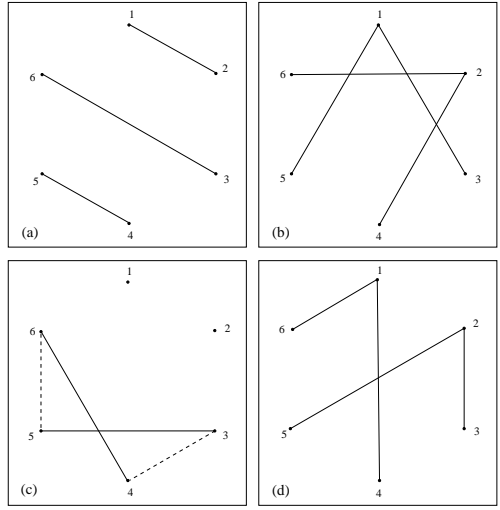}
\caption{ Diagrammatic representation of groups of longitudinal observables 
determining the interference terms of $t$-matrix elements for the helicity 
basis. The nomenclature for the groups and the corresponding observables are 
listed in Table~\protect{\ref{tabgroupLhel}}.
\label{fig_long_tmatrix}}
\end{figure}

Another strategy, which leads in general to a smaller number 
of necessary observables, has been developed in~\cite{ArL00}. It is based on
an analysis of all interference terms with respect to the question, 
which and how many observables appear in the representation of an 
interference term by observables. 
Because a closer inspection of the explicit expressions reveals, that in 
general the interference terms can be divided into disjunct subgroups which 
are determined by a subgroup of observables. In order to visualize 
this grouping we have devised in~\cite{ArL00} a graphical representation. 
To this end one assembles the numbers ``1'' through ``$n$'', where $n$ 
denotes the number of $t$-matrix elements,  
by points on a circle and represents an interference term $t_{j'}^*t_j$ 
by a straight line joining the points ``$j$'' and 
``$j'$''. Interference terms belonging to the same subgroup are then 
represented by the same type of lines. An example for the longitudinal
matrix elements in the helicity basis is shown in Fig.~\ref{fig_long_tmatrix}. 

\begin{table}
\caption{Nomenclature for the diagrammatic representation of groups of 
longitudinal observables in Fig.~\protect\ref{fig_long_tmatrix} 
determining the interference terms of $t$-matrix elements for the 
helicity basis, where $f_L^{IM}(X)$ is represented by $X^{IM}$.}
\begin{ruledtabular}
\begin{tabular}{ccl}
Panel & Line type & Observables 
\\
\colrule
(a) & solid & $x_1^{10}$, $x_2^{22}$, $y_1^{00}$, $y_1^{20}$, $xz^{00}$, $xz^{20}$\\
(b) & solid & $1^{11}$, $1^{21}$, $z_1^{11}$, $z_2^{11}$, $z_1^{21}$, $z_2^{21}$, 
    $zz^{11}$, $zz^{21}$ \\
(c) & solid & $1^{22}$, $z_1^{22}$, $z_2^{22}$, $zz^{22}$ \\
    & dashed & $x_2^{10}$, $x_1^{22}$, $y_1^{22}$, $xz^{22}$ \\
(d) & solid & $x_1^{11}$, $x_2^{11}$, $x_1^{21}$, $x_2^{21}$, 
    $y_1^{11}$, $y_1^{21}$, $xz^{11}$, $xz^{21}$\\
\end{tabular}
\end{ruledtabular}
\label{tabgroupLhel}
\end{table}

We will call a set of interference 
terms {\it connected} if they 
generate a pattern of connected lines so that any point belonging to one of 
the considered interference terms is connected to any other point of the 
set either directly or via $k$ other intermediate 
points of that set. For example, 
in Fig.~\ref{fig_long_tmatrix}(a) one notes three disconnected lines, 
(b) and (d) contain two different groups of connected lines, whereas (c)
contains one connected group of four lines. In such a connected set, 
any matrix element $t_{j'}$ can be expressed in terms of any other matrix 
element $t_j$ of that set by one of the two forms
\begin{equation}
t_{j'}=\left\{\begin{array}{ll} \frac{T_{j_1 j'}}{T_{j_1 j_2}}
\frac{T_{j_3 j_2}}{T_{j_3 j_4}}\cdots 
\frac{T_{j_{k-2} j_{k-3}}}{T_{j_{k-2} j_{k-1}}}
\frac{T_{j_k j_{k-1}}}{T_{j_k j}}\,t_j & \mbox{ for }k\,\mbox{ odd},\\
\frac{T_{j_1 j'}}{T_{j_1 j_2}}
\frac{T_{j_3 j_2}}{T_{j_3 j_4}}\cdots 
\frac{T_{j_{k-3} j_{k-4}}}{T_{j_{k-3} j_{k-2}}}
\frac{T_{j_{k-1} j_{k-2}}}{T_{j_{k-1} j_k}}\frac{T_{j j_k}}{t_j^*} 
& \mbox{ for }k\,\mbox{ even},\end{array}\right.\label{connect}
\end{equation}
depending on whether the number $k$ of intermediate points 
connecting $j'$ with $j$ via the points $j_1$ through $j_k$ is odd or even. 
The proof of these equations is simple and given in \cite{ArL00}.

A special feature of the evolving geometric pattern is a closed loop,
see e.g. Fig.~\ref{fig_long_tmatrix}(c). 
If such a closed loop has an even number of points then one finds from
(\ref{connect}) for $j'=j$ and $k$ odd the following condition 
\begin{equation}
T_{j_k j_{k-1}}T_{j_{k-2} j_{k-3}}\cdots T_{j_3 j_2}T_{j_1 j}=
T_{j_1 j_2}T_{j_3 j_4}\cdots T_{j_{k-2} j_{k-1}}T_{j_k j}\,,\label{evenpoints1}
\end{equation}
which means that in such a closed loop any interference term is completely 
determined by the other remaining interference terms of that loop.  
This condition thus constitutes a complex relation between the participating 
observables which allows one to eliminate two observables. 
On the other hand, a closed loop through an odd number of points 
(\ref{connect}) yields for $j'=j$ and $k$ even 
\begin{equation}
T_{j j_{k}}T_{j_{k-1} j_{k-2}}\cdots T_{j_3 j_2}T_{j_1 j}=
T_{j j}T_{j_1 j_2}T_{j_3 j_4}\cdots T_{j_{k-3} j_{k-2}}T_{j_{k-1} j_k}\,,
\label{oddpoints1}
\end{equation}
which again allows the elimination of two observables. It means furthermore
that the modulus of each of the participating $t$-matrix elements is 
completely determined, namely one has
\begin{equation}
|t_{j}|^2=\frac{T_{j_1 j}}{T_{j_1 j_2}}
\frac{T_{j_3 j_2}}{T_{j_3 j_4}}\cdots 
\frac{T_{j_{k-3} j_{k-4}}}{T_{j_{k-3} j_{k-2}}}
          \frac{T_{j_{k-1} j_{k-2}}}{T_{j_{k-1} j_k}}\,T_{j j_k}\,.
\label{oddpoints}
\end{equation}
Thus one may choose one matrix element of that loop as real and 
non-negative, fix its modulus according to (\ref{oddpoints}) and then 
all other matrix elements of that loop are uniquely determined. 
As a side remark, we would like to point out, that the conditions in
(\ref{evenpoints1}) and (\ref{oddpoints1}) constitute particular 
nonlinear relations between observables, which follow from the ones 
discussed in the Appendix~\ref{quadrel}. In fact, these conditions 
are obtained by applying successively the condition in (\ref{tijlma}) 
of this appendix to the left-hand sides of (\ref{evenpoints1}) and 
(\ref{oddpoints1}) yielding then the corresponding right-hand sides.

As next step one has to choose from the total number of all interference 
terms $t_{j'}^*t_j$  with $j'>j$ which is $\frac{1}{2}n(n-1)$ -- not 
counting $t_{j'}^*t_j$ with $j'<j$, because $(t_{j'}^*t_j)^*=t_{j}^*t_{j'}$ 
-- a set of $n-1$ independent interference terms. Hereby we define a set of 
{\it independent interference terms} by the property that they generate
a geometric pattern which does not contain any closed loop. From this
definition follows that a set of $n-1$ independent interference terms 
is represented by a pattern of $n-1$ lines in such a fashion that 
(i) each of the $n$ points is endpoint of at 
least one line, and (ii) each point is connected to all other points not 
necessarily in a direct manner but via intermediate points. It is 
obvious that in such a pattern no closed loops can be present, 
because one cannot construct from $n-1$ lines a pattern which
contains a closed loop and which still connects all $n$ points. 
For such a set of $n-1$ independent interference terms 
all matrix elements can be expressed by one arbitrarily chosen 
matrix element, say $t_{j_0}$ according to (\ref{connect}). In order to 
fix the remaining undetermined matrix element $t_{j_0}$ one has to choose 
one additional observable $f_0$. From (\ref{f_0})
one obtains in general an equation of the type 
\begin{equation}
f_0= a + b\, |t_{j_0}|^{2} + c\,|t_{j_0}|^{-2}\,,\label{f_00}
\end{equation}
from which $t_{j_0}$ can be obtained, although not uniquely in general. 
The ideal 
situation would be such that one finds $n-1$ independent interference terms 
each of them represented by only two observables. Because in this case one 
employs just $2n-1$ observables. On the 
other hand, analyzing the grouping of observables mentioned above, one will
in general not find such a situation, either the number of observables for 
a set of $n-1$ independent interference terms is larger than $2n-2$, or 
the grouping is such, that the choice of $n-1$ independent interference terms 
involves observables which govern at least one additional interference term 
leading to one or several closed loops. However, in that case those loops
lead to the elimination of superfluous observables. For example, 
considering Fig.~\ref{fig_long_tmatrix}, one notes that combination of
the three lines of (a) with two connected lines of the connected groups in
(b) or (d) results in a set of independent interference terms. According
to Table~\ref{tabgroupLhel}, such a choice involves 14 structure functions,
which means, that all six longitudinal matrix elements can be expressed in 
terms of 15 observables. In order to eliminate four of them, one can include
the other group of two connected lines of (c) or (d), yielding two closed
loops which allow the elimination. Thus, at the end one has a complete set 
of eleven observables. Further illustrative examples 
for such an analysis are discussed in some detail in~\cite{ArL00}.

\newpage
\section{Ingredients of calculation}
\label{ingredients}

In this section we will review briefly the various ingredients
which go into the calculation of the structure functions
presented in this work. The structure functions are determined by 
hermitean quadratic forms in the matrix elements 
of the e.m.\ current operator between the initial and final states. 
As already mentioned in Sect.~\ref{genform}, the
calculation is done with respect to the c.m.\ frame of the final
hadronic $NN$ state. Thus for a momentum transfer $\vec q$ 
the initial deuteron moves in
this reference frame with a momentum $-\vec q$, and one has to take
into account the transformation to the lab frame as governed
by the $\beta$-factor of (\ref{betaxieta}). 

Structure functions are calculated
within a nonrelativistic framework from the $t$-matrix defined in 
(\ref{Tmatrix}). Therefore, the wave functions
are purely nonrelativistic and are obtained by solving the two-body
{\sc Schr\"odinger} equation with a realistic $NN$-potential for the bound as
well as for the scattering states.
Thus one of the principal ingredients of our calculation is a 
realistic potential model in order to generate the bound and scattering 
$np$-wave functions.

The leading current contribution is provided by the nonrelativistic 
one-body nucleon current. As another important ingredient, we consider  
subnuclear degrees of freedom (d.o.f.) related to meson exchange 
currents (MEC) and isobar configurations (IC). Because of the 
increasing importance of relativistic effects with increasing energy 
and momentum transfer, we include also 
relativistic contributions of leading order beyond the nonrelativistic 
current. 

In view of the fact that some realistic NN-potentials are defined
in $r$-space while others are in $p$-space, we have employed two
separate corresponding codes.  The $r$-space code is an outgrowth
of the work of {\sc Fabian} and {\sc Arenh\"ovel}~\cite{FaA79} 
incorporating improvements and additions, particularly with respect 
to the most important relativistic 
spin-orbit current and other leading order relativistic contributions.
In this code, isobar configurations are treated in a perturbative
approach~\cite{ArD71} even though we had developed in the past 
a coupled channel 
$r$-space code~\cite{LeA87}. However, in that code the 
$\Delta$-propagator could not be treated as exactly as 
in a $p$-space code. 

The $p$-space code is described in detail in~\cite{RiG97}. Besides 
inclusion of the $N\Delta$-configuration as the most important isobar 
configuration in a coupled channel approach,
it allows furthermore an exact treatment of all leading order 
relativistic contributions to one- and two-body currents as well
as the Lorentz boost for a one-boson-exchange $NN$-interaction model. 
In fact, this code was developed for the {\sc Bonn} $p$-space
potentials (OBEPQ models) of~\cite{MaH87} which are potentials 
of this type with short range cut-off form factors.

A further difference between the two codes lies in their treatment
of the electric multipoles. The $r$-space version incorporates
the {\sc Siegert} operators thereby insuring the inclusion of the
dominant MEC implicitly in electric transitions. In the $p$-space 
version, mainly for historical reasons, the {\sc Siegert} form of the 
electric multipoles is not used. In view of the fact that 
some potentials, e.g.\ the {\sc Bonn} $r$-space, are derived from their 
corresponding $p$-space versions, one can estimate the inherent 
numerical differences between the two codes by using in this case the same 
potential model in both codes. 

The calculation of the $t$-matrix elements is based on an expansion  
of the final state into partial waves with total angular momentum
$j$. The final state interaction (FSI) is taken into account by
solving the corresponding scattering equation for a given
partial wave $|\vec p,j\,m_j\rangle$. Then the electric and magnetic
multipole transitions into this state are evaluated explicitly. For a
given $j$ the contributing multipoles are $L=j-1,\,j,\,j+1$. However,
at some point one has to truncate the series at a maximal angular
momentum $j_{\mathrm{max}}$ for the explicit inclusion of FSI. On the
other hand, the convergence of the partial wave expansion depends on
the kinematics, in fact the convergence is quite slow for the
quasi-free case. The solution to this dilemma is based on the fact that
for higher partial waves the influence of FSI becomes increasingly
unimportant with growing $j$ so that for these the undistorted 
partial waves can be used instead. Thus we include all electric and magnetic
multipoles up to a maximal multipolarity $L_{\mathrm{max}}$ with
consideration of FSI up to $j_{\mathrm{max}}=L_{\mathrm{max}}+1$ and
subtract the corresponding transitions without FSI and add finally the
complete $t$-matrix obtained with a plane wave as 
final state, which we call the plane wave {\sc Born} approximation
(PWBA). This is described in detail in~\cite{FaA79}. For all
kinematics considered in this work $L_{\mathrm{max}}=4$ was found
to be sufficient, i.e.\ the final result did not change if we
increased $L_{\mathrm{max}}$. As already mentioned, the higher  
partial waves are needed only in the vicinity of the quasi-free 
ridge because of the then slow convergence of the partial wave expansion. 
The formal expression for the $t$-matrix in PWBA is given in the 
Appendix~\ref{Born}. We will now explain some details of the separate 
ingredients.

\subsection{$NN$-potential models}\label{potmod}

As mentioned above, the two-body wave functions, needed for the calculation of 
the observables, are based on realistic $NN$-potentials. In past work 
\cite{LeT91,ArL92,ArL95} we 
have employed a number of realistic potentials, such as 
{\sc Nijmegen}~\cite{NaR78}, {\sc Paris}~\cite{LaL80}, 
{\sc Bonn} ($r$- and $p$-space 
versions)~\cite{MaH87}, and {\sc Argonne} V$_{14}$~\cite{WiS84} potentials. 
In general, we found that the dependence of the observables on the choice 
of a realistic potential is rather 
moderate, in particular at low and medium energy and momentum transfers. 
For this reason we have chosen only one semi-modern potential, the 
{\sc Bonn} $p$-space model, for which a consistent meson exchange current has 
been constructed recently including all leading 
order relativistic contributions~\cite{RiG97} to the current operators, 
boost and internal dynamics as explained below in Sect.~\ref{icdof}. 
This potential, as most of the realistic 
$NN$-potentials, is  defined in purely nucleonic space 
without explicit $\Delta$-d.o.f. Since, however, we will also present 
results, where such $\Delta$-d.o.f.\ are treated explicitly in a coupled 
channel approach, we use in addition 
an interaction potential which has been constructed 
recently for this purpose~\cite{RiG97} and which is based on the {\sc Bonn} 
$p$-space potential. Furthermore, we consider as a prototype of recent 
high precision potentials the {\sc Argonne} V$_{18}$ potential~\cite{WiS95}
in the $r$-space code. 

\subsection{One-body currents}\label{OBC}

The one-nucleon current is derived from the nonrelativistic reduction 
of the {\sc Dirac} current, retaining the leading order relativistic 
contributions. The internal nucleon structure is taken care of by including 
the free on-shell nucleon e.m.\ form factors. In the $p$-space 
code we use the {\sc Dirac} and {\sc Pauli} form, $F_1(Q^2)$ and $F_2(Q^2)$, 
respectively, where $Q^2=-q_\mu^2$. Explicit expressions are listed in 
the Appendix of~\cite{RiG97}. The {\sc Sachs} form with $G_E(Q^2)$ for the 
electric and $G_M(Q^2)$ for the magnetic form factor is used in the 
$r$-space code. It is obtained from the {\sc Dirac-Pauli} form by the
transformation
\begin{subequations}
\beqa
F_1(Q^2)&=&\tau(Q^2) (G_E(Q^2)+\frac{Q^2}{4\,M^2}\,G_M(Q^2))\,,\\
F_2(Q^2)&=&\tau(Q^2) (G_E(Q^2)-G_M(Q^2))\,,
\eeqa
\end{subequations}
with $\tau(Q^2)=(1+{Q^2}/{4\,M^2})^{-1}$ 
and neglecting terms of higher order beyond the leading relativistic order. 
Since the difference between these two forms of the one-body current
is of higher relativistic order, the different treatments in the two codes 
does not matter as long as the leading order relativistic contributions 
are included and as long as the kinematics stays within the limits of 
validity of the truncated expansion keeping only the terms of leading 
relativistic order~\cite{BeA92}.

A variety of form factor parametrizations is 
available~\cite{GaK71,BlZ74,HoP76,GaK92,HaM96}.
The largest uncertainty exists for $G_{En}$. Indeed, one of the early 
motivations for investigating deuteron electrodisintegration with polarization 
degrees of freedom was that it could provide a nearly model independent 
method of determining the neutron electric form factor $G_{En}$ 
\cite{Are87,ToA88,ArL88}. Since in these studies the sensitivity of various
observables with respect to $G_{En}$ has been investigated extensively, we
have employed here only one form factor model, namely the dipole model  
including a nonvanishing electric form factor of the neutron in the 
{\sc Galster} parametrization~\cite{GaK71} (with p=5.6).

In principle, one should consider also off-shell effects in the 
one-body current in view of the fact, that the nucleons are 
not free but subject to the hadronic interaction. In such a situation, 
the form factors would acquire an additional dependence on the initial
and final squared four-momenta of the nucleons and, moreover, additional
currents with more off-shell form factors would appear. That such effects 
potentially may be non-negligible has been shown recently for deuteron 
photodisintegration~\cite{ScA01}. In that study off-shell effects were 
evaluated using a simple, but not very realistic pion cloud model for the 
nucleon structure. At present, however, no realistic treatment of such 
off-shell form factors exists, and thus we neglect them here. 

\subsection{Meson exchange currents}\label{MEC}

An important property of realistic $NN$-interaction models is that they
induce two-body meson exchange currents (MEC). Indeed, any isospin and/or 
momentum dependence of an $NN$-potential requires on a formal basis 
the existence of an interaction current in order to satisfy the continuity 
equation. The physics underlying such MECs is related to the coupling of 
the hadronic interaction diagrams to the e.m.\ field. In the case that the 
$NN$-potential is explicitly derived from a meson-exchange model, this 
connection is obvious and one obtains straightforwardly the associated 
nonrelativistic MEC, of which the $\pi$-MEC is the most important one,
as well as the leading order relativistic two-body charge and current 
contributions.  

However, for potential models, which in their  
medium range part use a phenomenological parametrization, the construction 
of a proper exchange current for the isospin dependent potential part is 
not unique, because the connection to the underlying physical process
is obscured. For such cases, a recipe has been developed in the past 
independently by {\sc Riska}~\cite{Ris85} and by {\sc Buchmann} 
{\it et al.}~\cite{BuL85}, 
which is inspired by the genuine meson exchange models. This recipe is 
based on the observation that the spin-isospin dependent central and 
tensor parts of a given $NN$-potential can be split 
into a pion- and rho-exchange-like potential for which the 
corresponding meson exchange currents are known. While the approach of 
{\sc Riska}
is based on the momentum space representation of the potential and thus 
can be applied to any phenomenological potential, the method of 
{\sc Buchmann} {\it et al.}\ was conceived for an $r$-space 
representation of the 
potential as a superposition of appropriate {\sc Yukawa} 
functions, and thus its 
application appeared to be limited to such type of potentials. But recently, 
this approach was extended to potentials with a more general radial 
behaviour by applying a {\sc Laplace} transform~\cite{ArS01}. 

This phenomenological method works for the nonrelativistic MEC
reasonably well, but one should be aware, that its construction
contains some inherent arbitrariness, because the 
$\pi$- and $\rho$-MEC contain purely transverse pieces, which are not
constrained by current conservation and thus can be modified
arbitrarily without destroying the consistency. An example for such a
modification is given in~\cite{BuL85}. Thus, extending this recipe to
the construction of relativistic MEC contributions in a corresponding
manner has to be considered as purely heuristic. 
Fortunately, an important part of MEC
can be incorporated model independently by the use of {\sc Siegert}
operators for the transverse 
electric multipoles~\cite{Are81,GaH81}. In fact, in this way the major
MEC contribution to electric transitions is consistently incorporated
implicitly, at least for low and medium energy and momentum
transfers. In fact, the results in our work~\cite{LeT91,ArL92,ArL95,ArL98} 
labeled as the normal part (N) contain MEC contributions via the 
electric {\sc Siegert} operators implicitly, 
but otherwise no {\em explicit} MEC in either the electric or 
magnetic transitions. However, as already mentioned, 
in the $p$-space code, based on the work of {\sc Ritz} et al.~\cite{RiG97}, 
no {\sc Siegert} operators are used. In this case the results labeled 
``N'' do not contain any MEC implicitly. With respect to the isobar 
configurations, which will be discussed in the next section, we also 
include those isobar-MEC which are induced by the transition potentials. 

A last remark concerns the question of 
e.m.\ form factors for the MEC. Here we take the heuristic approach 
multiplying the isoscalar and isovector pieces by the appropriate isoscalar 
and isovector nucleon form factors. In this way, current conservation holds 
also in the presence of such form factors. 

\subsection{Isobar contributions}\label{icdof}

Isobar d.o.f., describing phenomenologically internal nucleon d.o.f., 
can be incorporated either in the form of effective nonlocal
two-body operators, describing intermediate excitations of one or two 
isobars, or by allowing explicit isobar configurations (IC) in the nuclear 
wave functions, where one or several nucleons are replaced by an isobar, 
and with appropriate strong and e.m.\ operators~\cite{WeA78}. 
In the present work the latter approach is used by admitting
isobars as explicit constituents in the two-body system. This allows 
one also to handle in a natural manner the real excitation of a 
$\Delta$(1232)-resonance for high enough energy transfers above pion 
production threshold. It furthermore 
avoids the often applied static approximation of the effective MEC 
induced by the intermediate excitation of an isobar, 
which has very limited value only. In the present work, we have included 
the $N\Delta$, $NN(1440)$, and $\Delta\Delta$ configurations in the 
$r$-space code while in the $p$-space code only the $N\Delta$ configuration 
is included as the most important one for energy transfers 
up to about 400 MeV. 

The corresponding wave function components, called isobar configurations, 
are obtained either in a perturbative treatment or in a coupled channel 
approach. In the perturbative approach, an isobar configuration, consisting, 
for example of a $N\Delta$ configuration, is generated by just one 
$NN$-collision via a transition potential $NN\rightarrow N\Delta$ for 
which a simple one-boson exchange model is used~\cite{WeA78}. On the other 
hand, in a coupled channel approach one has to renormalize the original 
$NN$-interaction as mentioned above, because being 
fitted in pure $NN$-space to experimental scattering data, it contains 
implicitly already the effect of such intermediate configurations,
 e.g.\ $N\Delta$. This means that in principle one would need to redo 
the fit of the potential parameters if such isobar configurations are 
included explicitly. In order to avoid such involved work, a reliable 
box-subtraction method, first proposed by {\sc Green} and 
{\sc Sainio}~\cite{GrS82}, 
is available and was applied in~\cite{LeA87,RiG97}. In most cases, 
however, we use the simpler perturbative calculation~\cite{WeA78}. 
Only in case of the $N\Delta$-configuration do we also consider a coupled 
channel calculation for the {\sc Bonn} $p$-space potential~\cite{RiG97} as 
mentioned above, because in momentum space the $\Delta$ propagation 
can be treated in a more exact manner compared to an $r$-space 
calculation~\cite{LeA87}. Finally, one- and two-body current operators 
involving isobars have to be considered. In view of the fact, 
that these currents are less well known, we neglect relativistic terms and 
restrict ourselves to their nonrelativistic 
expressions which are given in~\cite{WeA78} for the one-body terms and 
in~\cite{Lei80} for the MEC contributions, again with appropriate e.m.\ 
form factors. 

\subsection{Relativistic contributions}\label{relcon}

Relativistic contributions arise from three sources. These are 
(i) the internal relativistic dynamics in the rest frame of the nucleus, 
(ii) the boost of the intrinsic wave function from the nuclear 
rest frame, here the lab frame, to a moving one because of the nonvanishing 
momentum transfer, and finally (iii) relativistic contributions to the 
interaction operators, here the current operators. In the present 
approach we resort to a $p/M$-expansion, retaining only the leading 
order relativistic terms beyond the nonrelativistic limit. 
The boost of a wave function is described by a unitary operator for 
which the $p/M$-expansion yields in leading order two separate contributions, 
a purely kinematic part, which can be interpreted as the effect of 
{\sc Lorentz} 
contraction and {\sc Thomas-Wigner} spin rotation, and a potential 
dependent part, which only is present for pseudoscalar meson 
exchange~\cite{Fri77}. Relativistic contributions to the one-body current
have been discussed already in Sect.~\ref{OBC}. With respect to MEC, 
consistent treatments are available in~\cite{TrA89,GoA92,TaN92,AdA97}
which are based on a meson-theoretical one-boson-exchange potential 
as nuclear interaction. The construction is more questionable for 
semi-phenomenological potentials like the {\sc Argonne} $V_{18}$, as pointed 
out in Sect.~\ref{MEC}. Even a consistent nonrelativistic 
MEC is ambiguous for such potentials~\cite{Ris85}. Although one can 
proceed also for the relativistic MEC in analogy to purely one-boson-exchange 
models, one should be aware of the inherent ambiguities of such an 
approach. 

A consistent treatment of all three types of contributions is given in 
the work of {\sc G\"oller} and {\sc Arenh\"ovel}~\cite{GoA92} for a pure 
one-pion-exchange model and by {\sc Tamura} et al.~\cite{TaN92} 
for a more general potential type. The work of~\cite{GoA92} has been 
generalized in~\cite{RiG97} to a consistent leading order relativistic
treatment for the $p$-space {\sc Bonn} potentials, 
and our $p$-space code is based on this work. 
In view of the problems associated with the construction of a 
consistent MEC for a semi-phenomenological potential, we
include in the $r$-space code as relativistic current contribution only
the one-body part, containing the most important spin-orbit current,
and the kinematic boost as described in~\cite{WiB93}.
\section{Discussion of Results}\label{examples}

In the following discussion we will use as shorthand ``FSI'' for final 
state interaction, ``PWBA'' for nonrelativistic one-body current with 
plane wave final state, ``RPWBA''  for relativistic one-body current 
with plane wave final state, ``N'' for the 
nonrelativistic normal theory, i.e.\ without explicit meson exchange currents 
and isobar configurations. This means in the case that {\sc Siegert} operators 
are used, that a part of MEC is implicitly included in the electric 
multipoles for ``N'' as is the case for the $r$-space code, whereas 
for the $q$-space code no MEC contributions in ``N'' appear. 
Furthermore, ``MEC''  and ``IC'' stand for the contributions 
of explicit meson exchange currents and isobar configurations, respectively, 
``RC'' for the inclusion of relativistic contributions 
and ``T=N+MEC+IC+RC'' for the complete calculation.

\subsection{Inclusive Observables}

\begin{figure}
\includegraphics[scale=0.6]{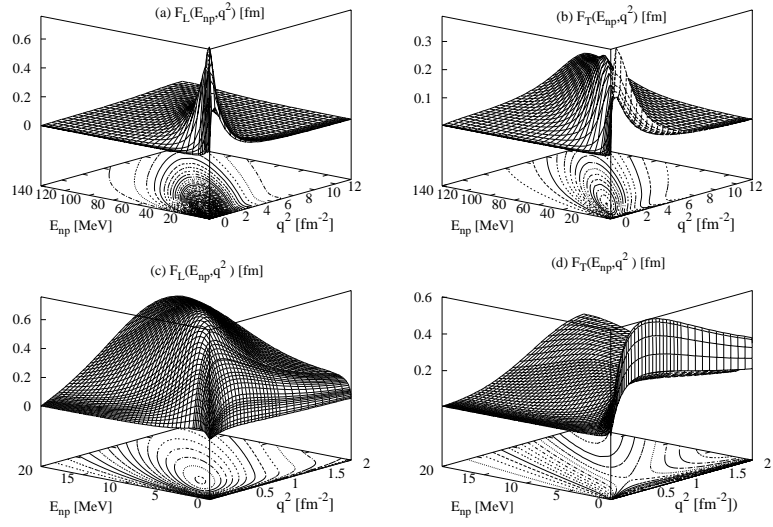}
\caption{Surface and contour plots of longitudinal and transverse 
form factors as function of $E_{np}$ and $q^2$ for
$E_{np}=0-300$~MeV and $q^2=0-25$~fm$^{-2}$ using the 
{\sc Argonne} $V_{18}$ potential.}
\label{fig_ff_unpol}
\end{figure}

\begin{figure}
\includegraphics[scale=0.5]{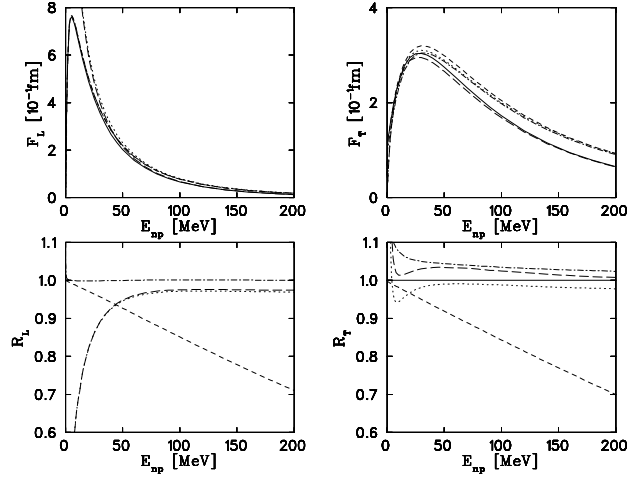}
\caption{Upper panels: 
Longitudinal and transverse form factors along the quasi-free
ridge as function of $E_{np}$ calculated using the {\sc Argonne} V$_{18}$ 
potential. Notation: dotted: PWBA;
long dashed: RPWBA; dash-dot: normal (N), i.e.\ nonrelativistic
approach with FSI included; short dashed:
nonrelativistic MEC and IC included (N+MEC+IC); solid: complete
calculation (T=N+MEC+IC+RC).
Lower panels: Form factor ratios $R_{L/T}=F_{L/T}$(N)$/F_{L/T}$(PWBA) 
(dotted), $R_{L/T}=F_{L/T}$(T)$/F_{L/T}$(RPWBA) (long dashed)
$R_{L/T}=F_{L/T}$(N+MEC+IC)$/F_{L/T}$(N) (dash-dot), and 
$R_{L/T}=F_{L/T}$(T)$/F_{L/T}$(N+MEC+IC) (short dashed).} 
\label{fig_ff_qf_ridge}
\end{figure}

\begin{figure}
\includegraphics[scale=0.6]{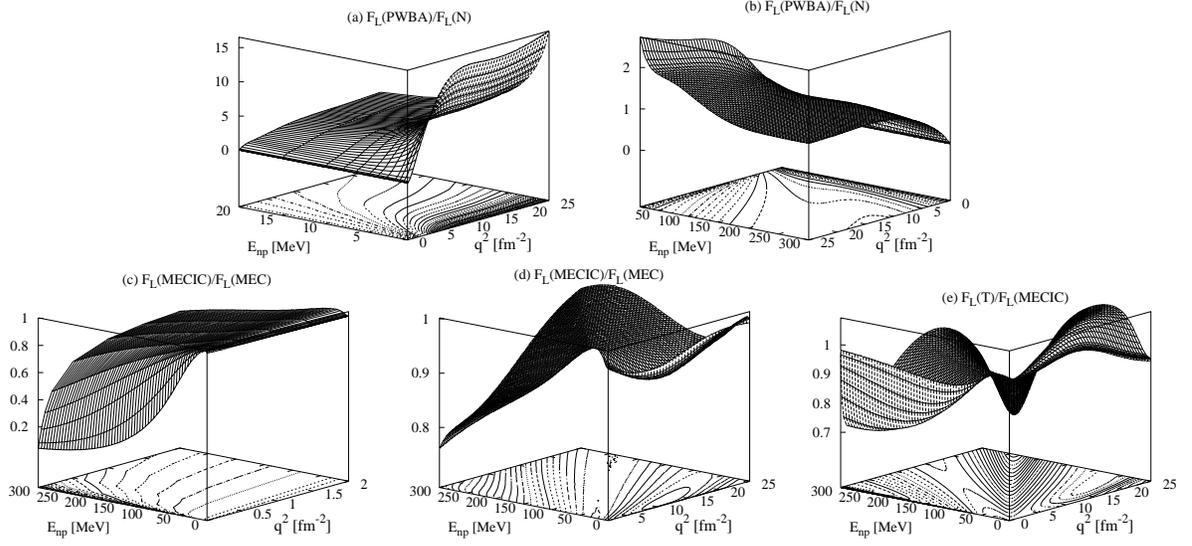}
\caption{Surface and contour plots over the $E_{np}$-$q^2$-plane 
of ratios of longitudinal form factors
for different contributions: (a) influence of FSI, $F_L$(PWBA)/$F_L$(N),
for low $E_{np}=0-20$~MeV and in (b) for $E_{np}=20-300$~MeV; (c) influence 
of IC, $F_L$(MECIC)/$F_L$(MEC), for low $q^2=0-2$~fm$^{-2}$ and in 
(d) for $q^2=2-25$~fm$^{-2}$; (e) influence of RC, 
$F_L$(T)/$F_L$(MECIC).}
\label{fig_ff_ratio_L}
\end{figure}

\begin{figure}
\includegraphics[scale=0.6]{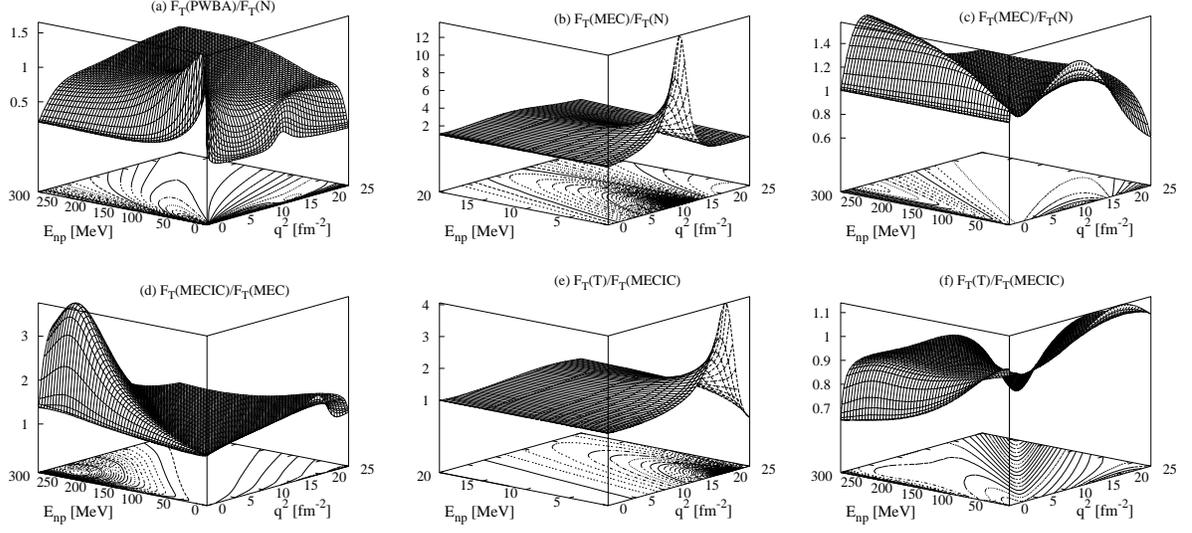}
\caption{Surface and contour plots over the $E_{np}$-$q^2$-plane 
of ratios of transverse form factors
for different contributions: (a) influence of FSI, $F_T$(PWBA)/$F_T$(N); 
(b) influence of MEC, $F_T$(MEC)/$F_T$(N) for low $E_{np}=0-20$~MeV and 
in (c) for $E_{np}=20-300$~MeV; (d) influence of IC, 
$F_T$(MECIC)/$F_T$(MEC); (e) influence of RC, $F_T$(T)/$F_T$(MECIC) 
for low $E_{np}=0-20$~MeV and in (f) for $E_{np}=20-300$~MeV.}
\label{fig_ff_ratio_T}
\end{figure}

\begin{figure}
\includegraphics[scale=0.6]{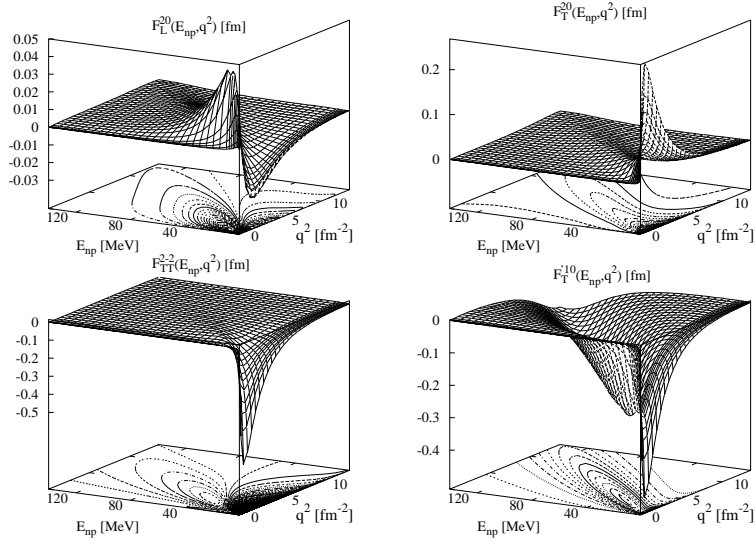}
\caption{Surface and contour plots of polarization form factors 
$F_L^{20}$, $F_T^{20}$, $F_T^{\prime 10}$, and 
$F_{TT}^{2-2}$ for polarized beam and target as function of 
$E_{np}$ and $q^2$ for
$E_{np}=0-160$~MeV and $q^2=0-16$~fm$^{-2}$.}
\label{fig_ff_polarized_L_T}
\end{figure}

\begin{figure}
\includegraphics[scale=0.6]{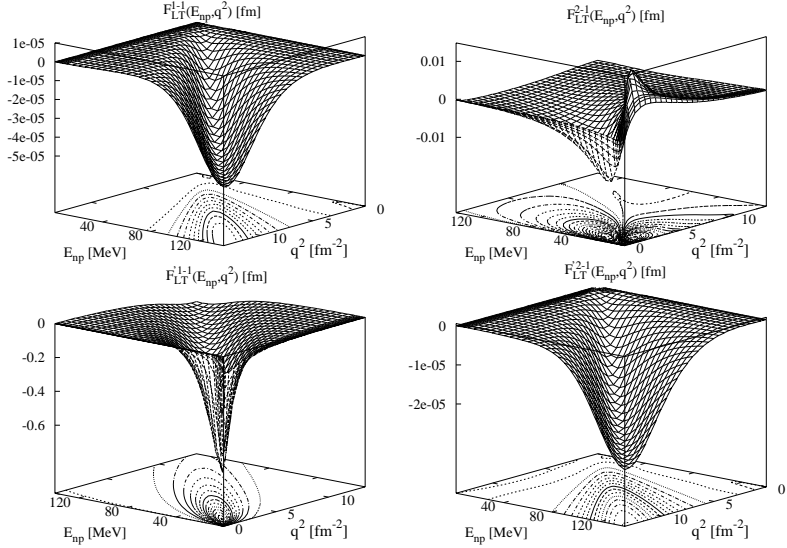}
\caption{Surface and contour plots of polarization 
$LT$-interference form factors for polarized beam and target 
as function of $E_{np}$ and $q^2$ for
$E_{np}=0-160$~MeV and $q^2=0-16$~fm$^{-2}$.}
\label{fig_ff_polarized_LT}
\end{figure}

\begin{figure}
\includegraphics[scale=0.6]{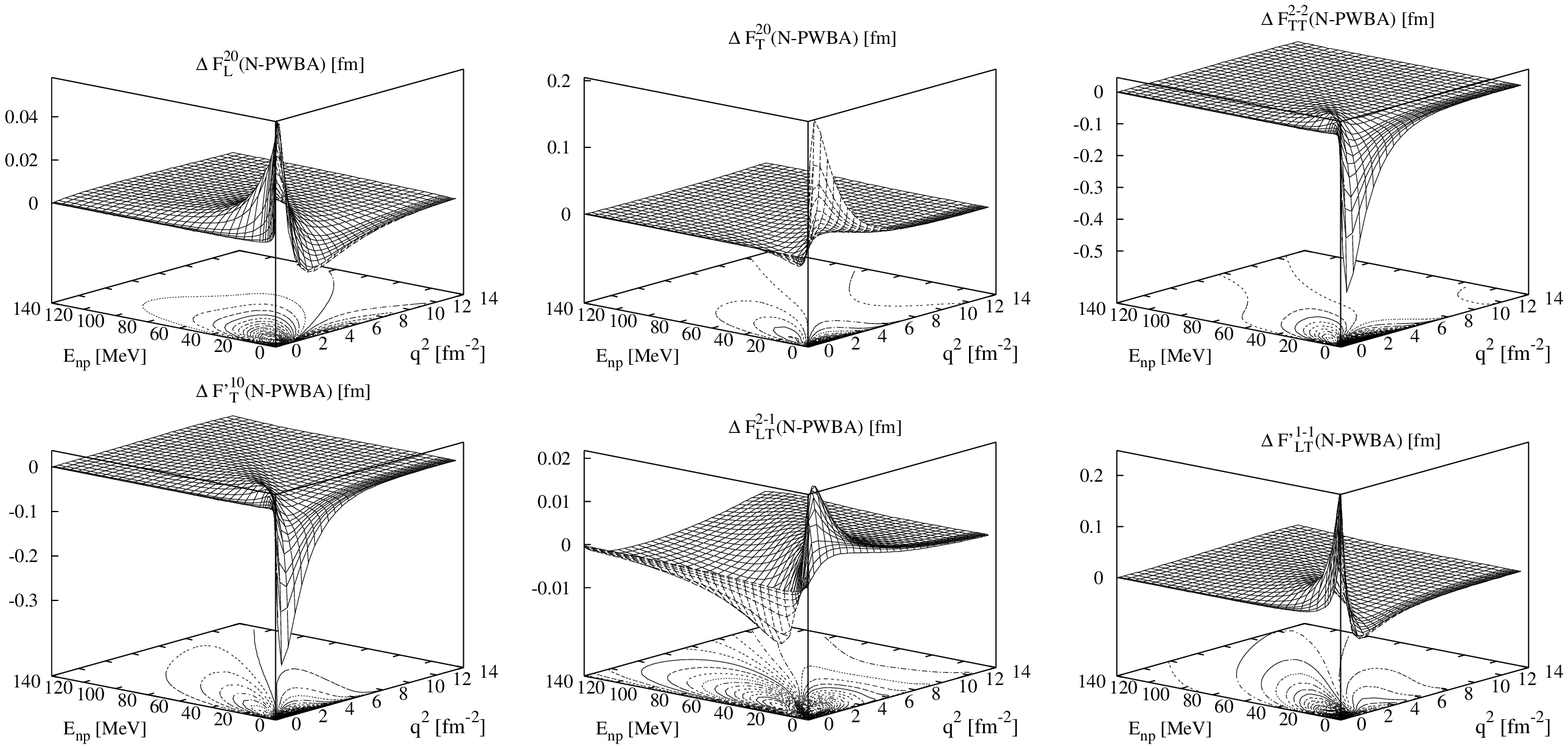}
\caption{Influence of FSI: 
Surface and contour plots of differences of polarization 
form factors calculated in normal nonrelativistic theory (N) and in PWBA
as function of $E_{np}$ and $q^2$ for
$E_{np}=0-140$~MeV and $q^2=0-14$~fm$^{-2}$.}
\label{fig_ff_polarized_diff_n_b}
\end{figure}

\begin{figure}
\includegraphics[scale=0.6]{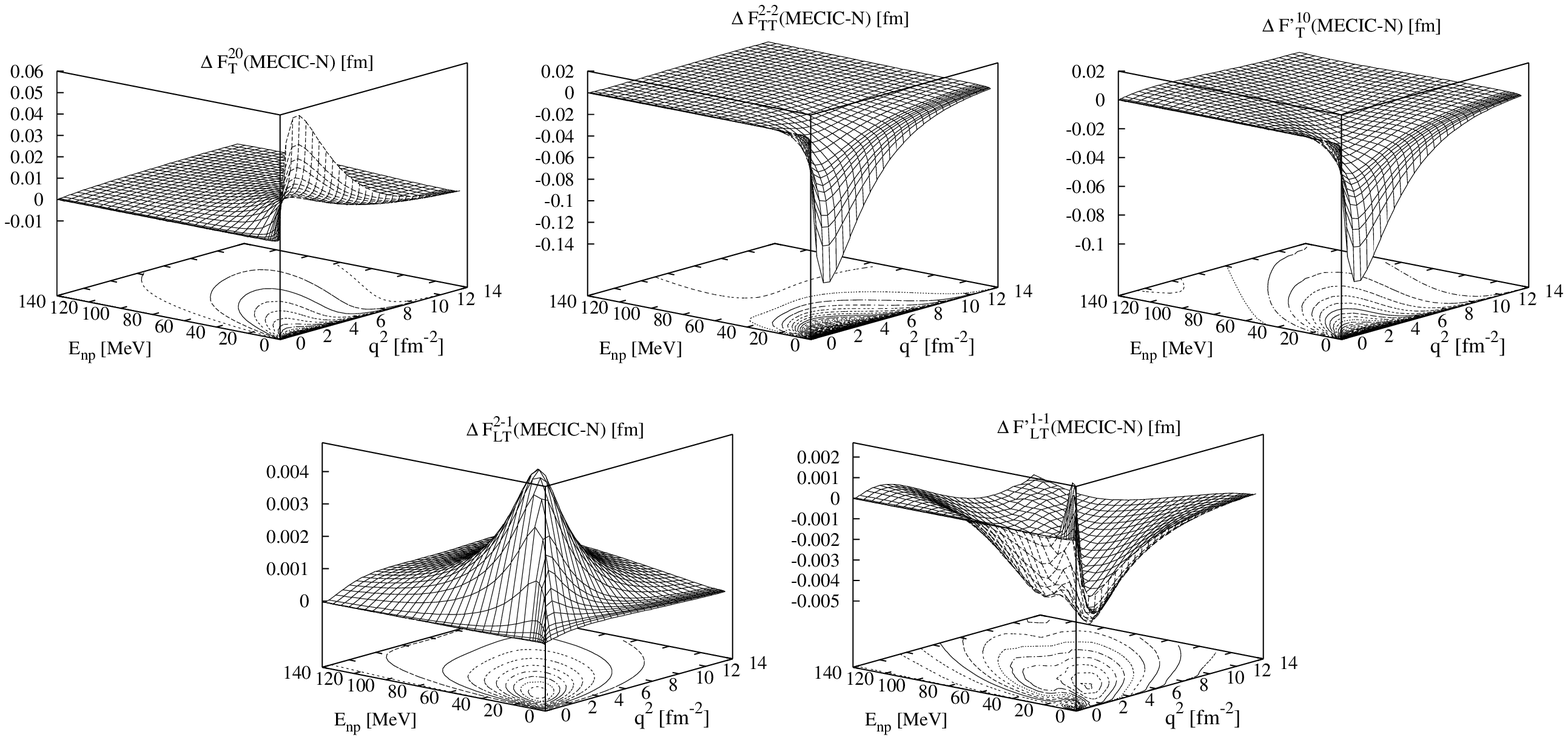}
\caption{Influence of MEC and IC: 
Surface and contour plots of differences of polarization 
form factors calculated with MEC and IC and in normal nonrelativistic 
theory as function of $E_{np}$ and $q^2$ for
$E_{np}=0-140$~MeV and $q^2=0-14$~fm$^{-2}$.}
\label{fig_ff_polarized_diff_mecic_n}
\end{figure}

\begin{figure}
\includegraphics[scale=0.6]{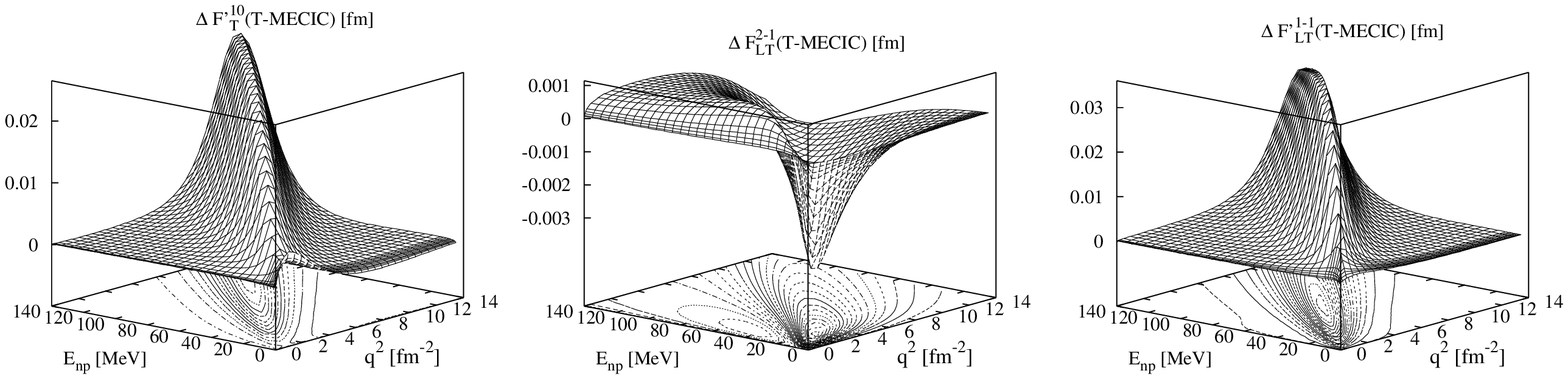}
\caption{Influence of RC: 
Surface and contour plots of differences of polarization 
form factors calculated with MEC and IC and for complete 
theory as function of $E_{np}$ and $q^2$ for
$E_{np}=0-140$~MeV and $q^2=0-14$~fm$^{-2}$.}
\label{fig_ff_polarized_diff_t_mecic}
\end{figure}

The inclusive reaction $d(e,e')np$ for unpolarized beam and target 
is governed by a longitudinal form factor $F_L$ and a transverse one
$F_T$, whereas eight
additional form factors appear if one allows for beam and target
polarization. These form factors depend on two variables, for which we
choose the final state c.m.\ excitation energy $E_{np}$ (see
Eq.~(\ref{E_np})) and the squared c.m.\ three-momentum transfer $\vec
q^{\,2}$ (from now on we use the notation $\vec q=\vec
q^{\,\mathrm{c.m.}}$), i.e.\ $F_{L/T}=F_{L/T}(E_{np},\,\vec
q^{\,2})$. All results presented in 
Figs.~\ref{fig_ff_unpol} through \ref{fig_ff_polarized_diff_t_mecic}
are obtained using the {\sc {\sc Argonne}} $V_{18}$ potential.
In order to give an overview, the upper panels of
Fig.~\ref{fig_ff_unpol} show $F_L$ and $F_T$, calculated with the complete 
theory, in combined surface and
contour plots over the $E_{np}$-$\,q^2$-plane for
$E_{np}=0-300$~MeV and $q^2=0-25$~fm$^{-2}$. 
For a more detailed view of the form factors in the near threshold region 
the lower panels display them in a smaller part of
the $E_{np}$-$\,q^2$-plane, namely for $E_{np}=0-20$~MeV and 
$q^2=0-2$~fm$^{-2}$. 

One readily notes that the quasi-free ridge along $E_{np}$/MeV$\approx
10\,q^{2}$/fm$^{-2}$ is the dominant feature of these two form
factors, where the quasi-free
kinematic is defined by the requirement that the virtual exchanged
photon is absorbed by only one nucleon, which is emitted in the forward
direction, with energy and momentum transfer such that the spectator
nucleon remains at rest in the lab system. This yields the condition
\begin{equation}
E_{np}^{\mathrm{lab}}=M+\sqrt{M^2+(q^{\mathrm{lab}})^2}\,,
\end{equation}
which gives for the invariant mass according to (\ref{Wnp_lab}) using
(\ref{qcm_lab}) 
\beqa
W_{np}&=&2\,M\,\sqrt{1+\frac{q^2}{M_d^2}}\approx M\,(2+\frac{q^2}{M_d^2})\,
\eeqa
or for the final state c.m.\ excitation energy
\beqa
E_{np}&=&2\,M\,(\sqrt{1+\frac{q^2}{M_d^2}}-1)\approx\frac{M}{M_d^2}\,q^2\,.
\eeqa
The latter relation gives the already mentioned rule of thumb 
$E_{np}$/MeV$\approx 10\,q^2$/fm$^{-2}$, a straight line in the 
$E_{np}$-$\,q^2$-plane as is obvious in Fig.~\ref{fig_ff_unpol}. 

The behaviour of the form factors along the quasi-free ridge is
displayed in the upper two panels of Fig.~\ref{fig_ff_qf_ridge} while
the lower two panels show form factor ratios with respect to the
various ingredients and interaction effects. The longitudinal form factor
rises steeply, reaches its maximum at quite low $E_{np}$ around 5~MeV,
and falls off very rapidly with increasing $E_{np}$. On the other
hand, $F_T$ rises considerably slower and reaches its maximum only
around $E_{np}=30$~MeV. Also the fall off is much slower compared to
$F_L$. One furthermore notes that for $F_L$ the effect of MEC and
IC is unimportant while for $F_T$ these are still sizeable above 
$E_{np}=20$~MeV and of the order of several percent decreasing slowly 
with growing $E_{np}$ as shown by the dash-dotted curves in the 
lower panels. Also the influence of FSI becomes quite unimportant 
above $E_{np}\approx 30$~MeV as seen in the upper panels by comparing the 
dotted curves (nonrelativistic PWBA) with the short dashed curves
(N+MEC+IC). Quantitatively one finds from the ratios in the lower panels 
(dotted curves) that above $E_{np}\approx 100$~MeV an almost constant 
difference of a few percent remains. The analogous ratios with relativistic
contributions included, i.e.\ $F_{L/T}$(T)$/F_{L/T}$(RPWBA) show
a similar behaviour (long dashed curves). The reason for the large 
overestimation of $F_L$ in PWBA at low $E_{np}$ has 
its origin in the fact that the final state plane wave is not
orthogonal to the deuteron bound state so that the charge monopole
transition is not suppressed near threshold. On the other hand,
$F_T$ is strongly underestimated in PWBA because of the absence 
of the resonance in the $^1S_0$ state. The only notable effect 
arises from relativistic contributions leading for both form factors 
to a sizeable reduction which increases almost linearly with $E_{np}$. 
However, these RC are
quite well accounted for in the relativistic RPWBA as is demonstrated
by the little difference between the complete calculation and the 
relativistic RPWBA their ratios approaching one with increasing $E_{np}$
(long dashed curves). 

We now will turn to the near threshold behaviour shown in the 
lower panels of Fig.~\ref{fig_ff_unpol}. 
One readily notes that for $E_{np}\rightarrow 0$ along $q^2$=const.\
$F_L$ runs first through a broad maximum and then decreases rapidly to
zero while $F_T$ rises dramatically resulting in a very
sharp peak right above threshold. The peak height grows first with 
increasing $q^2$, reaches its maximum around $q^2=0.5$~fm$^{-2}$ 
and then falls off. The rapid decrease of $F_L$ and the 
sharp peak of $F_T$ near $E_{np}\approx 0$ is a consequence of the fact, 
that close to break-up threshold the $^1S_0$-scattering state, the
so-called anti-bound state, dominates the final state into which the
Coulomb monopole transition is forbidden while 
one has a very strong magnetic dipole isovector transition which is
further enhanced by MEC and IC contributions. Above $E_{np}\approx 4$~MeV and
$q^2\approx 0.4$~fm$^{-2}$ both the surface plot as well as the
contour lines exhibit clearly the onset of the quasi-free ridge in both
form factors. 

The relative influence of the various interaction effects are shown as
ratios in Fig.~\ref{fig_ff_ratio_L} for the longitudinal form factor. 
Because of the strong influence of FSI on $F_L$ near
threshold, we show for this form factor the ratio $F_L$(PWBA)/$F_L$(N)
separately for the region $E_{np}\leq 20$~MeV in
Fig.~\ref{fig_ff_ratio_L}~(a), exhibiting the 
already noted very strong effect of FSI, resulting in a strong increase 
of this ratio for $q^2\rightarrow 0$ along
$E_{np}=$const., whereas $F_L$(PWBA)/$F_L$(N)$\rightarrow 0$ for
$E_{np}\rightarrow 0$ along $q^2=$const. For the remaining part
($20\leq E_{np}/\mathrm{MeV}\leq 300$) this ratio is shown in
Fig.~\ref{fig_ff_ratio_L}~(b). The influence of FSI 
is minimal on top of the quasi-free ridge while it leads
to an increase above this ridge, i.e.\ for 
$E_{np}/\mathrm{MeV}>10\,q^2/\mathrm{fm}^{-2}$,
and to a decrease below. MEC have almost no effect on $F_L$ and thus
are not shown here, because the dominant nonrelativistic $\pi$-MEC
does not contribute to the charge density. The effect of isobar
configurations on 
$F_L$ is shown in Fig.~\ref{fig_ff_ratio_L}~(c) for $q^2\leq
2$~fm$^{-2}$ and in (d) for $q^2\ge 2$~fm$^{-2}$. Above the quasi-free
ridge $F_L$ is reduced by IC. In particular for low $q^2$
close to zero and $E_{np}$ approaching the $\Delta$-excitation
region, $F_L$ decreases drastically by IC to about 10 percent. Going
below the quasi-free ridge, one again notes a reduction but of
smaller size which diminishes when approaching small $E_{np}$. Furthermore,
RC, shown in Fig.~\ref{fig_ff_ratio_L}~(e), exhibit an interesting
behaviour: along the quasi-free ridge one finds 
a distinctive valley describing the increasing reduction by RC 
with increasing $E_{np}$ or $q^2$ as was already apparent in
Fig.~\ref{fig_ff_qf_ridge} (upper right panel). 
Away from the quasi-free ridge the
influence of RC diminishes first on both sides to almost zero and
increases then again when approaching the regions of higher $E_{np}$
and lower $q^2$ or vice versa. 

The corresponding ratios for $F_T$ are displayed in
Fig.~\ref{fig_ff_ratio_T}. For this form factor the effect of FSI is
much more pronounced off the quasi-free ridge than in $F_L$ as is
shown by the ratio $F_T$(PWBA)/$F_T$(N) in part (a). One finds
a very strong decrease on both sides, i.e.\ at low $E_{np}$ with
increasing $q^2$ as well as at low $q^2$ but increasing $E_{np}$, 
which means a strong enhancement by FSI. For
the display of MEC effects we show in part (b) the region of low
$E_{np}$ and in part (c) the remaining region. As already mentioned,
along the quasi-free ridge one finds little influence. But going away
from this ridge, MEC lead to a sizeable increase as shown in part (c),
especially strong close to threshold (see part (b)) up to about
$q^2=15$~fm$^{-2}$. For higher momentum transfers MEC result in a
reduction. This behaviour is well known and in agreement with 
experimental data~\cite{Aue85,Boe90}. Isobar effects displayed in
Fig.~\ref{fig_ff_ratio_T}~(d) become quite pronounced only in the
region of $\Delta$ excitation near $E_{np}=260$~MeV and for not too
high momentum transfers. Finally, relativistic contributions lead 
for $E_{np}$ near threshold and $q^2$ around $15-20$~fm$^{-2}$ to 
quite a significant increase as
shown in Fig.~\ref{fig_ff_ratio_T}~(e). For the remaining region in
Fig.~\ref{fig_ff_ratio_T}~(f) one notes again a significant reduction
along the quasi-free ridge as in $F_L$ in the form of a pronounced
valley. 

A survey for the additional form factors for polarized beam and target,
calculated with the complete 
theory, is shown in Fig.~\ref{fig_ff_polarized_L_T} for $F_L^{20}$,
$F_T^{20}$, $F_T^{\prime 10}$, and $F_{TT}^{2-2}$ and in
Fig.~\ref{fig_ff_polarized_LT} for the $LT$-interference form factors 
as surface and contour plots over a smaller portion of the 
$E_{np}$-$\,q^2$-plane, i.e.\ $E_{np}=0-160$~MeV and 
$q^2=0-16$~fm$^{-2}$. The largest polarization form factors
are $F_T^{20}$, $F_{TT}^{2-2}$, 
$F_T^{\prime 10}$, and $F_{LT}^{\prime 1-1}$ which are of the same 
order of magnitude as the unpolarized form factors. An order of 
magnitude smaller are $F_L^{20}$ and $F_{LT}^{2-1}$. The remaining
two, $F_{LT}^{1-1}$ and $F_{LT}^{\prime 2-1}$, are three orders of
magnitude smaller although they increase slightly in size along the
quasi-free ridge. 

The influence of FSI, MEC, IC and RC cannot be displayed
as ratios because these form factors have zeros. For this reason we show 
only for the larger form factors the most important influences by plotting
differences for FSI (N-PWBA) in Fig.~\ref{fig_ff_polarized_diff_n_b}, for 
combined contribution from MEC and IC (MECIC-N) in 
Fig.~\ref{fig_ff_polarized_diff_mecic_n}, and for RC (T-MECIC) in 
Fig.~\ref{fig_ff_polarized_diff_t_mecic}. Final state interaction is 
very important for all polarization form factors shown in 
Fig.~\ref{fig_ff_polarized_diff_n_b} in the region of low $E_{np}=0-40$~MeV 
and low $q^2=0-4$~fm$^{-2}$. MEC and IC effects displayed in 
Fig.~\ref{fig_ff_polarized_diff_mecic_n} are substantial in 
the transverse form factors $F_T^{20}$, $F_{TT}^{2-2}$, and 
$F_T^{\prime 10}$, mostly below the quasi-free ridge, 
whereas they are of minor importance in the interference form factors 
$F_{LT}^{2-1}$ and $F_{LT}^{\prime 1-1}$, the largest effect being near the 
quasi-free ridge. Relativistic contributions show up 
in Fig.~\ref{fig_ff_polarized_diff_t_mecic} along the quasi-free ridge in 
the primed form factors $F_T^{\prime 10}$ and $F_{LT}^{\prime 1-1}$ as 
well as in $F_{LT}^{2-1}$. 

\begin{figure}
\begin{minipage}[t]{0.47\textwidth}
\includegraphics[scale=0.45]{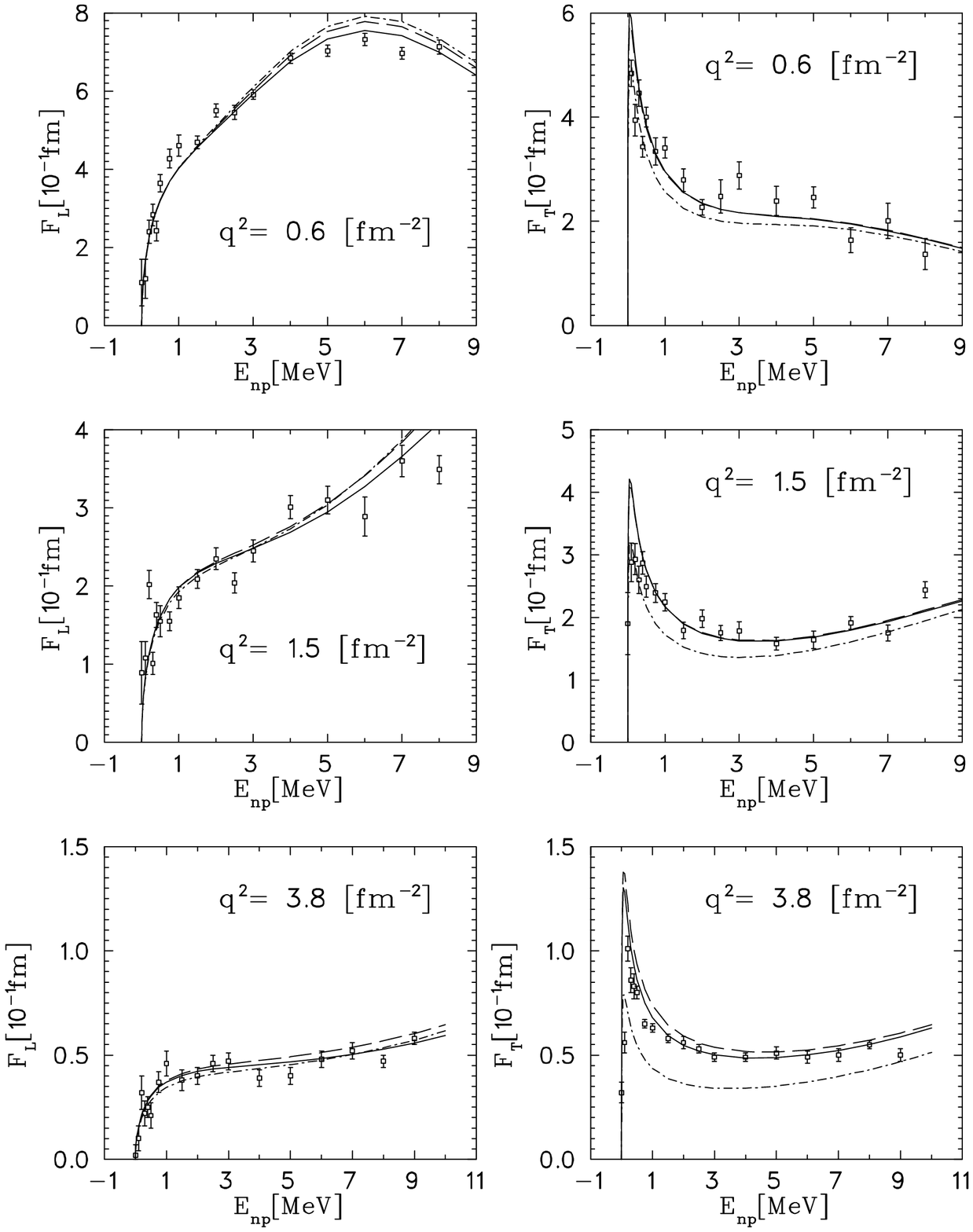}
\caption{Longitudinal and transverse form factors with various interaction
effects. Notation: dash-dot: N; dashed: N+MEC+IC; 
solid: N+MEC+IC+RC. Experimental data from {\sc Simon} et al.~\protect\cite{SiW79}.}
\label{fig_simon_bonn}
\end{minipage}\hfill
\begin{minipage}[t]{0.47\textwidth}
\includegraphics[scale=0.45]{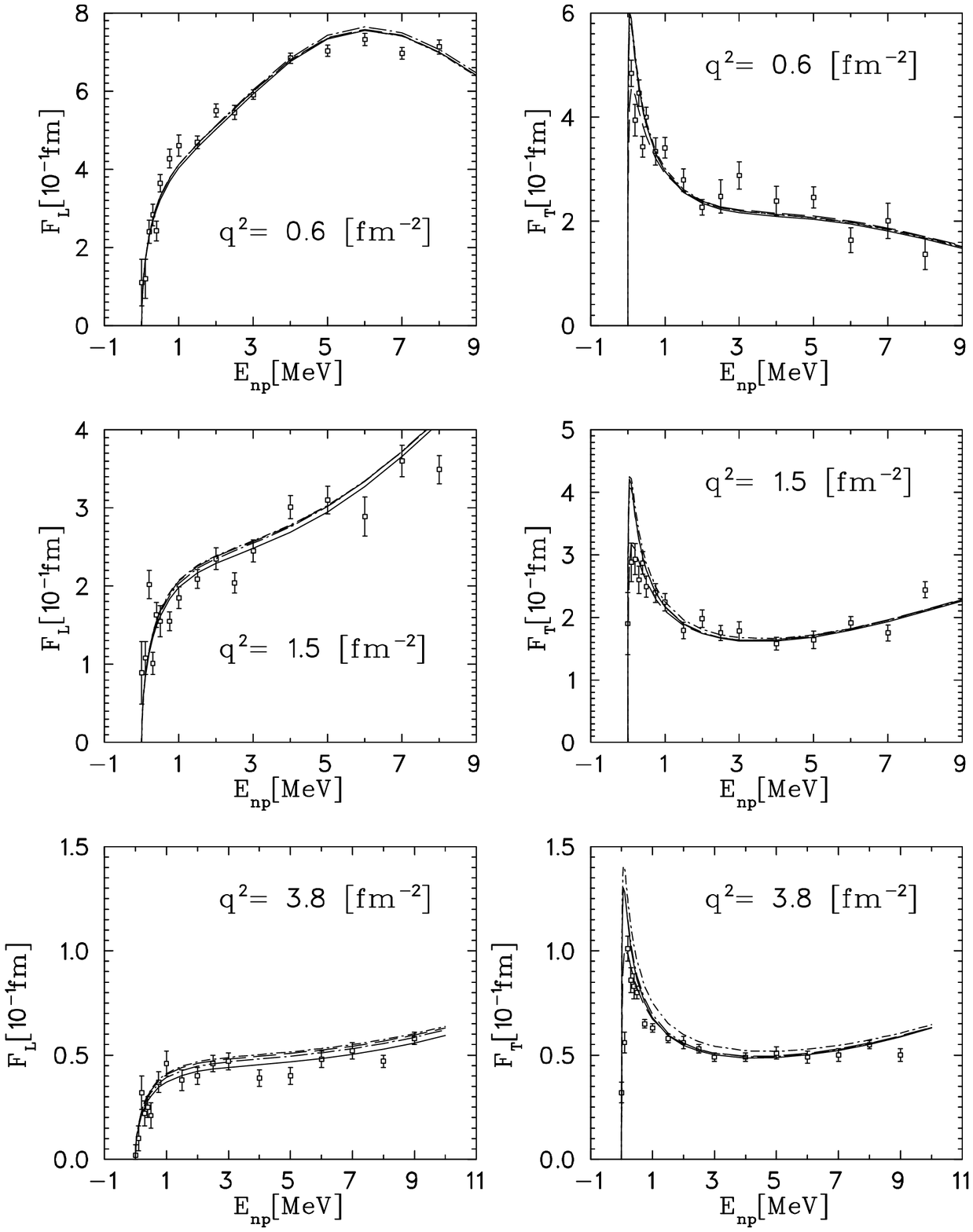}
\caption{Longitudinal and transverse form factors for various potential 
models. Notation: dashed: {\sc Paris}; long dash-dot: {\sc Argonne} V$_{14}$; 
short dash-dot: {\sc Argonne} V$_{18}$; solid: {\sc Bonn}-Qb. 
Experimental data from from {\sc Simon} et al.~\protect\cite{SiW79}.}
\label{fig_simon_pot}
\end{minipage}
\end{figure}

Now we will turn to a comparison with experimental data. Most
extensively discussed has been the inclusive electrodisintegration
near threshold, averaged over $E_{np}=0$ through 3-5~MeV, at
higher momentum transfers (see~\cite{Aue85,Boe90}) and we will not repeat the
discussion here. Instead we will compare the theory to the
experimental data for $F_L$ and $F_T$ as obtained by a Rosenbluth
separation in the near threshold region for various momentum transfers
by {\sc Simon} et al.~\cite{SiW79}. The only comparison to theory 
reported in that work was for a nonrelativistic treatment employing 
the {\sc Paris}
potential and $\pi$-MEC, and a satisfactory agreement was found. In
the light of the considerable progress which theory has achieved since
then, it appears timely to compare the modern approaches with those
data. Thus we confront in Fig.~\ref{fig_simon_bonn} three data sets
of~\cite{SiW79} corresponding to the squared momentum transfers
$q^2\approx 0.6,\,1.5$, and 3.8~fm$^{-2}$ for $E_{np}=0-9$~MeV with various
interaction effects, i.e.\ normal nonrelativistic approach (N), with
inclusion of meson exchange currents and isobar configurations
(N+MEC+IC), and the complete theory including relativistic
contributions (T=N+MEC+IC+RC) for the {\sc Bonn}-Qb potential. 

As expected,
for $F_L$ the interaction effects from MEC and IC are very small
though not completely negligible. Relativistic effects show some
influence resulting in a small reduction above $E_{np}\approx
5$~MeV which increases slightly in size with $E_{np}$. The
agreement of the full theory with the data for $F_L$ is quite satisfactory. 
One should keep in mind, that no open parameter has been fit in this
comparison. 

\begin{figure}
\includegraphics[scale=0.7]{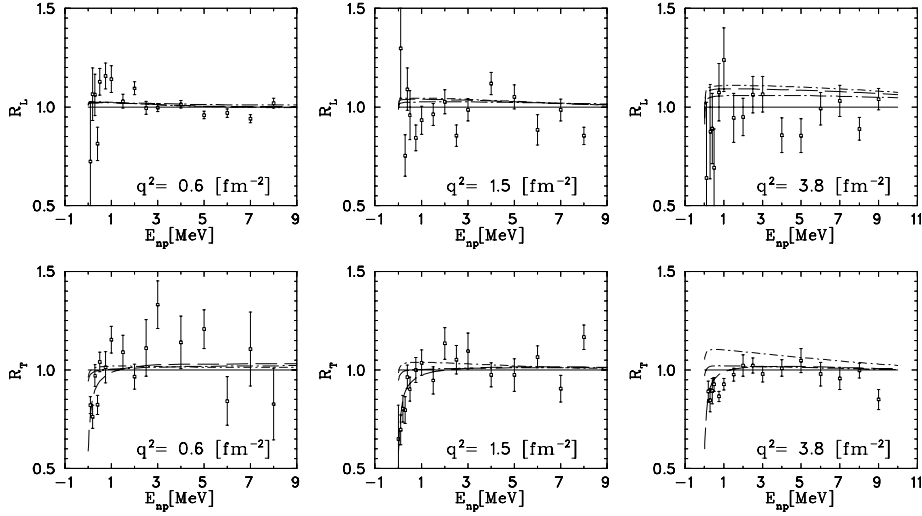}
\caption{Ratios of the longitudinal and transverse form factors 
for various potential models with respect to the {\sc Bonn}-Qb potential. 
Notation as in Fig.~\ref{fig_simon_pot}.
Experimental data from from {\sc Simon} et al.~\protect\cite{SiW79}.}
\label{fig_simon_pot_rel}
\end{figure}

The transverse form factor $F_T$ shows right above threshold the
already noted distinctive peak arising from the $M1$-transition into
the anti-bound $^1S_0$-state. Here the 
influence of MEC and IC is quite significant, especially in the peak
region, and it grows sizeably with increasing squared momentum
transfer. On the other hand, RC lead only to a slight reduction for 
$q^2=3.8$~fm$^{-2}$ while for the other two cases they show no effect at
all. The agreement with the data is in general quite good except 
for the peak where the theory lies above the data. It is very likely, 
however, that the experimental resolution was insufficient to resolve this
very sharp peak. The potential model dependence is shown in
Fig.~\ref{fig_simon_pot} using the {\sc Bonn}-Qb, the two {\sc Argonne}
potentials V$_{14}$ and V$_{18}$, and the {\sc Paris} potential. The
variation of the predicted form factors by these potentials is much
smaller in size than the size of interaction 
effects. The only exception is the prediction of the peak height for the
{\sc Paris} potential, which gives a lower value. The reason for this lies 
in the {\sc Paris} potential's prediction of too small a value for the 
$np$-scattering length. Finally,
in order to give a more detailed and quantitative 
comparison with experiment we show
in Fig.~\ref{fig_simon_pot_rel} the ratios of the data and the various 
theoretical model predictions to the results obtained with the {\sc Bonn}-QB 
potential. It is obvious that the variation with the potential model is
substantially smaller than the experimental errors. Certainly, much 
more precise data, in particular at the threshold peak, 
are needed in order to put the theory to a more 
critical test.

\begin{figure}
\begin{minipage}[t]{0.47\textwidth}
\includegraphics[scale=0.6]{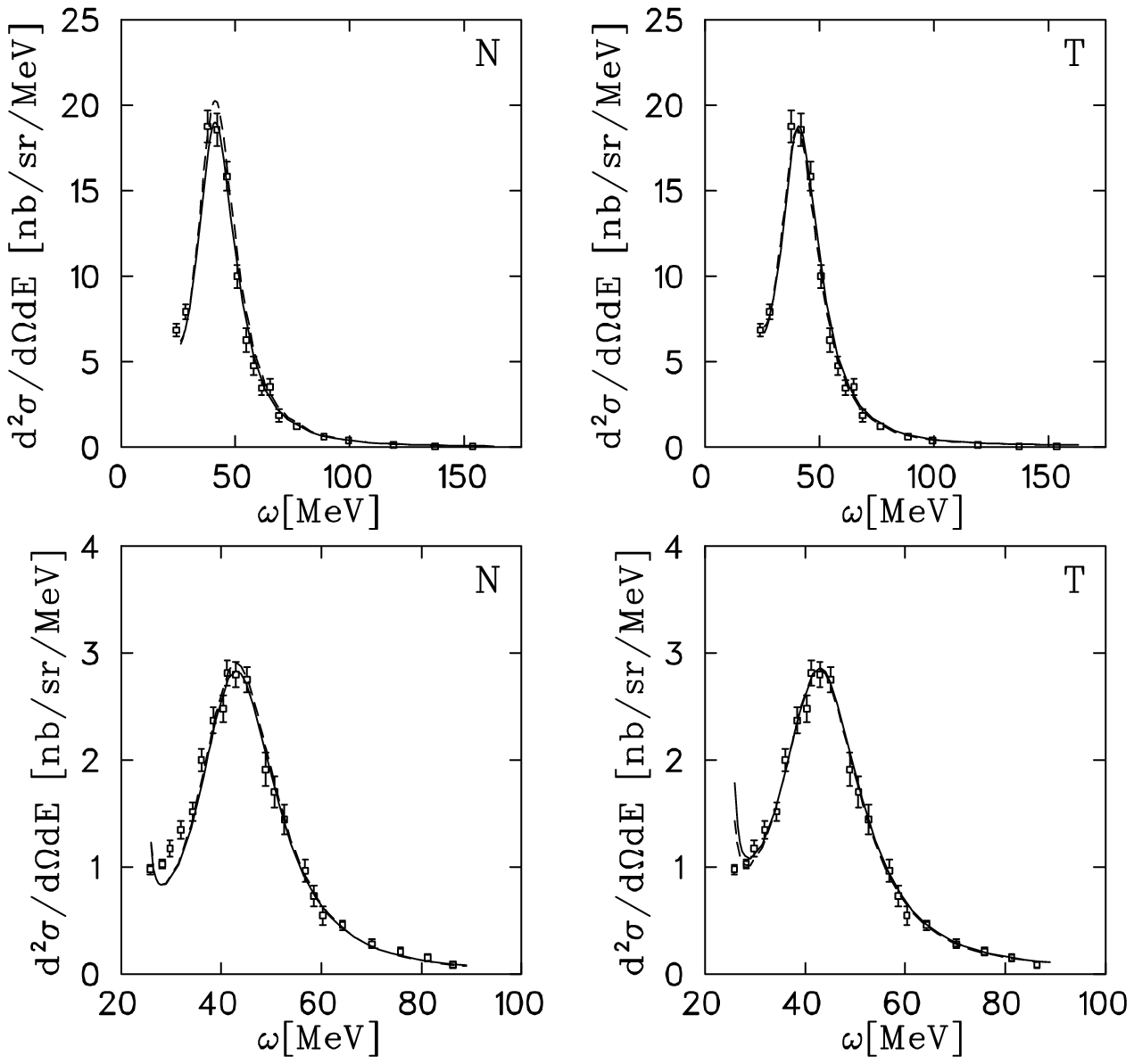}
\caption{Inclusive cross sections $d(e,e')$. Upper panels for 
$E_1^{\mathrm{lab}}=292.8$~MeV and $\theta_e^{\mathrm{lab}}=60^\circ$. 
Lower panels for $E_1^{\mathrm{lab}}=174.3$~MeV and 
$\theta_e^{\mathrm{lab}}=134.5^\circ$. Exp.\ from {\sc Quinn} et 
al.~\protect\cite{QuB88}. Notation of curves: Left panels for nonrelativistic 
normal theory (N) and right panels for complete theory (T) for {\sc Bonn}-Qb 
(dashed) and {\sc Argonne} V$_{18}$ (solid) potentials.}
\label{fig_quinn_cross1}
\end{minipage}\hfill
\begin{minipage}[t]{0.47\textwidth}
\includegraphics[scale=0.6]{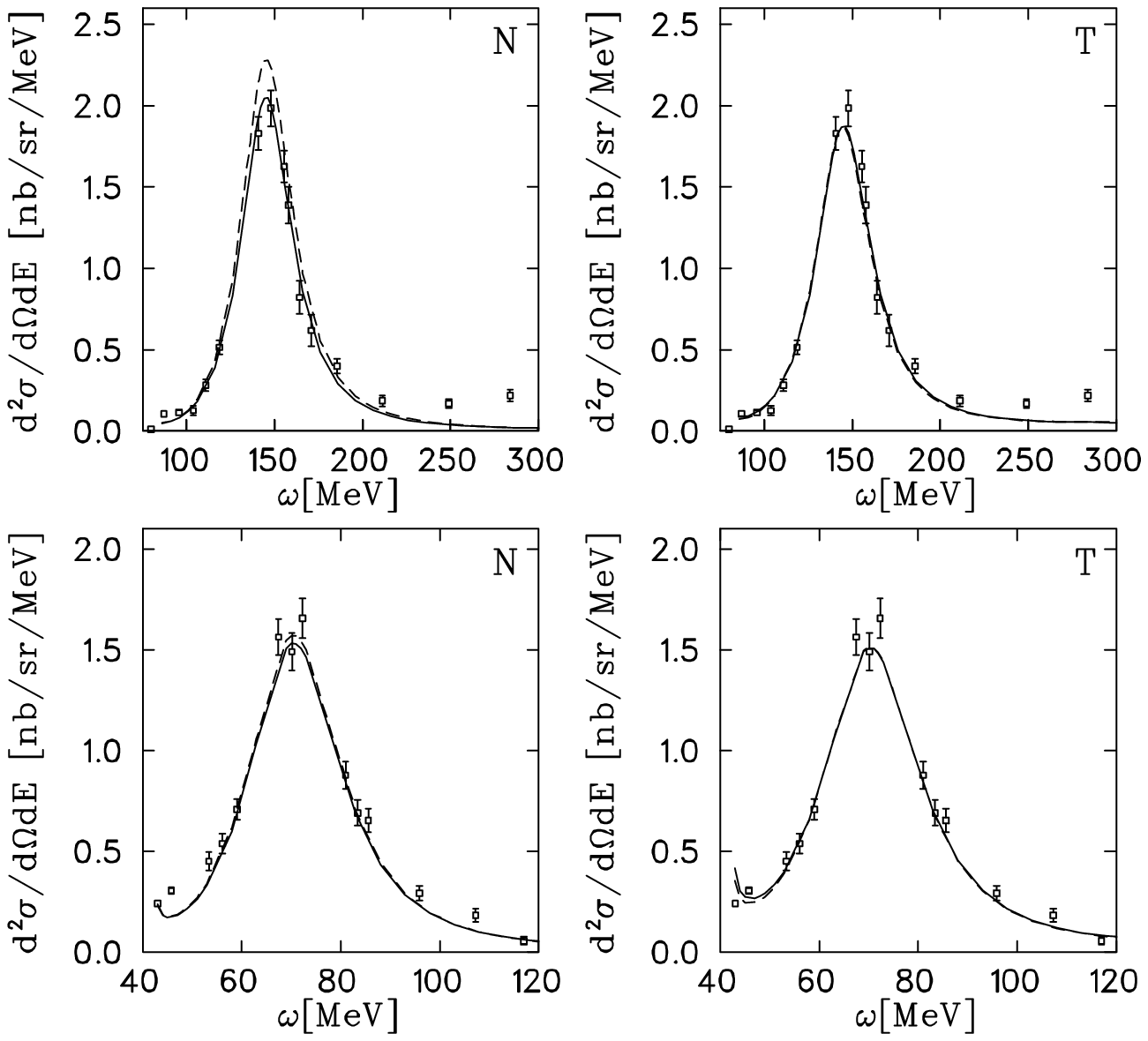}
\caption{Inclusive cross sections $d(e,e')$. Upper panels for 
$E_1^{\mathrm{lab}}=596.8$~MeV and $\theta_e^{\mathrm{lab}}=60^\circ$. 
Lower panels for $E_1^{\mathrm{lab}}=367.7$~MeV and 
$\theta_e^{\mathrm{lab}}=134.5^\circ$. Exp.\ from {\sc Quinn} et 
al.~\protect\cite{QuB88}. Notation of curves: Left panels for nonrelativistic 
normal theory (N) and right panels for complete theory (T) for {\sc Bonn}-Qb 
(dashed) and {\sc Argonne} V$_{18}$ (solid) potentials.}
\label{fig_quinn_cross2}
\end{minipage}
\end{figure}

Another set of data for the inclusive reaction at higher excitation
energy $E_{np}$ and higher momentum transfers is provided by {\sc Quinn} et
al.~\cite{QuB88}. We show first a comparison between theory and
experiment for some inclusive cross sections corresponding to several
different kinematics reported by {\sc Quinn} et al.\ which
exhibit nicely the quasi-free peak, namely at $E_{np}\approx
20$~MeV in Fig.~\ref{fig_quinn_cross1} and at $E_{np}\approx
70$~MeV in Fig.~\ref{fig_quinn_cross2}. The upper panels refer to 
forward angles (60$^\circ$) and the lower ones to backward angles
(134.5$^\circ$). In the left panels the predictions of the
nonrelativistic normal theory obtained with the {\sc Bonn}-Qb and {\sc Argonne}
V$_{18}$ potentials is compared to the data and in the right panels
the complete theory. The normal theory results in a slight
potential dependence as is apparent in the quasi-free peak with the
results for the {\sc Bonn}-Qb potential slightly higher and more pronounced
for the 60$^\circ$-data. However, this potential model dependence is
very much reduced for the complete theory. The most significant
improvement of the full calculation is seen in the near threshold
region at backward scattering angles (see lower panels). In general
the agreement is quite good. The underestimation above
$\omega=200$~MeV in the upper panels of Fig.~\ref{fig_quinn_cross2} has
its origin in the absence of pion production contributions in the theory 
which become significant in this region. 

\begin{figure}
\begin{minipage}[t]{0.47\textwidth}
\includegraphics[scale=.6]{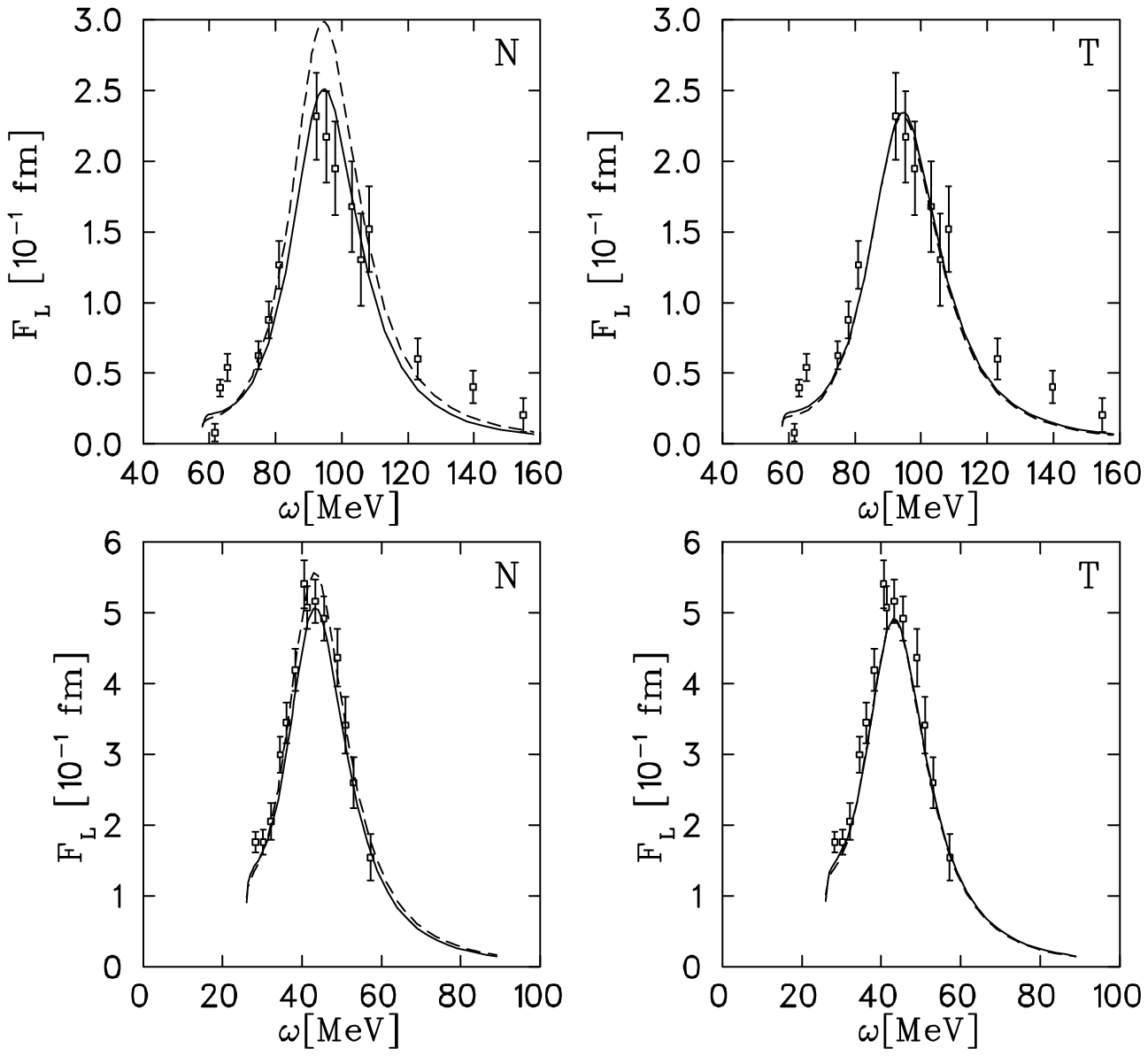}
\caption{Longitudinal form factors. Upper panels for
$E_1^{\mathrm{lab}}=278.5$~MeV and
$\theta_e^{\mathrm{lab}}=134.5^\circ$ and lower panels for 
$E_1^{\mathrm{lab}}=174.3$~MeV and
$\theta_e^{\mathrm{lab}}=134.5^\circ$. 
Exp.\  from {\sc Quinn} et al.~\protect\cite{QuB88}.
Notation of curves: Left panels for nonrelativistic 
normal theory (N) and right panels for complete theory (T) for {\sc Bonn}-Qb 
(dashed) and {\sc Argonne} V$_{18}$ (solid) potentials.}
\label{fig_quinn_ff_L}
\end{minipage}\hfill
\begin{minipage}[t]{0.47\textwidth}
\includegraphics[scale=.6]{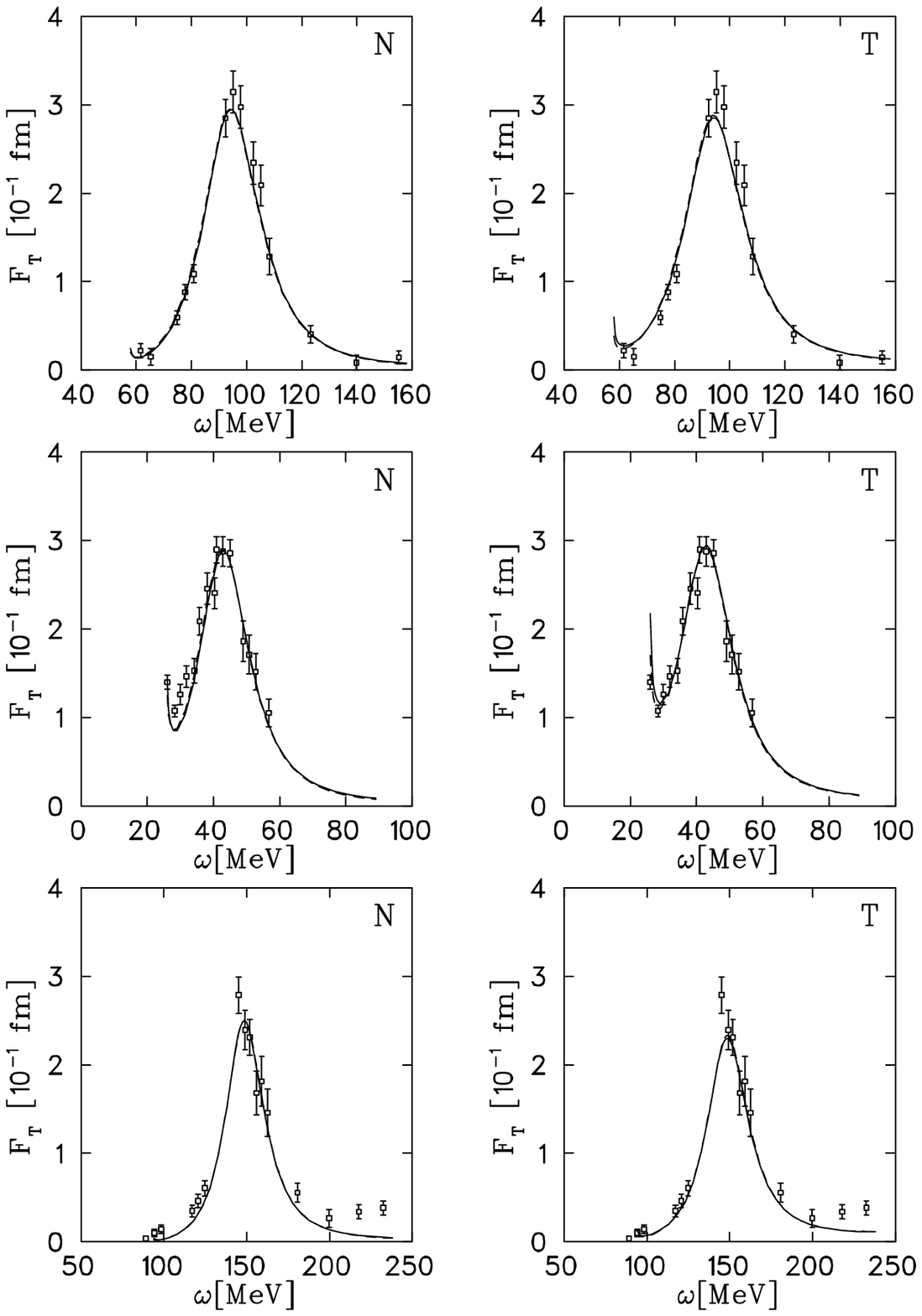}
\caption{Transverse form factors. Upper panels for
$E_1^{\mathrm{lab}}=278.5$~MeV and
$\theta_e^{\mathrm{lab}}=134.5^\circ$, middle panels for 
$E_1^{\mathrm{lab}}=174.3$~MeV and
$\theta_e^{\mathrm{lab}}=134.5^\circ$, and lower panels for 
$E_1^{\mathrm{lab}}=367.7$~MeV and
$\theta_e^{\mathrm{lab}}=134.5^\circ$. Exp.\  from {\sc Quinn} 
et al.~\protect\cite{QuB88}.
Notation of curves: Left panels for nonrelativistic 
normal theory (N) and right panels for complete theory (T) for {\sc Bonn}-Qb 
(dashed) and {\sc Argonne} V$_{18}$ (solid) potentials.}
\label{fig_quinn_ff_T}
\end{minipage}
\end{figure}

The fact, that the model
dependence of the nonrelativistic normal theory is stronger for
forward scattering angles, points to a stronger model dependence of $F_L$
compared to $F_T$. Indeed, this is confirmed by the comparison of the
theory with the 
experimentally determined longitudinal and transverse form factors in 
Figs.~\ref{fig_quinn_ff_L} and \ref{fig_quinn_ff_T}. However, this model 
dependence disappears almost completely for the full theory.

One should keep
in mind that a direct comparison with the response functions $R_{L/T}$
reported in~\cite{QuB88} is not possible because of different
definitions. However, the relations of $R_{L/T}$ to our form factors
is easily obtained by comparing the formal expressions for the
inclusive cross section. In Ref.~\cite{QuB88} the response functions
$R_{L/T}$ are defined by 
\begin{equation}
\frac{d \sigma}{dk _2 ^{\mathrm{lab}} d \Omega _e ^{\mathrm{lab}}}=
\sigma_{\mathrm{Mott}}\,\Big(\xi^2R_L+(\eta+\frac{\xi}{2})\,R_T\Big)\,,
\end{equation}
whereas we use according to (\ref{incl_cross}) in conjunction with 
(\ref{mott})
\beqa
\frac{d \sigma}{dk _2 ^{\mathrm{lab}} d \Omega _e ^{\mathrm{lab}}}&=&
6\,c(k_1^{\mathrm{lab}},\,k_2^{\mathrm{lab}})\,(\rho _L F_L + 
\rho _T  F _T)\nonumber\\
&=& \sigma_{\mathrm{Mott}}\,\Big(\xi^2\frac{\beta^2F_L}{2\,\pi^2\alpha}+
(\eta+\frac{\xi}{2})\,\frac{F_T}{2\,\pi^2\alpha}\Big)\,.
\eeqa
This yields finally
\begin{equation}
F_L=2\,\pi^2\alpha\,
\frac{q^2}{(q^{\mathrm{lab}})^2}\,R_L\quad\mathrm{and}
\quad F_T=2\pi^2\alpha R_T\,.
\end{equation}

For $F_L$ the normal nonrelativistic theory reveals a sizeable potential
model dependence according to the left panels of
Fig.~\ref{fig_quinn_ff_L} which, however, is strongly reduced for the
complete theory leading to a satisfactory agreement with the
data. On the other hand $F_T$ in Fig.~\ref{fig_quinn_ff_T} exhibits
much less sensitivity to the potential model for the normal
nonrelativistic theory. The improvement by the full calculation in the
near threshold region with respect to the data is clearly seen in the
middle panels. In the lower panel the absence of pion production in
the theory is again responsible for the underestimation above $\omega
=200$~MeV. 

\subsection{Exclusive Observables}

\begin{figure}
\includegraphics[scale=0.9]{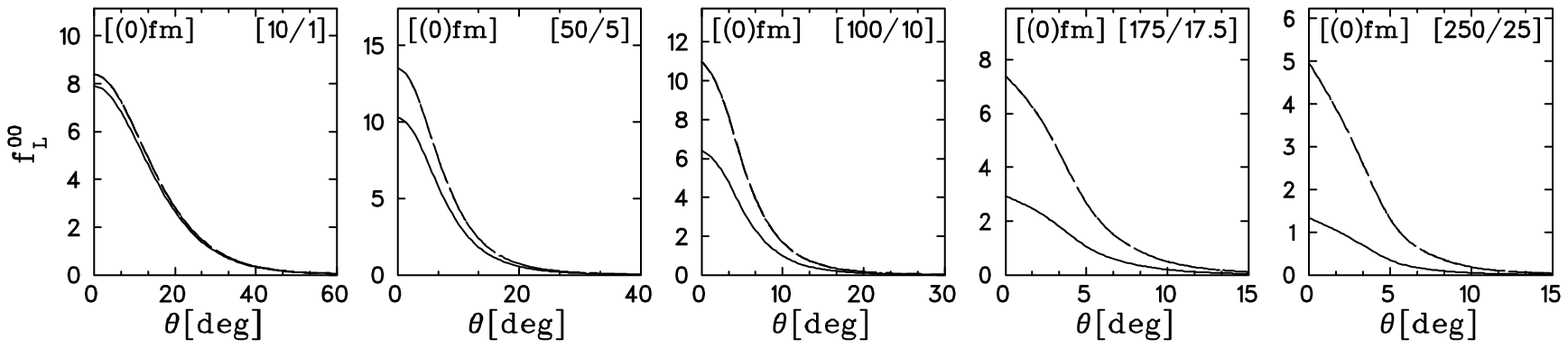}
\vskip .1cm
\includegraphics[scale=0.9]{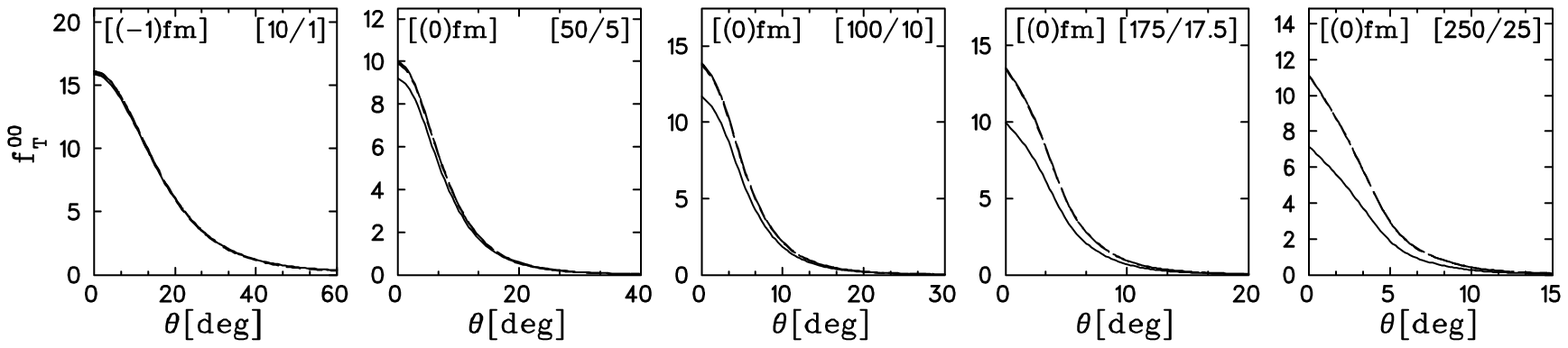}
\vskip .1cm
\includegraphics[scale=0.9]{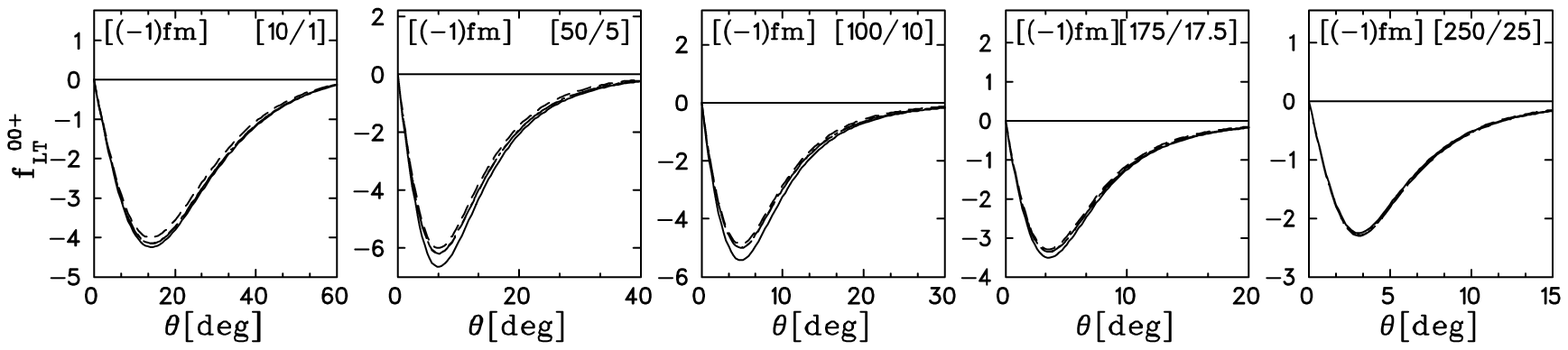}
\vskip .1cm
\includegraphics[scale=0.9]{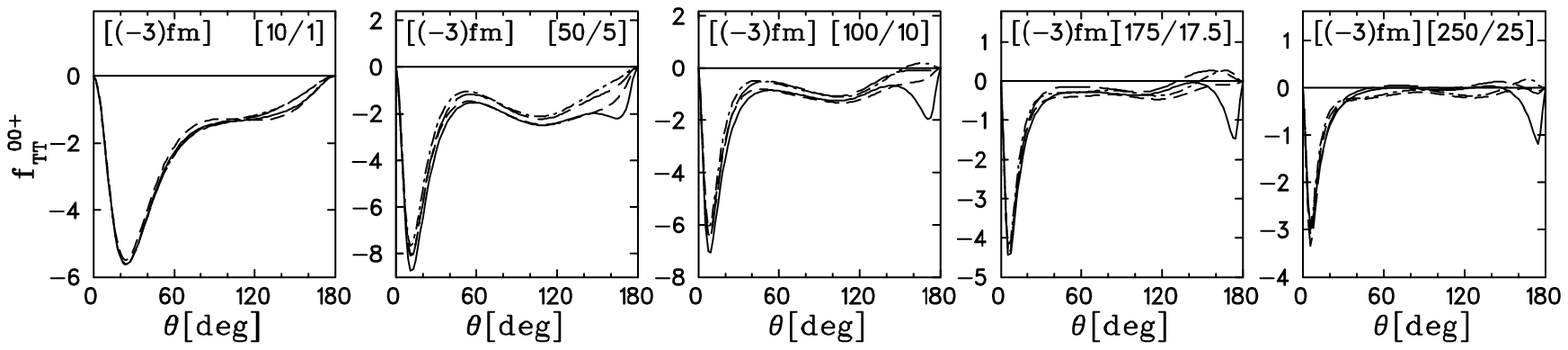}
\caption{Survey on the four unpolarized structure functions along
the quasi-free ridge, calculated for the {\sc Bonn}-Qb potential.
The top left inset ``[(-n) fm]'' indicates the unit [$10^{-n}$ fm] for 
the structure function and the top right inset ``[$E_{np}$/$q^2$]'', 
where $E_{np}$ in [MeV] and $q^2$ in [fm$^{-2}$], indicates the 
kinematic sector. Notation of the curves: N (short dashed), N+MEC 
(dash-dotted), N+MEC+IC (long dashed), 
total=N+MEC+IC+RC (solid).}
\label{fig_struc_fun_qe}
\end{figure}

Compared to the eight inclusive observables below pion threshold
one has 324 independent structure functions for the exclusive 
case.  This greater variety coupled with the angular dependence
of each observable allows a more detailed analysis 
of the reaction under study.  
For example, the differential cross section alone provides 41 structure
functions if one allows for beam and target polarizations 
(see Table~\ref{tab5}). 
Polarization analysis of one or both outgoing nucleons yields an even 
larger number of structure functions. However, as has been discussed 
in detail in Sect.~\ref{complsets}, in principle a set of 35 independent
observables should suffice for a complete determination of all 
reaction matrix elements. In practice, however, some of these observables
will be very difficult to measure experimentally with the required accuracy. 

For this reason
we have decided to discuss here only those structure functions  
which are relatively easily accessible, i.e.\ which require the measurement of
only one asymmetry (see Table~\ref{tab2} of Appendix~\ref{sepstrucfun}).
Our discussion will focus on the sensitivity of these structure
functions to various interaction effects.

To begin we will briefly give a survey on the four structure functions
of the unpolarized differential cross section. Since the structure functions
depend on $E_{np}$ and $q^2$ in addition 
to $\theta^{c.m.}$, we have
chosen to represent the various kinematical regions by
a grid in the $E_{np}-q^2$-plane, defined by 
$E_{np}=10,\,50,\,100,\,175,\,250$~MeV and 
$q^2=1,\,5,\,10,\,17.5,\,25$~fm$^{-2}$. Only for $E_{np}=250$~MeV
the lowest $q^2$-value was taken as 1.5~~fm$^{-2}$ because
of the photon line according to (\ref{photonline}).
Fig.~\ref{fig_struc_fun_qe} shows the four structure functions along 
the quasi-free ridge. The longitudinal structure function $f_L$ 
shows a pronounced peak in 
the forward direction essentially caused by the charge interaction of the 
virtual photon with a proton which is emitted preferentially along $q$. 
There is no corresponding peak at 180$^\circ$ for the neutron because of 
its very small electric form factor. For this reason we have restricted 
the angular range to the forward direction. 
With increasing momentum transfer, 
the width of this forward peak decreases markedly (one should note 
that the angular range differs for the various cases). 
Similarly, along the quasi-free ridge $f_T$ exhibits forward and
backward peaks with decreasing 
width for growing $q$. The forward peak
arises from the e.m.\ interaction with the dominant spin current 
of the proton at $\theta=0^\circ$ while the backward peak 
at $\theta=180^\circ$ (again not shown)
is similar in structure and corresponds to neutron emission along $q$. 
The ratio of the forward to backward 
peaks is essentially given by the square of the ratio of the proton
to neutron magnetic moments i.e. $(\mu_p/\mu_n)^2 \approx$ 2.  
As was already pointed out 
in the discussion of the form factors, along the quasi-free 
ridge most of the interaction effects are marginal, except for 
relativistic contributions which can result in a decrease of the 
forward and, in the case of $f_T$, backward peaks growing sizeably
with increasing momentum transfer. The decrease is particularly 
significant in $f_L$ amounting, for example, at $q^2=10$~fm$^{-2}$
to about 40~\% and at $q^2=25$~fm$^{-2}$ to even 70~\%. 
In $f_T$ the effect is smaller, roughly by a factor two. 

\begin{figure}
\includegraphics[scale=0.9]{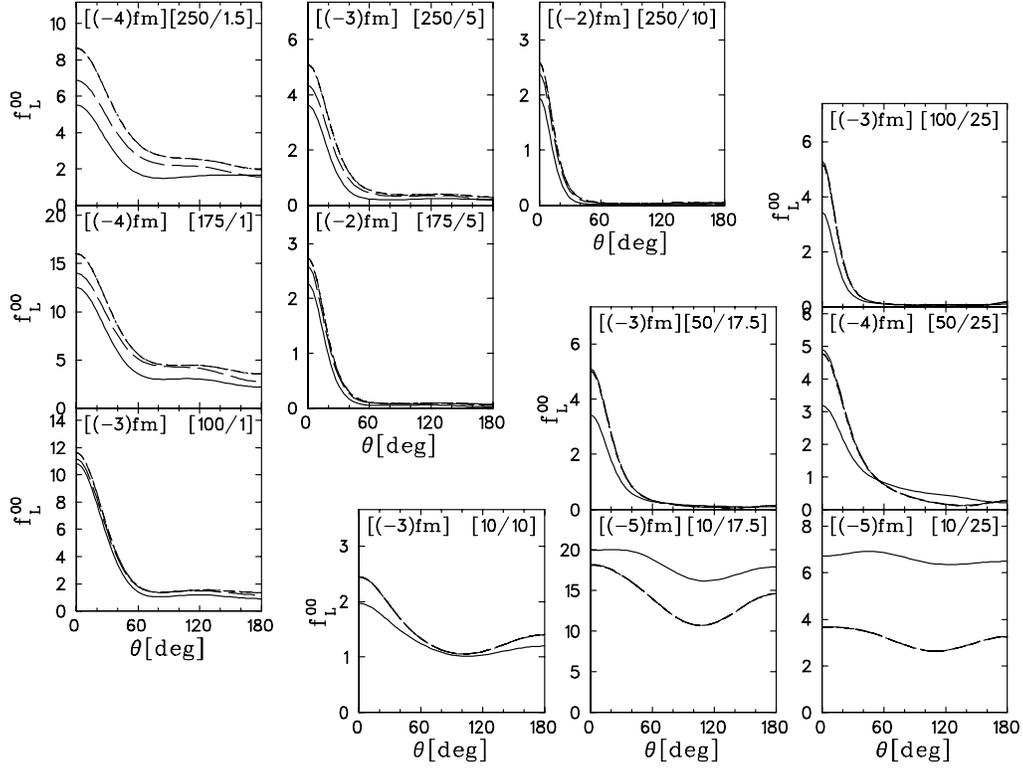}
\caption{Survey on longitudinal structure function $f_L$ off the 
quasi-free ridge, calculated for the {\sc Bonn}-Qb potential.
Notation of the curves as in Fig.~\ref{fig_struc_fun_qe}.}
\label{figqL_offqe}
\end{figure}

\begin{figure}
\includegraphics[scale=0.9]{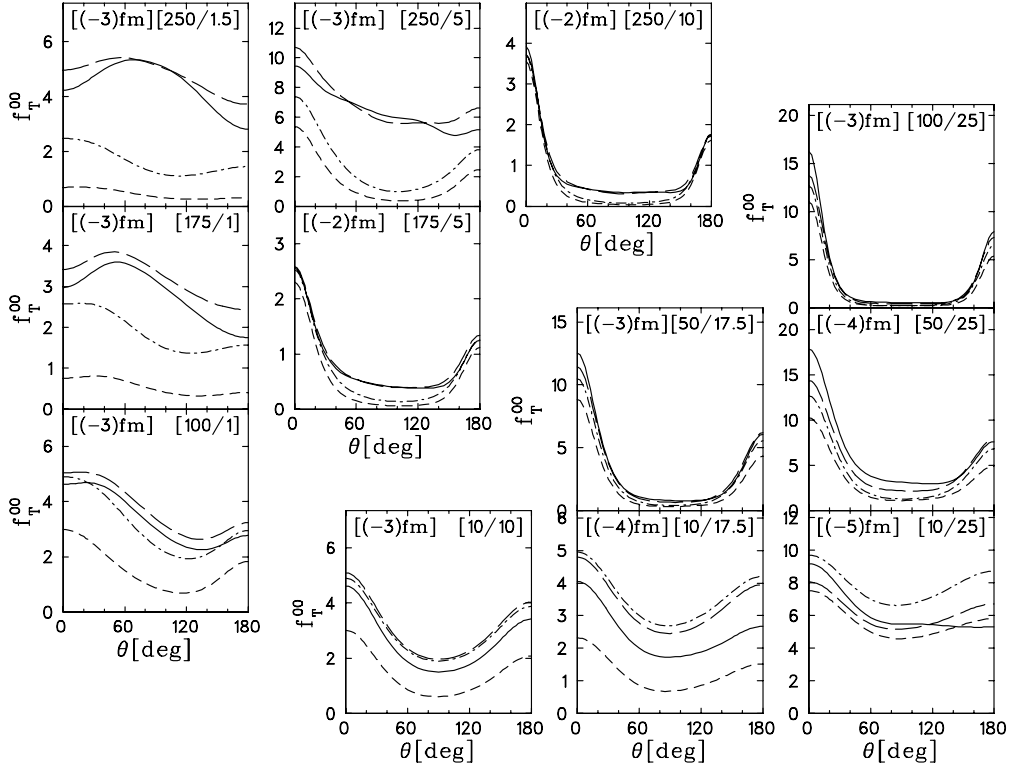}
\caption{Survey on transverse structure function $f_T$ off the 
quasi-free ridge, calculated for the {\sc Bonn}-Qb potential.
Notation of the curves as in Fig.~\ref{fig_struc_fun_qe}.}
\label{figqT_offqe}
\end{figure}

The interference structure function $f_{LT}$, which has to
vanish for $\theta=0^\circ$ and 180$^\circ$, 
exhibits a negative forward peak of broader 
distribution than the diagonal structure functions and is almost an order
of magnitude lower. A much smaller peak appears for neutron emission 
close to 180$^\circ$ with opposite sign which is not shown in 
Fig.~\ref{fig_struc_fun_qe}. The influence of MEC and IC is very tiny and 
even RC show up only slightly at intermediate $E_{np}=50$ to 100 MeV. The
interference structure function $f_{TT}$ is even smaller, about two
orders of magnitude compared to $f_{LT}$. Interaction effects from
MEC and IC play only a slightly more significant role. However, RC 
result in a distinct and sizeable peak at backward angles.

\begin{figure}
\includegraphics[scale=0.9]{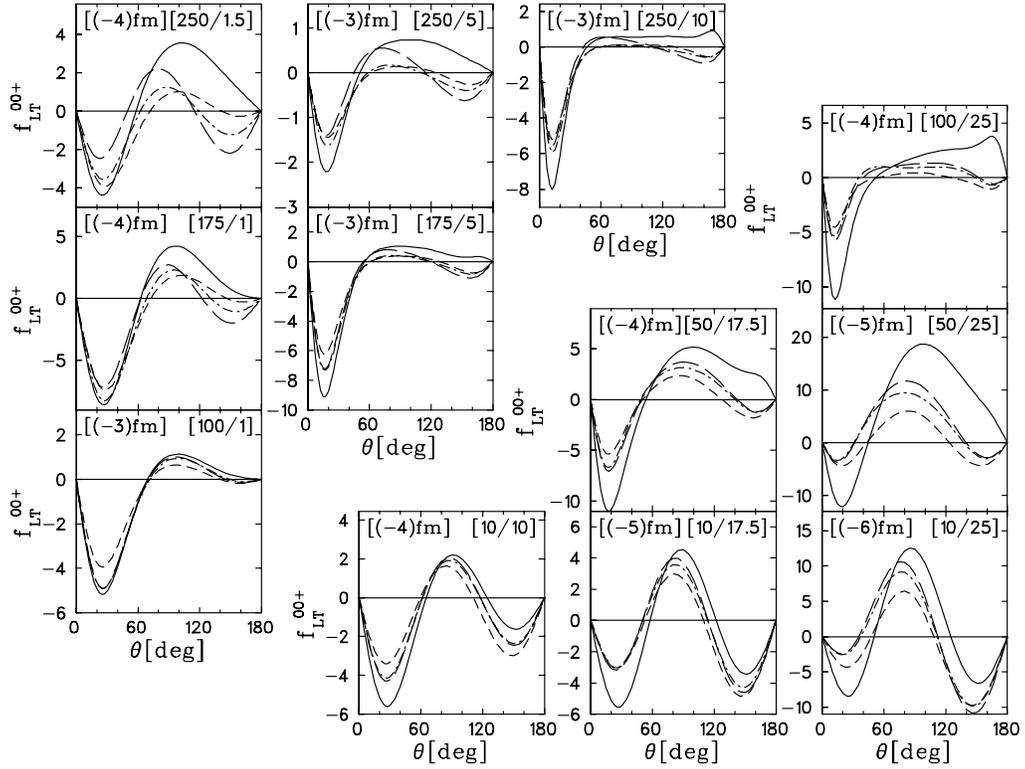}
\caption{Survey on interference structure function $f_{LT}$ off the 
quasi-free ridge, calculated for the {\sc Bonn}-Qb potential.
Notation of the curves as in Fig.~\ref{fig_struc_fun_qe}.}
\label{figqLT_offqe}
\end{figure}

\begin{figure}
\includegraphics[scale=0.9]{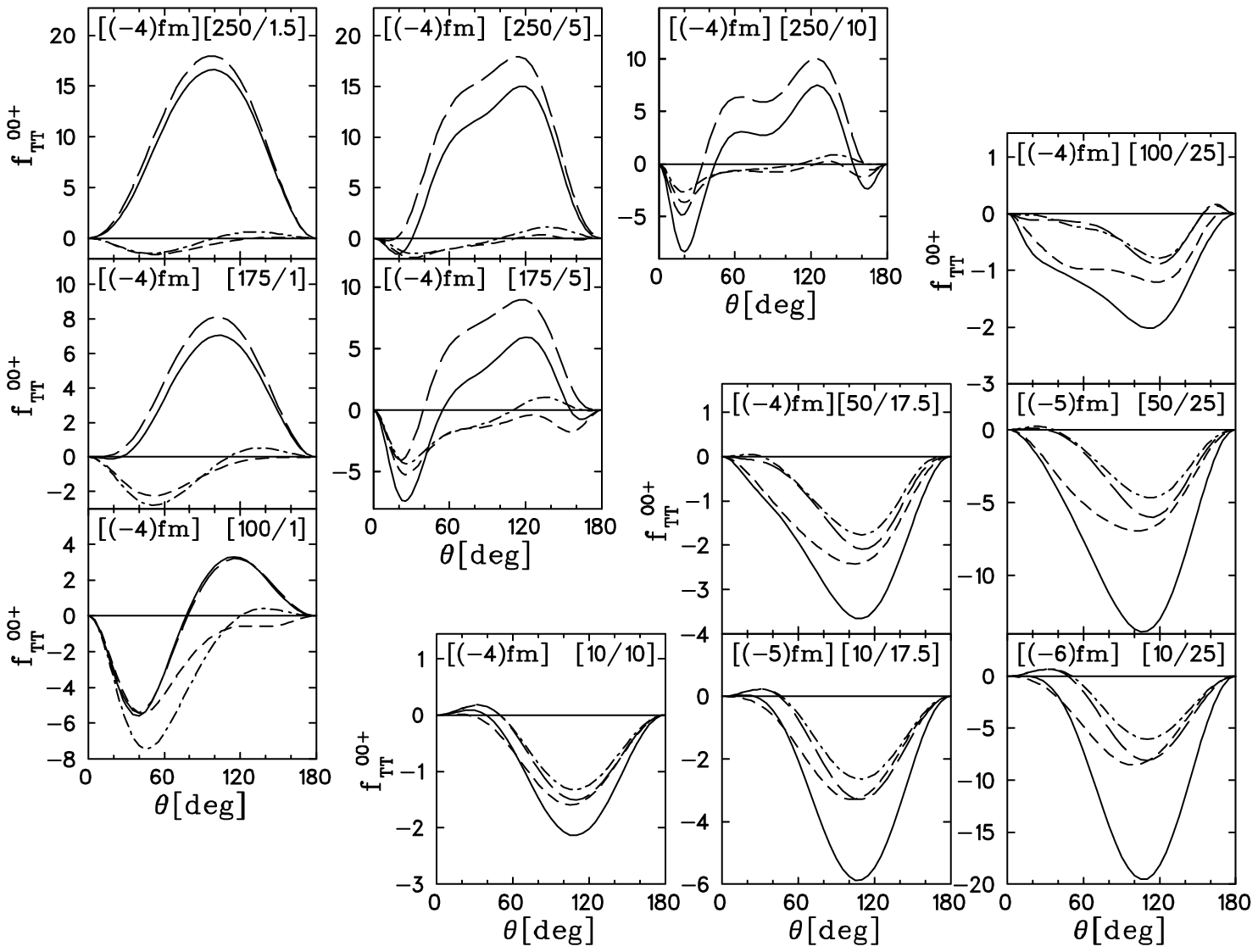}
\caption{Survey on interference structure function $f_{TT}$ off the 
quasi-free ridge, calculated for the {\sc Bonn}-Qb potential.
Notation of the curves as in Fig.~\ref{fig_struc_fun_qe}.}
\label{figqTT_offqe}
\end{figure}

In Figs.~\ref{figqL_offqe} through \ref{figqTT_offqe} we show the 
four structure functions for kinematical settings off the quasi-free 
ridge, i.e. six settings for $E_{np}/$MeV$>10q^2/$fm$^{-2}$ 
in the upper left panels and six settings for 
$E_{np}/$MeV$<10q^2/$fm$^{-2}$ in the lower right panels.
Away from the quasi-free ridge, the magnitudes of $f_L$ and $f_T$ drop rapidly.
For $f_L$, shown in Fig.~\ref{figqL_offqe}, 
the width of the forward peak increases. On the lower right 
side only RC show quite sizeable 
influences because of the absence of nonrelativistic MEC contributions 
and the neglect of a very small charge excitation of the $\Delta$ --
for this reason the short and long dashed and the dash-dot curves coincide in 
Fig.~\ref{figqL_offqe} --, 
whereas on the upper left side, i.e.\
for $E_{np}/$MeV$>10q^2/$fm$^{-2}$, also IC become 
increasingly important when approaching the $\Delta$-region while keeping 
the momentum transfer small. The influence of IC arises through the 
change of the normal wave function component via the dynamic coupling 
of the $NN$- and $N\Delta$-channels.

For the transverse structure function $f_T$ in Fig.~\ref{figqT_offqe},
which is slightly larger in size than $f_L$, the various interaction 
effects become much more pronounced off the quasi-free ridge than for 
$f_L$ because of the presence of nonrelativistic MEC and strong 
transverse excitation of the $\Delta$. Indeed, at low energy MEC 
provide the largest interaction effect followed by RC while IC remain 
small. That changes for energies in the region of $\Delta$-excitation 
above the quasi-free ridge, 
where naturally the IC contributions dominate, although MEC are also 
sizeable and even RC cannot be neglected, in particular in the forward 
and backward regions. 

\begin{figure}
\includegraphics[scale=1.]{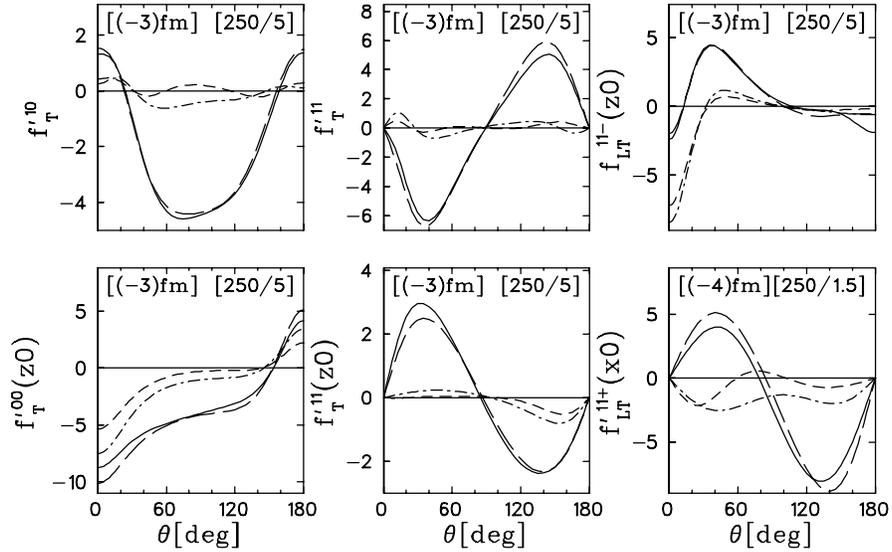}
\caption{Influence of IC on various structure functions calculated with
the {\sc Bonn}-Qb potential. 
Notation: N (short dashed), N+MEC (dash-dotted), N+MEC+IC (long dashed), 
T=N+MEC+IC+RC (solid).}
\label{fig_ic}
\end{figure}

\begin{figure}
\includegraphics[scale=1.]{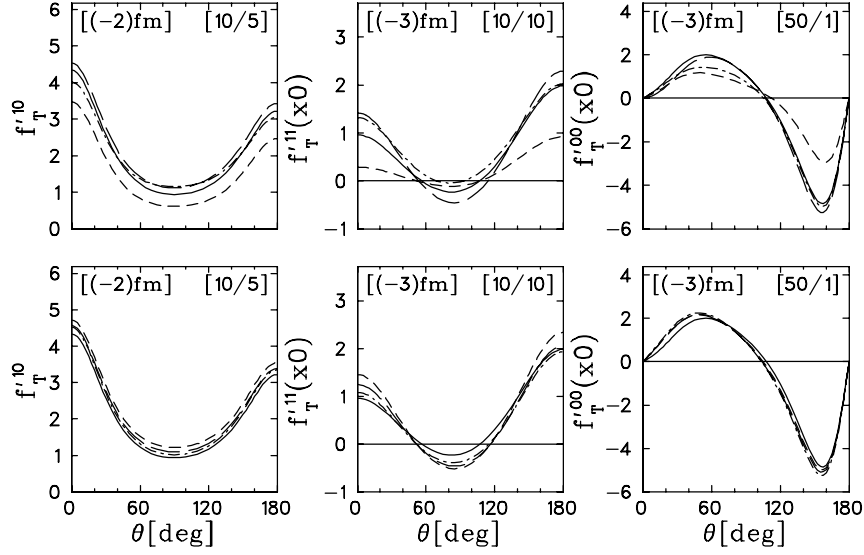}
\caption{Upper panels: Influence of MEC on various structure functions 
calculated with the {\sc Bonn}-Qb potential. 
Notation as in Fig.~\ref{fig_ic}. 
Lower panels: Potential model dependence of the
total result for the same structure functions. Notation: {\sc Bonn}-R 
(short dashed), {\sc Paris} (dash-dotted), {\sc Argonne} $V_{18}$  
(long dashed), {\sc Bonn}-Qb (solid).}
\label{fig_mec}
\end{figure}

\begin{figure}
\includegraphics[scale=1.]{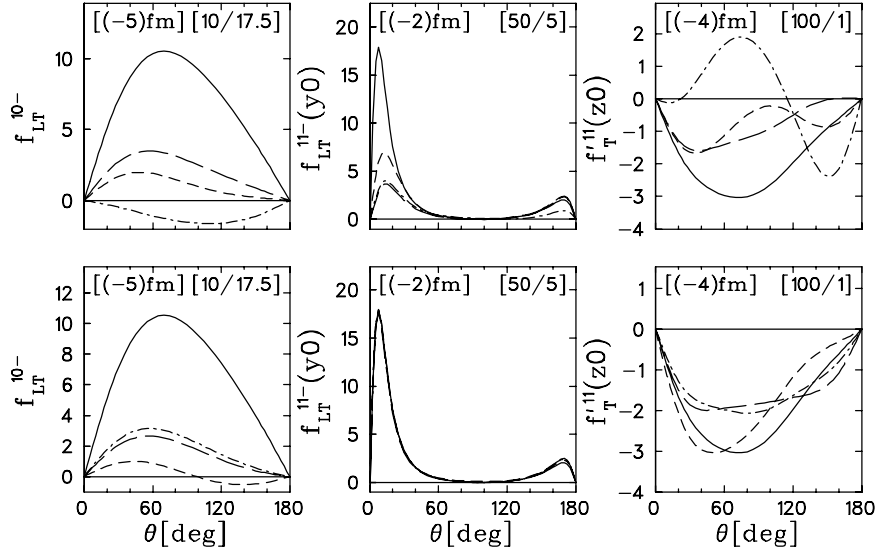}
\caption{Upper panels: Influence of RC on various structure functions 
calculated with the {\sc Bonn}-Qb potential.
Lower panels: Potential model dependence of the
total result for the same structure functions. 
Notation as in Fig.~\ref{fig_mec}.}
\label{fig_rc}
\end{figure}

The interference structure function $f_{LT}$ in Fig.~\ref{figqLT_offqe} 
is comparable in size to $f_L$, however, considerably more sensitive to
interaction effects. It shows a negative peak in forward direction, a 
maximum around 90$^\circ$ and another but much less pronounced negative 
peak at backward angles. RC produce by far the strongest effect, deepening
the forward negative peak, enhancing the maximum and weakening or even 
washing out completely the negative backward peak so that a small positive 
peak results for some kinematics. 

The other interference structure function $f_{TT}$, shown in 
Fig.~\ref{figqTT_offqe} is of comparable size to $f_{LT}$ off the 
quasi-free ridge. This is in contrast to their behavior along the 
quasi-free ridge where $f_{TT}$ is about two orders of magnitude smaller 
than $f_{LT}$ as depicted in Fig.~\ref{fig_struc_fun_qe}. Below the 
quasi-free ridge it shows a distinct minimum around 110$^\circ$ which 
is decreased by MEC but then deepened slightly by IC and, but more strongly,
by RC. Above the quasi-free ridge the influences of interaction effects
show in a certain sense an opposite behaviour. Here MEC deepen the 
minimum, lying more at forward angles, but produce a sign change 
in the backward direction. This effect is drastically counterbalanced
at forward angles by IC and amplified at larger angles. The 
additional RC lead then finally to a smaller reduction, deepening the 
small forward minimum, which exists for higher $E_{np}$ and $q^2$.

Now we turn to the question of which structure functions exhibit the most 
significant sensitivity with respect to the various interaction effects.
As already mentioned we take into account only those cases which require 
a single 
asymmetry measurement (see Table~\ref{tab2} of Appendix~\ref{sepstrucfun}). 
In addition, we consider as observables the cross sections and 
single proton polarizations ($x0$, $y0$, $z0$) only.  
We have studied these selected structure functions in the same kinematic 
regions as taken for the survey of the unpolarized structure functions but
present here only those kinematic cases where we found the strongest 
signatures.

In Fig.~\ref{fig_ic} we show the most relevant cases for the IC 
contribution. It is not 
surprising that they are all found at an excitation energy in the 
$\Delta$-resonance region at lower momentum transfer. As pointed out 
before, appropriate potential models with $\Delta$ degrees of freedom are 
lacking and thus we only show results for the coupled channel calculation with 
the renormalized {\sc Bonn}-Qb potential. 
The figure shows that there are quite a number of 
different structure functions 
which are essentially dominated by their IC contribution and thus are ideal 
cases for the study of such IC effects. It it worthwhile to note 
that there are not only transverse ($f_T^{\prime 10}$, $f_T^{\prime 11}$,
$f_T^{\prime 00}(z0)$, $f_T^{\prime 11}(z0)$), but also LT-type structure 
functions ($f_{LT}^{\prime 11-}(z0)$, $f_{LT}^{\prime 11+}(x0)$) among 
the cases presented.

Observables which exhibit sensitivity to the MEC contribution are shown in 
Fig.~\ref{fig_mec}. While for $f_T^{\prime 10}$ the strongest relative MEC
effect is found around $\theta=90^\circ$, 
for $f_T^{\prime 11}(x0)$ and $f_T^{\prime 00}(x0)$
one has the most pronounced effects at backward angles. The potential model
dependence is not very important and is particularly small for  
$f_T^{\prime 00}(x0)$. 
We would like to mention that the $f_T^{\prime 10}$ result 
is very similar to that of $f_T^{00}$ for the same kinematics 
(see Fig.~\ref{figqT_offqe}). In 
fact, $f_T^{00}$ is very sensitive to the MEC contribution and can in 
principle be determined in a single measurement (scattered electron 
at backward angle and $\phi=45^\circ$).

In Fig.~\ref{fig_rc} we show those structure functions 
where there are major effects due to relativistic
contributions. A very strong effect is seen for $f_{LT}^{10-}$, although
this structure function shows a rather severe potential model dependence
as well. For $f_{T}^{\prime 11}(z0)$ 
one finds a similar situation, but with somewhat 
smaller relativistic influence and a less pronounced variation with 
the potential model. On the contrary, for $f_{LT}^{11-}(y0)$
one observes essentially no potential model dependence and a large relativistic
contribution at forward angles.

\begin{figure}
\includegraphics[scale=1.]{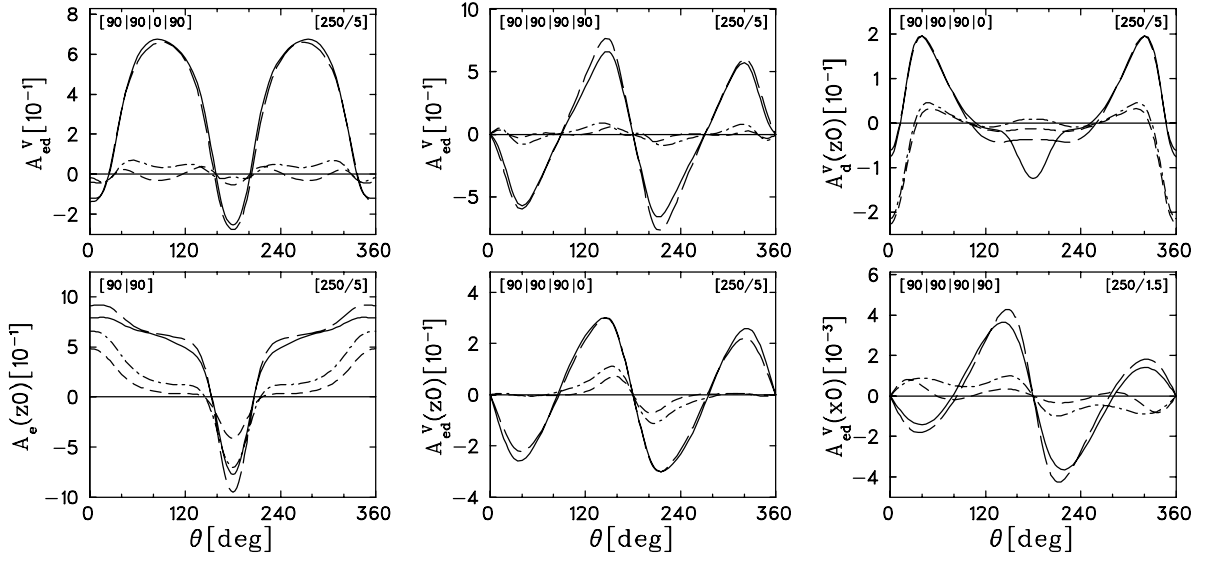}
\caption{Influence of IC on asymmetries for kinematic settings for 
the separation of the various structure functions in Fig.~\ref{fig_ic}
calculated with the {\sc Bonn}-Qb potential. 
Notation: N (short dashed), N+MEC (dash-dotted), N+MEC+IC (long dashed), 
T=N+MEC+IC+RC (solid). The upper left inset lists the 
various angles according to $[\theta_e,\,\phi,\,\theta_d,\,\phi_d]$.}
\label{fig_asy_ic}
\end{figure}

\begin{figure}
\includegraphics[scale=1.]{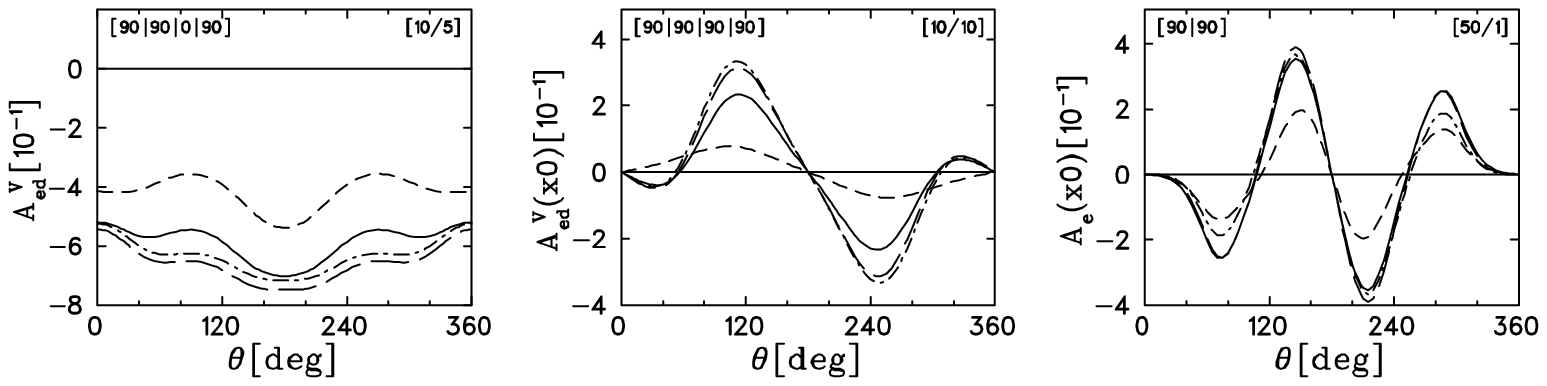}
\caption{Influence of MEC on asymmetries for kinematic settings for 
the separation of the various structure functions in Fig.~\ref{fig_mec}
calculated with the {\sc Bonn}-Qb potential. 
Notation as in Fig.~\ref{fig_asy_ic}.}
\label{fig_asy_mec}
\end{figure}

\begin{figure}
\includegraphics[scale=1.]{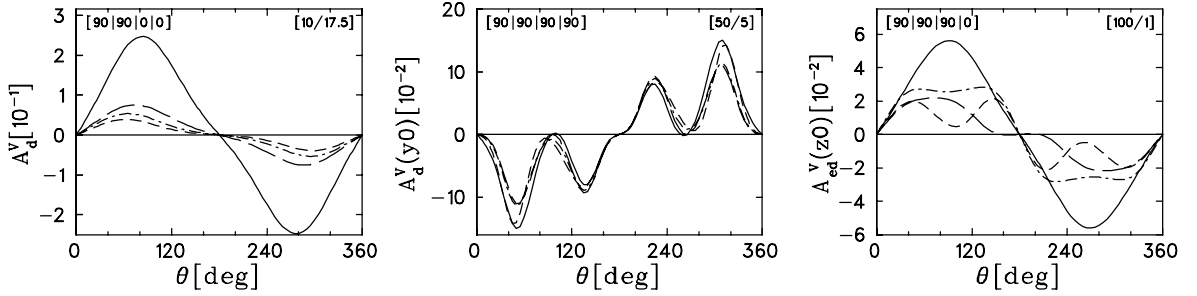}
\caption{Influence of RC on asymmetries for kinematic settings for 
the separation of the various structure functions in Fig.~\ref{fig_rc}
calculated with the {\sc Bonn}-Qb potential. 
Notation as in Fig.~\ref{fig_asy_ic}.}
\label{fig_asy_rc}
\end{figure}

In view of the fact that, besides the unpolarized differential 
cross section, the quantities which one measures experimentally 
are asymmetries, we present 
in Figs.~\ref{fig_asy_ic} through \ref{fig_asy_rc} the asymmetries 
corresponding to the structure functions of Figs.~\ref{fig_ic} 
through \ref{fig_rc} in order to see which of them produce sizeable, i.e.\
easily accessible asymmetries. In order to emphasize the interaction 
and relativistic effects in the numerator, the following asymmetries 
always refer to the unpolarized differential cross section in which 
all contributions from MEC, IC and RC are included, i.e.\ $S_0$ in 
Eq.~(\ref{asyall}) is $S_0(T)$.

With respect to IC effects, one readily notes
in Fig.~\ref{fig_asy_ic} three sizeable asymmetries of the order of one, 
namely $A_{ed}^V$ for two different settings ($\theta_d=0^\circ$ and 
90$^\circ$ with $\phi_d=90^\circ$), 
requiring electron and deuteron vector polarization, 
and $A_{e}(z0)$, requiring electron polarization and a proton polarimeter. 
Two other asymmetries, $A_{d}^V(z0)$ and $A_{ed}^V(z0)$, are of the 
order of 0.2. Only $A_{ed}^V(x0)$ is too small being of the order of 0.004. 
Of the asymmetries sensitive to MEC in Fig.~\ref{fig_asy_mec} the largest
one is $A_{ed}^V$ being of the order of 0.7, whereas the other two are
smaller, of the order of 0.4. With respect to RC, two asymmetries in 
Fig.~\ref{fig_asy_rc} are sufficiently large, namely $A_{d}^V$ and 
$A_{d}^V(y0)$ with a magnitude of about 0.1 to 0.2, but only $A_{d}^V$ 
exhibits a large RC effect.

\begin{figure}
\includegraphics[scale=1.]{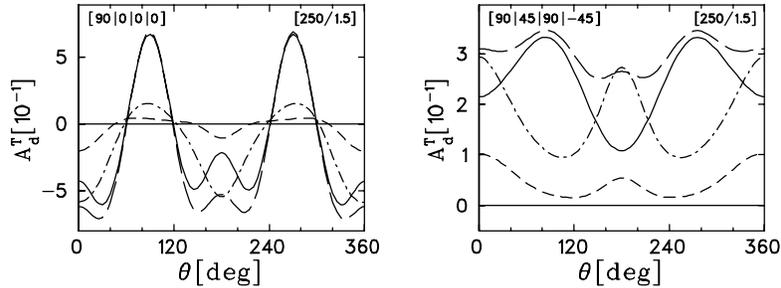}
\caption{Tensor asymmetry $A_d^T$ for $E_{np}=250$~MeV, $q^2=1.5$~fm$^{-2}$
calculated with the {\sc Bonn}-Qb potential. Notation as in
Fig.~\ref{fig_asy_ic}.}
\label{fig_asy_ic_a}
\end{figure}

\begin{figure}
\includegraphics[scale=1.]{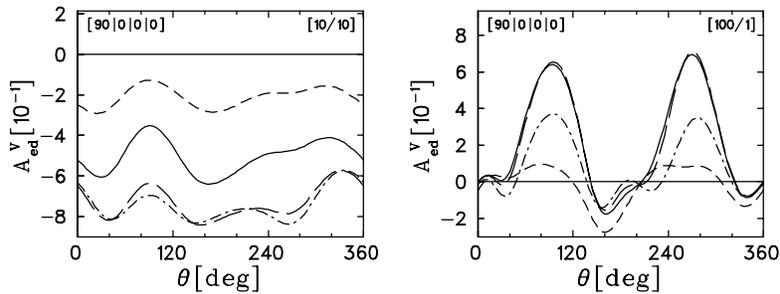}
\caption{Vector asymmetry $A_{ed}^V$ at two different off-quasi-free
kinematics, calculated with the {\sc Bonn}-Qb potential. Left panel for
$E_{np}=10$~MeV, $q^2=10$~fm$^{-2}$ and right panel for
$E_{np}=100$~MeV, $q^2=1$~fm$^{-2}$. Notation as in Fig.~\ref{fig_asy_ic}.}
\label{fig_asy_mec_ic}
\end{figure}

Finally we would like to briefly discuss irrespective of the problem 
of separating a specific structure function the question which 
observables show in specific asymmetries of sizeable magnitude either 
a large influence or none by interaction contributions. 
We have investigated for a large 
number of observables various asymmetries for a variety of kinematic 
settings. Some results of this search are presented in
Figs.~\ref{fig_asy_ic_a} and \ref{fig_asy_mec_ic}. 

Fig.~\ref{fig_asy_ic_a} exhibits the tensor
asymmetry $A_d^T$ for two different settings of the angles $\phi$,
$\theta_d$ and $\phi_d$. Without interaction effects the asymmetry
would be quite small, about 0.1 or less. In the left panel, MEC lead
to a strong enhancement around 0$^\circ$ and 180$^\circ$ which are
partially canceled by relativistic contributions. On the other hand, IC
show a large and dominant influence around 90$^\circ$ and 270$^\circ$
where the other effects are very small. The right panel is an
instructive example on how the the asymmetry and the relative size of
the various contributions change with a change of
the out-of-plane angle $\phi$ and the deuteron orientation angles.
In the left panel of Fig.~\ref{fig_asy_mec_ic}, the vector asymmetry
$A_{ed}^V$ exhibits at low energy but higher momentum transfer a large
MEC effect which is cut down to almost one half by relativistic
contributions whereas IC is negligible. The same asymmetry at a
different kinematics, $E_{np}=100$~MeV and $q=1$~fm$^{-2}$, in the
right panel of Fig.~\ref{fig_asy_mec_ic} shows a 
strong influence from MEC and IC of almost equal size and interfering
constructively, while RC effects are tiny. 

\begin{figure}
\includegraphics[scale=1.]{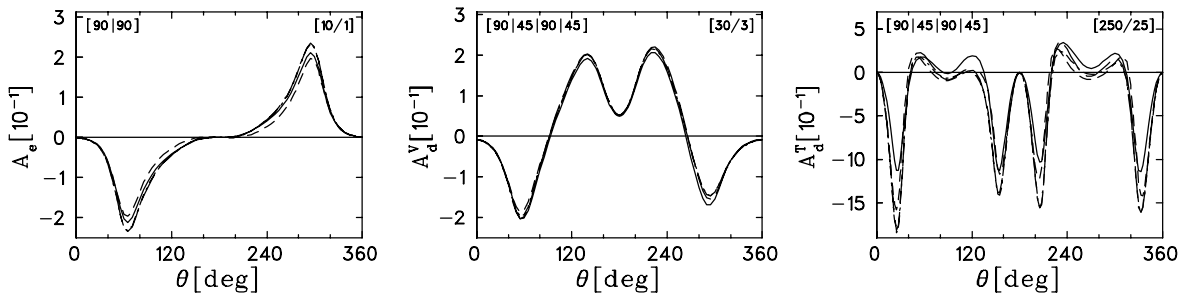}
\caption{Asymmetries at three different on-quasi-free
kinematics, calculated with the {\sc Bonn}-Qb potential. Left panel
$A_{e}$ for $E_{np}=10$~MeV, $q^2=1$~fm$^{-2}$, middle panel
$A_{d}^V$ for $E_{np}=30$~MeV, $q^2=3$~fm$^{-2}$, and right panel
$A_{d}^T$ for $E_{np}=250$~MeV, $q^2=25$~fm$^{-2}$. Notation as in
Fig.~\ref{fig_asy_ic}.} 
\label{fig_asy_no_effect}
\end{figure}

As last examples we show in
Fig.~\ref{fig_asy_no_effect} three asymmetries for on-quasi-free
kinematics, two at low energies and one in the $\Delta$-region which
exhibit a nearly interaction independent behaviour. The two asymmetries 
at quite low energies, $A_{e}$ in the left panel and in particular 
$A_{d}^V$ in the middle panel show very little influences from 
interaction and relativistic effects. In general these effects increase with 
growing energy $E_{np}$. But even for the example in 
the right panel at considerably high energy such influences are still
relatively small. It is not surprising, 
that one finds such a behaviour for kinematics belonging to the quasi-free 
ridge, because one expects that influences from FSI, MEC and IC will be 
minimal there. Because of their model independence, such cases provide 
consistency checks for theory and experiment. 

With this we will close the discussion of our results. We will not 
compare to experimental data on the exclusive reaction~\cite{BuA95,PeA97,
BlB98,KaH98,DoA99,Ost99,NiA01,ZhA01,ZhC01,vNR02,UlA02,Mul03,NiA03,MaS03,MaS03a}
because they have already been compared to our approach and thus nothing 
new can be said. 

\section{Summary and conclusion}
In this work we have presented a thorough and detailed survey on
polarization observables in electrodisintegration of the deuteron. 
It contains a general review of the basic formal aspects with 
respect to kinematics, the definition of observables and structure functions, 
their multipole decomposition, the question of completeness and 
the construction of independent sets of observables including analytic
solutions of the $t$-matrix elements in terms of structure functions.

Furthermore, a detailed account of the dynamic ingredients of
the theoretical framework is given 
with respect to the basic $NN$-force, the associated meson exchange currents,
isobar configurations for the consideration of internal nucleon dynamics, and
leading order relativistic contributions. 

As results we first have presented a general survey on the unpolarized 
and polarized form factors of the inclusive process indicating the various
kinematic regions with respect to energy and momentum transfer where 
sensitivities to MEC, IC and RC according to the theoretical framework
are to be expected. Furthermore, a comparison with experimental data for 
the near threshold region and for quasi-free kinematics is discussed. 
For the threshold region the influence of interaction effects is 
quantitatively confirmed. Also the quasi-free peak is well reproduced 
provided the reduction by relativistic contributions is included. 

For the exclusive reaction an extensive survey on the unpolarized 
structure function for a representative grid of energy and momentum 
transfers is given including a detailed discussion of the various 
interaction effects. Furthermore, for a selected set of polarization
structure functions whose determination requires only one asymmetry 
measurement, we have chosen those kinematic regions in which either IC or
MEC or RC play a major role with subsequent discussion of the 
associated asymmetries. 

Finally we hope that this survey will stimulate further experimental 
and theoretical research allowing one to test more thoroughly the
underlying basic dynamics. This is of particular importance with 
respect to the question how well do we understand the strong interaction 
in terms of effective degrees of freedom, i.e.\ in terms of meson, 
nucleon and isobar degrees of freedom, and where will we need to introduce
explicitly the basic quark and gluon degrees of freedom of QCD.

\acknowledgments                                                      
This work was supported by the Deutsche Forschungsgemeinschaft (SFB 443) and 
by the National Science and Engineering Research Council of Canada.

\appendix

\setcounter{equation}{0}
\section{Explicit expressions for the ${\cal U}^{\lambda' \lambda I M}_{X}$ 
in the standard representation of the $t$-matrices}\label{Uexplicit}

In view of the angular momentum algebra, it is useful 
to switch to a spherical representation replacing the cartesian 
components of the nucleon spin operators by their spherical ones  
\beqa
\sigma_{\alpha}(i)&=&\sum_{\tau=0,1}\sum_{\nu=-\tau}^{\tau}
   s_{\alpha}^{\tau\nu}\,\sigma_{\nu}^{[\tau]}(i)\,\,\quad \mbox{for } 
\alpha=0,\dots,3\,,
\eeqa
where we have introduced the equivalent complete set of 
$2\times 2$-matrices in $s=1/2$  space 
\begin{equation}
\sigma_{\nu}^{[\tau]}(i)
\,\qquad \mbox{for } \tau = 0,\,1\,\mbox{ and } |\nu|\leq \tau,,\label{sphcom}
\end{equation}
defining $\sigma_0^{[0]}=\sigma_0={\mathbb 1}_2$ and 
$\sigma^{[1]}_\nu=\sigma_\nu$, the spherical components of the Pauli spin 
matrices. The transformation matrix is given by 
\begin{equation}
s_{\alpha}^{\tau\nu}=\bar c(\alpha)\,\delta_{\tau,\widetilde \tau(\alpha)}\,
  (\delta_{\nu,\widetilde\nu (\alpha)}+
   \hat c(\alpha)\,\delta_{\nu, -\widetilde\nu (\alpha)})\,,\label{s_alpha}
\end{equation}
with 
\begin{equation}
\begin{array}{ll}
\hat c(\alpha) = \delta_{\alpha 2} - \delta_{\alpha 1}\,, & 
\bar c(\alpha) = 
  \left\{\begin{matrix}
1 & \mbox{for }\alpha=0,\,3\cr
      \frac{i^{-\alpha-1}}{\sqrt{2}} & \mbox{for }\alpha=1,\,2\cr\end{matrix}
\right. \,,\\ & \\
\widetilde \tau (\alpha) = 1- \delta_{\alpha 0}\,, &
\widetilde \nu (\alpha) =   
  \left\{\begin{matrix}
0 & \mbox{for }\alpha=0,\,3\cr
       1 & \mbox{for }\alpha=1,\,2\cr\end{matrix}
\right.\,.\end{array}
\end{equation}
For the inverse transformation from spherical to cartesian components one
easily finds
\begin{equation}
\sigma_{\nu}^{[\tau]}(i)=\sum_{\alpha=0}^3 c^{\alpha}_{\tau\nu}\,
\sigma_{\alpha}(i)\,,\label{sphcart}
\end{equation}
where we have introduced 
\begin{equation}
c_{\tau \nu}^{\alpha} = c(\nu)
       (\delta_{\alpha, a(\tau, \nu)}
 +i\nu\,\delta_{\alpha, b(\tau, \nu)})\,,
\end{equation}
and
\begin{equation}
c\,(\nu) = -\frac{\nu}{\sqrt{2}}\,\delta_{|\nu| 1}+\delta_{\nu 0}\,,\quad
a\,(\tau, \nu) = 3\tau - 2|\nu|\,,\quad
b\,(\tau, \nu) = 3\tau - |\nu|\,.
\end{equation}

Then the transformation of the ${\cal U}$'s to spherical components is 
given by
\begin{equation}
{\cal U}_{\alpha'\alpha}^{\lambda' \lambda I M}=
\sum_{\tau'\nu'\tau\nu}s_{\alpha'}^{\tau'\nu'}s_{\alpha}^{\tau\nu}\,
{\cal U}^{\lambda' \lambda I M}_{\tau'\nu'\tau\nu}\,,
\label{cartsph}
\end{equation}
where ${\cal U}_{\tau'\nu'\tau\nu}^{\lambda' \lambda I M}$ is defined as in
(\ref{ulamcart}) with $\sigma_{x_j}(i)$ being replaced by the spherical
components according to (\ref{sphcom}). 
In the coupled representation one has the following explicit form for the 
${\cal U}$'s
\begin{equation}
{\cal U}^{\lambda'\lambda I M}_{\tau'\nu'\tau\nu}=\sum_{S\sigma}
(-)^{\tau'+\tau+\sigma}\hat\tau'\hat \tau 
\hat S\left( \begin{matrix}
\tau'&\tau &S \cr \nu'&\nu&-\sigma \cr\end{matrix}\right)\,
V_{\lambda'\lambda I M}^{\tau'\tau S\sigma}\,,\label{Ustandard}
\end{equation}
where the quantities $V_{\lambda'\lambda I M}^{\tau'\tau S\sigma}$ are 
given by 
\begin{eqnarray}
V_{\lambda'\lambda I M}^{\tau'\tau S\sigma}&=&2\hat S
\sum_{s's}\hat s'\hat s 
\left\{ \begin{matrix}
\frac{1}{2}&\frac{1}{2}&\tau'\cr 
\frac{1}{2}&\frac{1}{2}&\tau \cr s'&s &S \cr\end{matrix} \right\}
u^{s's S\sigma}_{\lambda'\lambda I M}\,,\label{defv}
\end{eqnarray}
with 
\begin{eqnarray}
u^{s's S\sigma}_{\lambda'\lambda I M}&=&\hat I \sqrt{3}\sum_{m_s' m_s m'm}
(-)^{1-m+s'-m_s'}\left( \begin{matrix}
1&1&I \cr m'&-m&M \cr\end{matrix} \right)
\left( \begin{matrix}
s'&s&S \cr m_s'&-m_s&-\sigma  \cr\end{matrix} \right)
t_{s'm_s'\lambda' m'}^*t_{s m_s\lambda m}\,.\label{defu}
\eeqa

Specifying the observable $X$, one has in detail:\\
(i) differential cross section ($X=1=(00)$) 
\begin{equation}
{\cal U}^{\lambda'\lambda I M}_{1}=\sum_s\hat s 
u_{\lambda'\lambda IM}^{ss00}\,.
\end{equation}
(ii) single nucleon polarization ($X=(i0)$ for the proton or $X=(0i)$ 
for the neutron) 
\begin{eqnarray}
{\cal U}^{\lambda'\lambda I M}_{x_p}&=&-\sqrt{\frac{3}{2} }
\left(V_{\lambda'\lambda I M}^{1011}-V_{\lambda'\lambda I M}^{101-1}\right)
\,,\\
{\cal U}^{\lambda'\lambda I M}_{x_n}&=&-\sqrt{\frac{3}{2} }
\left(V_{\lambda'\lambda I M}^{0111}-V_{\lambda'\lambda I M}^{011-1}\right)
\,,\\
{\cal U}^{\lambda'\lambda I M}_{y_p}&=&i\sqrt{\frac{3}{2} }
\left(V_{\lambda'\lambda I M}^{1011}+V_{\lambda'\lambda I M}^{101-1}\right)
\,,\\
{\cal U}^{\lambda'\lambda I M}_{y_n}&=&i\sqrt{\frac{3}{2} }
\left(V_{\lambda'\lambda I M}^{0111}+V_{\lambda'\lambda I M}^{011-1}\right)
\,,\\
{\cal U}^{\lambda'\lambda I M}_{z_p}&=& \sqrt{3}
V_{\lambda'\lambda I M}^{1010}\,,\\
{\cal U}^{\lambda'\lambda I M}_{z_n}&=& \sqrt{3}
V_{\lambda'\lambda I M}^{0110}\,,
\end{eqnarray}
where
\begin{eqnarray}
V_{\lambda'\lambda I M}^{1 0 1\sigma}&=&\sqrt{2}
\sum_{s's}(-)^{s}\hat s'\hat s \left\{ \begin{matrix}
s'&s &1 \cr
\frac{1}{2}&\frac{1}{2}&\frac{1}{2} \cr \end{matrix} \right\}
u^{s's 1\sigma}_{\lambda'\lambda I M}\,,\\
V_{\lambda'\lambda I M}^{0 1 1\sigma}&=&\sqrt{2}
\sum_{s's}(-)^{s'}\hat s'\hat s \left\{ \begin{matrix}
s'&s &1 \cr
\frac{1}{2}&\frac{1}{2}&\frac{1}{2} \cr \end{matrix} \right\}
u^{s's 1\sigma}_{\lambda'\lambda I M}\,.
\end{eqnarray}
(iii) double nucleon polarization ($X=(ij)$)
\begin{eqnarray}
{\cal U}^{\lambda'\lambda I M}_{xx/yy}&=&-\sqrt{3}\Big[
V_{\lambda'\lambda I M}^{1 1 00}+\frac{1}{\sqrt{2}}
V_{\lambda'\lambda I M}^{1 1 20}\mp\frac{\sqrt{3}}{2}
\Big(V_{\lambda'\lambda I M}^{1 1 22}+V_{\lambda'\lambda I M}^{1 1 2-2}
\Big)\Big]\,,\\
{\cal U}^{\lambda'\lambda I M}_{zz}&=&-\sqrt{3}\Big[
V_{\lambda'\lambda I M}^{1 1 00}-\sqrt{2}
V_{\lambda'\lambda I M}^{1 1 20}\Big]\,,\\
{\cal U}^{\lambda'\lambda I M}_{xy/yx}&=&-\frac{3i}{2}\Big[\pm\sqrt{2}
V_{\lambda'\lambda I M}^{1 1 10}+
\Big(V_{\lambda'\lambda I M}^{1 1 22}-V_{\lambda'\lambda I M}^{1 1 2-2}
\Big)\Big]\,,\\
{\cal U}^{\lambda'\lambda I M}_{xz/zx}&=&-\frac{3}{2}\Big[\pm
\Big(V_{\lambda'\lambda I M}^{1 1 11}+V_{\lambda'\lambda I M}^{1 1 1
-1}\Big)+
\Big(V_{\lambda'\lambda I M}^{1 1 21}-V_{\lambda'\lambda I M}^{1 1 2-1}
\Big)\Big]\,,\\
{\cal U}^{\lambda'\lambda I M}_{yz/zy}&=&\frac{3i}{2}\Big[\pm
\Big(V_{\lambda'\lambda I M}^{1 1 11}-V_{\lambda'\lambda I M}^{1 1 1-1}
\Big)+
\Big(V_{\lambda'\lambda I M}^{1 1 21}+V_{\lambda'\lambda I M}^{1 1 2-1}
\Big)\Big]\,.
\end{eqnarray}


\setcounter{equation}{0}
\section{Quadratic Relations between Observables}\label{quadrel}
In this appendix we will show that for a set of $n$ independent $t$-matrix 
elements $\{t_j;\,j=1\dots n\}$ one finds exactly $(n-1)^2$ quadratic 
relations between observables by which the $n^2$ linearly independent 
observables are reduced to a set of $2n-1$ independent ones. To this end 
we introduce the bilinear form in the $t$-matrix elements
\begin{equation}
T_{j'j}=t_{j'}^*t_j\,,
\end{equation}
which can be expressed as a linear form of the observables ${\cal O}^\alpha$ 
\begin{equation}
T_{j'j}=\sum_\alpha \tau^\alpha_{j'j}\,{\cal O}^\alpha\,,
\end{equation}
with appropriate coefficients $\tau^\alpha_{j'j}$. They have the property
\begin{equation}
\tau^\alpha_{j'j}=\tau^{\alpha\,*}_{jj'}\,,
\end{equation}
which follows from $T_{j'j}=T_{jj'}^*$ and the fact that the observables 
are real quantities. 
It is straightforward to show that these bilinear forms obey the relation 
\begin{equation}
T_{j'j}T_{lm}=T_{j'm}T_{lj}\,,\label{tijlma}
\end{equation}
which, expressed in terms of observables, yields quadratic relations between 
the latter. In particular, choosing $k=l=m$, one finds
\begin{equation}
T_{j'j}=\frac{T_{j'k}T_{kj}}{T_{kk}}\,,\label{tijlmb}
\end{equation}
where $k$ can be chosen arbitrarily. It is also clear that from (\ref{tijlmb})
one can recover the relation (\ref{tijlma}). Thus we only need to consider 
the latter relation, and the question is, how many independent quadratic 
relations one can find. 

We first note, it is sufficient to consider only one specific $k$, because 
from (\ref{tijlmb}) one can derive straightforwardly the analogous relation for
any other $k'$. Second, it is sufficient to consider only the cases 
$j'\leq j$, 
because $T_{jj'}=T_{j'j}^*$. The remaining relations certainly are independent 
because of the independency of the $t$-matrix elements. Choosing then first 
$j'=j$, the case $j'=k$ yields the identity, whereas for $j'\neq k$ one finds 
\begin{equation}
T_{j'j'}T_{kk}=|T_{j'k}|^2\,,
\end{equation}
which constitute $(n-1)$ real quadratic relations
\begin{equation}
\sum_{\alpha \alpha'} \tau^\alpha_{j'j'}\,\tau^{\alpha'}_{kk}\,{\cal O}^\alpha 
\,{\cal O}^{\alpha'}=\sum_{\alpha \alpha'} \tau^{\alpha\,*}_{j'j}\,
\tau^{\alpha'}_{j'j}\,{\cal O}^\alpha \,{\cal O}^{\alpha'}\,.
\end{equation}
As next we consider the case $i<j$ for which one has $N=n(n-1)/2$ different 
pairs. Again one can discard the cases $j'=k$ or $j=k$, because they do not 
result in quadratic relations, thus ruling out $n-1$ relations. Therefore, one 
finds in this case ($j'<j$) a total number of 
\begin{equation}
N-(n-1)=\frac{1}{2}(n-1)(n-2)
\end{equation}
different complex quadratic relations of the form
\begin{equation}
\sum_{\alpha \alpha'} \tau^\alpha_{j'j}\,\tau^{\alpha'}_{kk}\,{\cal O}^\alpha 
\,{\cal O}^{\alpha'}=\sum_{\alpha \alpha'} \tau^{\alpha}_{j'k}\,
\tau^{\alpha'}_{kj}\,{\cal O}^\alpha \,{\cal O}^{\alpha'}\,.
\end{equation}
Separating these into real and imaginary parts, one finds as total number 
of independent real quadratic relations between observables
\begin{equation}
(n-1)+2\frac{1}{2}(n-1)(n-2)=(n-1)^2\,,
\end{equation}
which is just the required number of relations in order to reduce the 
number of $n^2$ linearly independent observables to $n^2-(n-1)^2=2n-1$ 
independent ones.


\setcounter{equation}{0}
\section{Multipole expansion of structure functions and form factors}
\label{formfactors}

Here we will list more explicit expressions for the multipole expansion of 
the structure functions of the differential cross section 
\begin{equation}
f^{(\prime)\,IM(\pm)}_a= \sum_{K}
f^{(\prime)\,IM(\pm),\,K}_a\,
d^K_{-M-\beta(a),0}(\theta)\,.\label{fdiffmultipole}
\end{equation}
As shown in detail in~\cite{ArL02}, one obtains 
for the coefficients of the structure functions 
\beqa
f_L^{IM,\,K}&=&\frac{8\sqrt{3}}{1+\delta_{M0}}\,\pi\,
\hat I\, {\hat K}^2\,\sum_{L' \mu' j' L \mu j}(-)^{L}
\sum_{J} {\hat J}^2 \, 
\left(\begin{matrix}
J&I&K \cr 0 &M& -M\cr\end{matrix}\right)
\left(\begin{matrix}
L'&L& J \cr 0 &0&0 \cr\end{matrix}\right) 
\left\{ \begin{matrix}
j'&j&K \cr L'&L&J \cr 1&1&I \cr\end{matrix} \right\}
\nonumber\\\label{strucfunmultL}
&&\hspace*{2cm}\Big((-)^{L'+\mu'+j'}+1\Big)\Big((-)^{L+\mu+j}+1\Big)
\widetilde{{\cal D}}_{00}^{K0}(\mu' j' \mu j)\,
\Re e\Big(i^{\delta_{I1}}
\widetilde C^{L'\ast}(\mu'j')\,\widetilde C^l(\mu j)\Big)\,,\\
f_T^{IM,\,K}&=&
-\frac{32\sqrt{3}}{1+\delta_{M0}}\,\pi\,
\hat I\, {\hat K}^2\,\sum_{L' \mu' j' L \mu j}(-)^{L}
\sum_{J} {\hat J}^2 \, 
\left(\begin{matrix}
J&I&K \cr 0 &M& -M\cr\end{matrix}\right)
\left(\begin{matrix}
L'&L& J \cr 1 &-1&0 \cr\end{matrix}\right) 
\left\{ \begin{matrix}
j'&j&K \cr L'&L&J \cr 1&1&I \cr\end{matrix} \right\}
\nonumber\\\label{strucfunmultT}
&&\hspace*{3.5cm}\widetilde{{\cal D}}_{00}^{K0}(\mu' j' \mu j)\,
\Re e\Big(i^{\delta_{I1}}
\widetilde N^{L'\ast}_{1}(\mu' j')\,\widetilde N^L_1(\mu j)
\Big)\,,\\
f_{LT}^{IM\pm,\,K}&=&
\frac{16\sqrt{6}}{1+\delta_{M0}}\,\pi\,
\hat I\, {\hat K}^2\,\sum_{L' \mu' j' L \mu j}(-)^{L}
\sum_{J} {\hat J}^2 \, 
\,\Big[
\left(\begin{matrix}
J&I&K \cr 1 &M& -M-1\cr\end{matrix}\right)\pm (-)^{I+M}
\left(\begin{matrix}
J&I&K \cr 1 &-M& M-1\cr\end{matrix}\right)\Big]
\nonumber\\\label{strucfunmultLT}&&
\left(\begin{matrix}
L'&L& J \cr 0 &-1&1 \cr\end{matrix}\right) 
\left\{ \begin{matrix}
j'&j&K \cr L'&L&J \cr 1&1&I \cr\end{matrix} \right\}
\Big((-)^{L'+\mu'+j'}+1\Big)\widetilde{{\cal D}}_{00}^{K0}(\mu' j' \mu j)\,
\Re e\Big(i^{\delta_{I1}}
\widetilde C^{L'\ast}(\mu'j')\,\widetilde N^L_1(\mu j)\Big)\,,\\
f_{TT}^{IM\pm,\,K}&=&-
\frac{16\sqrt{3}}{1+\delta_{M0}}\,\pi\,
\hat I\, {\hat K}^2\,\sum_{L' \mu' j' L \mu j}(-)^{L}
\sum_{J} {\hat J}^2 \,\Big[
\left(\begin{matrix}
J&I&K \cr 2 &M& -M-2\cr\end{matrix}\right)\pm (-)^{I+M}
\left(\begin{matrix}
J&I&K \cr 1 &-M& M-2\cr\end{matrix}\right)\Big]
\nonumber\\& & 
\left(\begin{matrix}
L'&L& J \cr -1 &-1&2 \cr\end{matrix}\right) 
\left\{ \begin{matrix}
j'&j&K \cr L'&L&J \cr 1&1&I \cr\end{matrix} \right\}
\widetilde{{\cal D}}_{00}^{K0}(\mu' j' \mu j)\,
\Re e\Big(i^{\delta_{I1}}
\widetilde N^{L'\ast}_{-1}(\mu' j')\,\widetilde N^L_1(\mu j)
\Big)\,,\label{strucfunmultTT}\\
f_T^{\prime\,IM,\,K}&=&
\frac{32\sqrt{3}}{1+\delta_{M0}}\,\pi\,
\hat I\, {\hat K}^2\,\sum_{L' \mu' j' L \mu j}(-)^{L}
\sum_{J} {\hat J}^2 \, 
\left(\begin{matrix}
J&I&K \cr 0 &M& -M\cr\end{matrix}\right)
\left(\begin{matrix}
L'&L& J \cr 1 &-1&0 \cr\end{matrix}\right) 
\left\{ \begin{matrix}
j'&j&K \cr L'&L&J \cr 1&1&I \cr\end{matrix} \right\}
\nonumber\\
&&\hspace*{3.5cm}\widetilde{{\cal D}}_{00}^{K0}(\mu' j' \mu j)\,
\Im m\Big(i^{\delta_{I1}}
\widetilde N^{L'\ast}_{1}(\mu' j')\,\widetilde N^L_1(\mu j)
\Big)\,,\label{strucfunmultTs}\\
f_{LT}^{\prime\,IM\pm,\,K}&=&
-\frac{16\sqrt{6}}{1+\delta_{M0}}\,\pi\,
\hat I\, {\hat K}^2\,\sum_{L' \mu' j' L \mu j}(-)^{L}
\sum_{J} {\hat J}^2 \, 
\,\Big[
\left(\begin{matrix}
J&I&K \cr 1 &M& -M-1\cr\end{matrix}\right)\pm (-)^{I+M}
\left(\begin{matrix}
J&I&K \cr 1 &-M& M-1\cr\end{matrix}\right)\Big]
\nonumber\\&&
\left(\begin{matrix}
L'&L& J \cr 0 &-1&1 \cr\end{matrix}\right) 
\left\{ \begin{matrix}
j'&j&K \cr L'&L&J \cr 1&1&I \cr\end{matrix} \right\}
\Big((-)^{L'+\mu'+j'}+1\Big)\widetilde{{\cal D}}_{00}^{K0}(\mu' j' \mu j)\,
\Im m\Big(i^{\delta_{I1}}
\widetilde C^{L'\ast}(\mu'j')\,\widetilde N^L_1(\mu j)\Big)\,.
\label{strucfunmultLTs}
\eeqa
Note, that the tilde indicates the incorporation of the hadronic phase
factor $e^{i\delta^j_\mu}$ in the multipole matrix element. 
Now we specialize further in order to obtain the angular coefficients for 
the unpolarized differential cross section, 
\begin{equation}
S_0= c(k_1^{\mathrm{lab}},\,k_2^{\mathrm{lab}})\,
\sum_K\Big([\rho _L f_L^K + \rho_T f_T^K]\,d^K_{00}(\theta) + 
\rho_{LT} {f}_{LT}^K\,d^K_{-10}(\theta)\, \cos \phi
+\rho _{TT} {f}_{TT}^K\,d^K_{-20}(\theta)\, \cos2 \phi\Big)
\,,\label{multxsection}
\end{equation}
by setting $I=M=0$ in (\ref{strucfunmultL}) through 
(\ref{strucfunmultLTs}). Writing for simplicity
$f_{L/T}^K$ and $f_{LT/TT}^{(\prime)\,K}$ instead of $f_{L/T}^{00,\,K}$ 
and $f_{LT/TT}^{(\prime)\,00+,\,K}$, respectively, one obtains
\beqa
f_L^{K}=-4\,\pi\,
{\hat K}^2\,\sum_{L' \mu' j' L \mu j}&&(-)^{L'+L+j}
\left(\begin{matrix}
L'&L& K \cr 0 &0&0 \cr\end{matrix}\right) 
\left\{ \begin{matrix}
j'&j&K \cr L&L'&1\cr\end{matrix} \right\}
\nonumber\\
&&\Big((-)^{L'+\mu'+j'}+1\Big)\Big((-)^{L+\mu+j}+1\Big)
\widetilde{{\cal D}}_{00}^{K0}(\mu' j' \mu j)\,
\Re e\Big(i^{\delta_{I1}}
\widetilde C^{L'\ast}(\mu'j')\,\widetilde C^l(\mu j)\Big)\,,\\
f_T^{K}=
16\,\pi\,
{\hat K}^2\,\sum_{L' \mu' j' L \mu j}&&(-)^{L'+L+j}
\left(\begin{matrix}
L'&L& K \cr 1 &-1&0 \cr\end{matrix}\right) 
\left\{ \begin{matrix}
j'&j&K \cr L&L'&1 \cr\end{matrix} \right\}
\nonumber\\
&&\widetilde{{\cal D}}_{00}^{K0}(\mu' j' \mu j)\,
\Re e\Big(i^{\delta_{I1}}
\widetilde N^{L'\ast}_{1}(\mu' j')\,\widetilde N^L_1(\mu j)
\Big)\,,\\
f_{LT}^{K}=
16\sqrt{2}\,\pi\,
{\hat K}^2\,\sum_{L' \mu' j' L \mu j}&&(-)^{L'+L+j}
\left(\begin{matrix}
L'&L& K \cr 0 &-1&1 \cr\end{matrix}\right) 
\left\{ \begin{matrix}
j'&j&K \cr L&L'&1\cr\end{matrix} \right\}
\nonumber\\
&&
\Big((-)^{L'+\mu'+j'}+1\Big)\widetilde{{\cal D}}_{00}^{K0}(\mu' j' \mu j)\,
\Re e\Big(i^{\delta_{I1}}
\widetilde C^{L'\ast}(\mu'j')\,\widetilde N^L_1(\mu j)\Big)\,,\\
f_{TT}^{K}=
16\,\pi\,{\hat K}^2\,\sum_{L' \mu' j' L \mu j}&&(-)^{L'+L+j}
\left(\begin{matrix}
L'&L& K \cr -1 &-1&2 \cr\end{matrix}\right) 
\left\{ \begin{matrix}
j'&j&K \cr L&L'&1 \cr\end{matrix} \right\}
\nonumber\\
&&
\widetilde{{\cal D}}_{00}^{K0}(\mu' j' \mu j)\,
\Re e\Big(i^{\delta_{I1}}
\widetilde N^{L'\ast}_{-1}(\mu' j')\,\widetilde N^L_1(\mu j)
\Big)\,.
\eeqa

At the end of this appendix, 
we will give the explicit multipole decomposition of the 
various inclusive form factors of $d(e,e')np$, which can be obtained from the 
($K=0$)-coefficients of (\ref{strucfunmultL})-(\ref{strucfunmultLTs}) 
according to
\begin{equation}
F_{a} ^{(\prime)I-M} =  (-)^{I+M}\,(1+\delta_{M0})\,
\frac{\pi}{3} \,(f_{a} ^{(\prime )IM+,\,0}-
f_{a} ^{(\prime )IM-,\,0})\,.
\end{equation}
The unpolarized form factors are given by
\beqa
F_L &=&\frac{16 \pi^2}{3} \sum_{Lj \mu }\frac{e^{-2\rho_\mu^j}}{2L+1}
|C^L (\mu j) |^2, \\
F_T &=&\frac{16 \pi^2}{3} \sum_{Lj \mu }\frac{e^{-2\rho_\mu^j}}{2L+1}
(|E^L (\mu j) |^2 +|M^L (\mu j) |^2)\,, \label{a2}
\eeqa
the vector polarization form factors by
\begin{eqnarray}
F_{LT} ^{\prime 1-1}&=& 32 \pi^2 \sqrt 2 \sum_{LL'j \mu } (-)^{j}
\left(\begin{matrix}
L'&L& 1 \cr 0 &-1&1 \cr\end{matrix}\right)
\bigg\{ \begin{matrix}
L'&L&1 \cr 1&1&j \cr\end{matrix} \bigg\}e^{-2\rho_\mu^j}\,
\Re e [C^{L'} (\mu j)^* N^L_1 (\mu j)],\\
F_{T} ^{\prime 10}&=& 16 \pi^2  \sum_{LL'j \mu } (-)^{j}
\left(\begin{matrix}
L'&L& 1 \cr 1 &-1&0 \cr\end{matrix}\right)
\bigg\{ \begin{matrix}
L'&L&1 \cr 1&1&j \cr\end{matrix} \bigg\}e^{-2\rho_\mu^j}\, 
\Re e [N^{L'}_1 (\mu j)^*N^L_1 (\mu j)]\,,\label{a5}
\end{eqnarray}
and finally the tensor polarization form factors by
\begin{eqnarray}
F_{L} ^{20}&=& -16 \pi^2 \sqrt{\frac{5}{3}} \sum_{LL'j \mu } (-)^{j}
\left(\begin{matrix}
L'&L& 2 \cr 0 &0&0 \cr\end{matrix}\right)
\bigg\{ \begin{matrix}
L'&L&2 \cr 1&1&j \cr\end{matrix} \bigg\}e^{-2\rho_\mu^j}\, 
\Re e [C^{L'} (\mu j)^* C^L (\mu j)], \\
F_{LT} ^{2-1}&=& 32 \pi^2 \sqrt{\frac{10}{3}} \sum_{LL'j \mu } (-)^{j}
\left(\begin{matrix}
L'&L& 2 \cr 0 &-1&1 \cr\end{matrix}\right)
\bigg\{ \begin{matrix}
L'&L&2 \cr 1&1&j \cr\end{matrix} \bigg\}e^{-2\rho_\mu^j}\, 
\Re e [C^{L'} (\mu j)^*N^L_1 (\mu j)], \\
F_{T} ^{20}&=& 16 \pi^2 \sqrt{\frac{5}{3}} \sum_{LL'j \mu } (-)^{j}
\left(\begin{matrix}
L'&L& 2 \cr 1 &-1&0 \cr\end{matrix}\right)
\bigg\{ \begin{matrix}
L'&L&2 \cr 1&1&j \cr\end{matrix} \bigg\}e^{-2\rho_\mu^j}\,
\Re e [N^{L'}_1 (\mu j)^*N^L_1 (\mu j)], \label{a9}\\
F_{TT} ^{2-2}&=& 16 \pi^2 \sqrt{\frac{5}{3}} \sum_{LL'j \mu } (-)^{j}
\left(\begin{matrix}
L'&L& 2 \cr -1 &-1&2 \cr\end{matrix}\right)
\bigg\{ \begin{matrix}
L'&L&2 \cr 1&1&j \cr\end{matrix} \bigg\}e^{-2\rho_\mu^j}\,
\Re e [N^{L'}_{-1} (\mu j)^*N^L_1 (\mu j)]\,. \label{a10}
\end{eqnarray}
These expressions have already been reported before in~\cite{LeT91}. 
The following two additional form factors vanish below pion threshold 
if time reversal invariance holds. However, if one considers the partial
inclusive reaction $d(e,e')np$, they become nonvanishing above pion 
threshold in a coupled channel approach including isobars without explicit 
treatment of the $NN\pi$-channels~\cite{LeT91}.
\beqa
F_{LT} ^{1-1}&=& -32 \pi^2 \sqrt 2 \sum_{LL'j \mu } (-)^{j}
\left(\begin{matrix}
L'&L& 1 \cr 0 &-1&1 \cr\end{matrix}\right)
\bigg\{ \begin{matrix}
L'&L&1 \cr 1&1&j \cr\end{matrix} \bigg\}e^{-2\rho_\mu^j}\,
\Im m [C^{L'} (\mu j)^* N^L_1 (\mu j)]\,,\\
F_{LT} ^{\prime 2-1}&=&-32 \pi^2 \sqrt{\frac{10}{3}} \sum_{LL'j \mu }(-)^{j}
\left(\begin{matrix}
L'&L& 2 \cr 0 &-1&1 \cr\end{matrix}\right)
\bigg\{ \begin{matrix}
L'&L&2 \cr 1&1&j \cr\end{matrix} \bigg\}e^{-2\rho_\mu^j}\, 
\Im m [C^{L'} (\mu j)^*N^L_1 (\mu j)]\,.
\end{eqnarray}
We would like to emphasize, that if one considers the completely 
inclusive process $d(e,e')X$, the corresponding additional form factors 
will vanish as long as time reversal invariance holds. 


\setcounter{equation}{0}
\section{Separation of a specific structure function }
\label{sepstrucfun}

The first step in the determination of any structure function is the 
measurement of that asymmetry to which it contributes. This requires 
already a number of measurements with different electron and deuteron 
polarization parameters, i.e., two for $A_0(X)$, $A_e(X)$, $A_d^V(X)$, 
and $A_d^T(X)$ and four for $A_{ed}^V(X)$ and $A_{ed}^T(X)$. Next one must 
determine how many different settings of the angles $\phi$, $\phi_d$ and 
$\theta_d$ 
are necessary for the final separation. As one will see in the following, 
the structure functions can be divided into different classes according to 
the minimum number of asymmetry measurements required for their extraction. 
A quick glance at (\ref{asyall}) shows that the six asymmetries contain
differing numbers of structure functions, taking for example an observable 
of type $A$, one finds four in $A_0(X)$, one in $A_e(X)$,
8 in $A_d^V(X)$, 16 in $A_d^T(X)$, 5 in $A_{ed}^V(X)$, and 7 in
$A_{ed}^T(X)$. Similar numbers are found for an observable of type $B$.

First we will consider an observable of type $A$. 
Obviously, the simplest case is the electron asymmetry $A_e(X)$ 
containing only $f_{LT}'(X)$. This means that $f_{LT}'(X)$ can be determined 
from just one asymmetry measurement. Similarly 
a close inspection of the dependence of the asymmetries on the angles
 $\phi$, $\phi_d$ and $\theta_d$ shows that four other structure functions 
and two combinations need only one asymmetry measurement. 
These are, introducing for convenience 
\begin{equation}
S_{d/ed}^{V/T}(X;\phi ,\tilde \phi, \theta_d)=S_0(\phi) 
A_{d/ed}^{V/T}(X;\phi ,\tilde \phi, \theta_d)\,
\end{equation}
in order to exhibit the angular dependence explicitly, 
\begin{eqnarray}
c(k_1^{\mathrm{lab}},\,k_2^{\mathrm{lab}})\,\rho_{LT}f_{LT}^{10}(X)& 
= & S_d^V(X;\frac{\pi}{2}, \tilde \phi,0)\,,\\
c(k_1^{\mathrm{lab}},\,k_2^{\mathrm{lab}})\,\rho_{LT}f_{LT}^{11,-}(X)& 
= &-\sqrt{2} S_d^V(X;\frac{\pi}{2}, 
0,\frac{\pi}{2})\,,\\
c(k_1^{\mathrm{lab}},\,k_2^{\mathrm{lab}})\,\rho_{T}'f_{T}^{\prime \,10}(X)& 
= & -S_{ed}^V(X;\frac{\pi}{2}, 0,0)\,,\\
c(k_1^{\mathrm{lab}},\,k_2^{\mathrm{lab}})\,\rho_{T}'f_{T}^{\prime \,11}(X)& 
= &\sqrt{2} S_{ed}^VX;(\frac{\pi}{2}
, 0,\frac{\pi}{2})\,,\\
c(k_1^{\mathrm{lab}},\,k_2^{\mathrm{lab}})\,
\rho_{LT}'f_{LT}^{\prime \,11,+}(X)& = -&\sqrt{2} S_{ed}^V(X;\frac{\pi}{2}
,\frac{\pi}{2},\frac{\pi}{2})\,,\\
c(k_1^{\mathrm{lab}},\,k_2^{\mathrm{lab}})\,
\rho_{LT}'f_{LT}^{\prime \,20}(X)& = & S_{ed}^T(X;\frac{\pi}{2},\tilde \phi,0)\,.
\end{eqnarray}

Nine other structure functions and seven combinations require only two 
measurements for their separation. These are
\begin{eqnarray}
c(k_1^{\mathrm{lab}},\,k_2^{\mathrm{lab}})\,\rho_{TT}f_{TT}^{10}(X)& = &
\frac{2}{\sqrt{3}}
                 S_d^V(X;\frac{\pi}{3}, \tilde \phi,0)
                -S_d^V(X;\frac{\pi}{2}, \tilde \phi,0)\,,\\
c(k_1^{\mathrm{lab}},\,k_2^{\mathrm{lab}})\,\rho_{LT}f_{LT}^{11,+}(X)& = &
-\Big[
                 S_d^V(X;\frac{\pi}{4},\frac{\pi}{2},\frac{\pi}{2})
                -S_d^V(X;\frac{3\pi}{4},\frac{\pi}{2},\frac{\pi}{2})\Big]\,,\\
c(k_1^{\mathrm{lab}},\,k_2^{\mathrm{lab}})\,\rho_{TT}f_{TT}^{11,-}(X)& = &
\sqrt{\frac{2}{3}}\Big[
                2S_d^V(X;\frac{\pi}{3},0,\frac{\pi}{2})
           -\sqrt{3}S_d^V(X;\frac{\pi}{2},0,\frac{\pi}{2}\Big]\,,\\
c(k_1^{\mathrm{lab}},\,k_2^{\mathrm{lab}})\,\rho_{LT}f_{LT}^{20}(X)& = &
\frac{1}{\sqrt{2}}\Big[
                 S_d^T(X;\frac{\pi}{4},\tilde \phi,0)
                -S_d^T(X;\frac{3\pi}{4},\tilde \phi,0)\Big]\,,\\
c(k_1^{\mathrm{lab}},\,k_2^{\mathrm{lab}})\,\rho_{LT}f_{LT}^{21+}(X)& = &
\sqrt{\frac{3}{2}}\Big[
                 S_d^T(X;\pi,\frac{\pi}{4},\theta_d^0)
                -S_d^T(X;0,\frac{\pi}{4},\theta_d^0)\Big]\,,\\
c(k_1^{\mathrm{lab}},\,k_2^{\mathrm{lab}})\,\rho_{LT}f_{LT}^{21,-}(X)& = &
\frac{\sqrt{3}}{2}\Big[
                 S_{d}^T(X;\frac{\pi}{2},\frac{\pi}{2},\theta_d^0)
                -S_{d}^T(X;\frac{\pi}{2},-\frac{\pi}{2},\theta_d^0)\Big]\,,\\
c(k_1^{\mathrm{lab}},\,k_2^{\mathrm{lab}})\,\rho_{LT}f_{LT}^{22+}(X)& = &
\sqrt{\frac{3}{2}}\Big[
                 S_d^T(X;\pi,\frac{\pi}{2},\theta_d^0)
                -S_d^T(X;0,\frac{\pi}{2},\theta_d^0)\Big]\,,\\
c(k_1^{\mathrm{lab}},\,k_2^{\mathrm{lab}})\,\rho_{LT}f_{LT}^{22,-}(X)& = &
-\sqrt{\frac{2}{3}}\Big[
                2S_d^T(X;\frac{\pi}{2},\frac{\pi}{4},\frac{\pi}{2})
                +S_d^T(X;\frac{3\pi}{2},\frac{\pi}{4},0)\Big]\,,\\
c(k_1^{\mathrm{lab}},\,k_2^{\mathrm{lab}})\,
\rho'_{LT}f_{LT}^{\prime \,10}(X)& = &
                 -\Big[S_{ed}^V(X;0, \tilde \phi,0)
                -S_{ed}^V(X;\frac{\pi}{2}, \tilde \phi,0)\Big]\,,\\
c(k_1^{\mathrm{lab}},\,k_2^{\mathrm{lab}})\,
\rho'_{LT}f_{LT}^{\prime \,11,-}(X)& = &\sqrt{2}\Big[
                 S_{ed}^V(X;0,0,\frac{\pi}{2})
                -S_{ed}^V(X;\frac{\pi}{2},0,\frac{\pi}{2})\Big]\,,\\
c(k_1^{\mathrm{lab}},\,k_2^{\mathrm{lab}})\,\rho'_{T}f_{T}^{\prime \,21}(X)& 
= &-\frac{\sqrt{3}}{2}\Big[
                 S_{ed}^T(X;0,\frac{\pi}{2},\theta_d^0)
                +S_{ed}^T(X;\pi,\frac{\pi}{2},\theta_d^0)\Big]\,,\\
c(k_1^{\mathrm{lab}},\,k_2^{\mathrm{lab}})\,\rho'_{T}f_{T}^{\prime \,22}(X)& 
= &\sqrt\frac{2}{3}\Big[
                2S_{ed}^T(X;\frac{\pi}{2},\frac{\pi}{4},\frac{\pi}{2})
                +S_{ed}^T(X;\frac{\pi}{2},\frac{\pi}{4},0)\Big]\,,\\
c(k_1^{\mathrm{lab}},\,k_2^{\mathrm{lab}})\,
\rho'_{LT}f_{LT}^{\prime \,21,-}(X)& = &-\frac{\sqrt{3}}{2}\Big[
                 S_{ed}^T(X;0,\frac{\pi}{2},\theta_d^0)
                -S_{ed}^T(X;\pi,\frac{\pi}{2},\theta_d^0)\Big]\,,\\
c(k_1^{\mathrm{lab}},\,k_2^{\mathrm{lab}})\,
\rho'_{LT}f_{LT}^{\prime \,21,+}(X)& = &-\frac{\sqrt{3}}{2}\Big[
                 S_{ed}^T(X;\frac{\pi}{2},0,\theta_d^0)
                -S_{ed}^T(X;\frac{\pi}{2},\pi,\theta_d^0)\Big]\,,\\
c(k_1^{\mathrm{lab}},\,k_2^{\mathrm{lab}})\,
\rho'_{LT}f_{LT}^{\prime \,22,+}(X)& = &\sqrt{\frac{3}{2}}\Big[
                 S_{ed}^T(X;\frac{\pi}{2},0,\theta_d^0)
                +S_{ed}^T(X;\frac{\pi}{2},\pi,\theta_d^0)\Big]\nonumber\\
                          &\stackrel{\rm or}{=} &\sqrt\frac{2}{3}\Big[
                2S_{ed}^T(X;\frac{\pi}{2},0,\frac{\pi}{2})
                +S_{ed}^T(X;\frac{\pi}{2},0,0)\Big]\,,\\
c(k_1^{\mathrm{lab}},\,k_2^{\mathrm{lab}})\,\rho_{LT}^\prime 
f_{LT}^{\prime\, 22-}(X)& = &\sqrt{{\frac{2}{3}}}[
                 S_{ed}^T(X;0,\frac{\pi}{4},\frac{\pi}{2})
                -S_{ed}^T(X;\pi,\frac{\pi}{4},\frac{\pi}{2})\Big]\,,
\end{eqnarray}
where $\theta_d^0=\mbox{arcos}\,(1/\sqrt{3})$ (see Sect.~\ref{completesets}).

The terms $\rho_Lf_L^{11}+\rho_Tf_T^{11}$ in $A_d^V$ and 
$\rho_Lf_L^{20}+\rho_Tf_T^{20}$ in $A_d^T$ can also be determined from two 
asymmetry measurements, i.e., 
\begin{eqnarray}
c(k_1^{\mathrm{lab}},\,k_2^{\mathrm{lab}})\,
(\rho_Lf_L^{11}+\rho_Tf_T^{11}(X))&=& -\frac{1}{\sqrt{2}}\Big[
                 S_d^V(X;\frac{\pi}{4},\frac{\pi}{2},\frac{\pi}{2})
                +S_d^V(X;\frac{3\pi}{4},\frac{\pi}{2},\frac{\pi}{2})\Big]\,,\\
c(k_1^{\mathrm{lab}},\,k_2^{\mathrm{lab}})\,
(\rho_Lf_L^{20}+\rho_Tf_T^{20}(X))&=& \frac{1}{2}\Big[
                 S_d^T(X;\frac{\pi}{4},\tilde \phi,0)
                +S_d^T(X;\frac{3\pi}{4},\tilde \phi,0)\Big]\,.
\end{eqnarray}
In order to separate 
the longitudinal from the transverse part one needs in addition a
{\sc Rosenbluth} analysis.

Increasing the number of asymmetry measurements to three allows 
one to determine only one 
further structure function and one combination, namely
\begin{eqnarray}
c(k_1^{\mathrm{lab}},\,k_2^{\mathrm{lab}})\,
\rho_{TT}f_{TT}^{11,+}(X)& = &-\frac{1}{\sqrt{2}}\Big[
                 S_d^V(X;\frac{\pi}{4},\frac{\pi}{2},\frac{\pi}{2})
                +S_d^V(X;\frac{3\pi}{4},\frac{\pi}{2},\frac{\pi}{2})
               -2S_d^V(X;\frac{\pi}{2},\frac{\pi}{2},\frac{\pi}{2})\Big]\,,\\
c(k_1^{\mathrm{lab}},\,k_2^{\mathrm{lab}})\,
\rho_{TT}f_{TT}^{20}(X)& = &\frac{1}{2}\Big[
                 S_d^T(X;\frac{\pi}{4},\tilde \phi,0)
                +S_d^T(X;\frac{3\pi}{4},\tilde \phi,0)\Big]
                -S_d^T(X;\frac{3\pi}{2},\tilde \phi,0)\,.
\end{eqnarray}

All thirteen vector structure functions are then determined and fifteen of 
the tensor ones or respective combinations. For the remaining combinations
$f_{TT}^{21,+}(X)$, $f_{TT}^{21,-}(X)$, $f_{TT}^{22,+}(X)$, 
$f_{TT}^{22,-}(X)$, 
($\rho_Lf_L^{21}+\rho_Tf_T^{21}(X)$), and ($\rho_Lf_L^{22}+\rho_Tf_T^{22}(X)$)
one needs four measurements in order to 
determine them. The last two combinations require a {\sc Rosenbluth} separation in 
addition. 

Now we turn to the separation of structure functions for $B$-type. 
Without target and electron polarization one has two
structure functions, one of which can be determined by one out-of-plane 
measurements, namely $(f_{LT}^{00-}(X),\phi = \frac{\pi}{2})$,
and the other needs two, 
$(f_{TT}^{00-}(X),\phi = \frac{\pi}{4}, \frac{\pi}{2})$. 
Electron polarization alone without deuteron polarization leads to 
two other structure functions: $f_{T}^{\prime\, 00-}(X)$
which can be obtained from one out-of-plane measurement at
$\phi = \frac{\pi}{2}$, and  
$f_{LT}^{\prime\, 00}(X)$, which requires two settings.

For the structure functions with target polarization we find that  
six of them can be determined by a single asymmetry
measurement. They are
\begin{eqnarray}
c(k_1^{\mathrm{lab}},\,k_2^{\mathrm{lab}})\,\rho_{LT}f_{LT}^{11-}(X)& = & \sqrt{2} S_d^V(X;\frac{\pi}{2}, 
\frac{\pi}{2},\frac{\pi}{2})\,,\\
c(k_1^{\mathrm{lab}},\,k_2^{\mathrm{lab}})\,\rho_{LT}f_{LT}^{20-}(X)& = & S_d^T(X;\frac{\pi}{2}, 
\tilde \phi,0)\,,\\
c(k_1^{\mathrm{lab}},\,k_2^{\mathrm{lab}})\,\rho_{T}'f_{T}^{\prime \,11}(X)& = &-\sqrt{2} S_{ed}^V(X;\frac{\pi}{2},
\frac{\pi}{2} ,\frac{\pi}{2})\,,\\
c(k_1^{\mathrm{lab}},\,k_2^{\mathrm{lab}})\,\rho_{T}'f_{T}^{\prime \,20}(X)& = &-S_{ed}^T(X;\frac{\pi}{2},\tilde\phi,0)\,,\\
c(k_1^{\mathrm{lab}},\,k_2^{\mathrm{lab}})\,\rho_{LT}'f_{LT}^{\prime \,10+}(X)& = & S_{ed}^V(X;\frac{\pi}{2}
, \tilde\phi,0)\,,\\
c(k_1^{\mathrm{lab}},\,k_2^{\mathrm{lab}})\,\rho_{LT}'f_{LT}^{\prime \,11+}(X)& = &-\sqrt{2} 
S_{ed}^V(X;\frac{\pi}{2} ,0,\frac{\pi}{2})\,.
\end{eqnarray}

Sixteen other structure functions require only two 
measurements for their separation. These are
\begin{eqnarray}
c(k_1^{\mathrm{lab}},\,k_2^{\mathrm{lab}})\,\rho_{LT}f_{LT}^{10+}(X)& = &\frac{1}{\sqrt{2}}\Big[
                 S_d^V(X;\frac{\pi}{4}, \tilde \phi,0)
                -S_d^V(X;\frac{3\pi}{4}, \tilde \phi,0)\Big]\,,\\
c(k_1^{\mathrm{lab}},\,k_2^{\mathrm{lab}})\,\rho_{LT}f_{LT}^{11+}(X)& = &-\Big[
                 S_d^V(X;\frac{\pi}{4},0,\frac{\pi}{2})
                -S_d^V(X;\frac{3\pi}{4},0,\frac{\pi}{2})\Big]\,,\\
c(k_1^{\mathrm{lab}},\,k_2^{\mathrm{lab}})\,\rho_{LT}f_{LT}^{21+}(X)& = &\sqrt{\frac{2}{3}}\Big[
                 S_{d}^T(X;\pi,\frac{\pi}{2},\frac{\pi}{4})
                -S_{d}^T(X;0,\frac{\pi}{2},\frac{\pi}{4})\Big]\,,\\
c(k_1^{\mathrm{lab}},\,k_2^{\mathrm{lab}})\,\rho_{LT}f_{LT}^{21-}(X)& = &-\sqrt{\frac{2}{3}}\Big[
                 S_{d}^T(X;\frac{\pi}{2},0,\frac{\pi}{4})
                -S_{d}^T(X;\frac{\pi}{2},0,\frac{3\pi}{4})\Big]\,,\\
c(k_1^{\mathrm{lab}},\,k_2^{\mathrm{lab}})\,\rho_{LT}f_{LT}^{22+}(X)& = &\sqrt{\frac{2}{3}}\Big[
                 S_{d}^T(X;0,\frac{\pi}{4},\frac{\pi}{2})
                -S_{d}^T(X;\pi,\frac{\pi}{4},\frac{\pi}{2})\Big]\,,\\
c(k_1^{\mathrm{lab}},\,k_2^{\mathrm{lab}})\,\rho_{LT}f_{LT}^{22-}(X)& = &2\sqrt{\frac{2}{3}}\Big[
                S_d^T(X;\frac{\pi}{2},0,\frac{\pi}{2})
                +\frac{1}{2}S_d^T(X;\frac{\pi}{2},\tilde\phi,0)\Big]\,,\\
c(k_1^{\mathrm{lab}},\,k_2^{\mathrm{lab}})\,\rho_{TT}f_{TT}^{11-}(X)& = &-
                S_d^V(X;\frac{\pi}{2},\frac{\pi}{2},\frac{\pi}{2})
           +\sqrt{2} S_d^V(X;\frac{\pi}{4},\frac{\pi}{2},\frac{\pi}{2})\,,\\
c(k_1^{\mathrm{lab}},\,k_2^{\mathrm{lab}})\,\rho_{TT}f_{TT}^{20-}(X)& = &S_d^T(X;\frac{\pi}{4},\tilde\phi,0)
                -\frac{1}{\sqrt{2}} S_d^T(X;\frac{\pi}{2},\tilde \phi,0) \,,\\
c(k_1^{\mathrm{lab}},\,k_2^{\mathrm{lab}})\,\rho'_Tf_T^{\prime \,21}(X)& = &\sqrt{\frac{3}{2}}\Big[
                 S_{ed}^T(X;0,\frac{\pi}{4},\theta_d^0)
                +S_{ed}^T(X;\pi,\frac{\pi}{4},\theta_d^0)\Big] \,,\\
c(k_1^{\mathrm{lab}},\,k_2^{\mathrm{lab}})\,\rho'_{T}f_{T}^{\prime \,22}(X)& = &\sqrt{\frac{3}{2}}\Big[
                 S_{ed}^T(X;0,\frac{\pi}{2},\theta_d^0)
                +S_{ed}^T(X;\pi,\frac{\pi}{2},\theta_d^0)\Big]\,,\\
c(k_1^{\mathrm{lab}},\,k_2^{\mathrm{lab}})\,\rho'_{LT}f_{LT}^{\prime \,11-}(X)& = &-\sqrt{2}\Big[
                 S_{ed}^V(X;0, \frac{\pi}{2},\frac{\pi}{2})
                -S_{ed}^V(X;\frac{\pi}{2}, \frac{\pi}{2},\frac{\pi}{2})\Big]\,,\\
c(k_1^{\mathrm{lab}},\,k_2^{\mathrm{lab}})\,\rho'_{LT}f_{LT}^{\prime \,20-}(X)& = &
                 S_{ed}^T(X;\frac{\pi}{2},\tilde\phi,0)
                -S_{ed}^T(X;0,\tilde\phi,0)\,,\\
c(k_1^{\mathrm{lab}},\,k_2^{\mathrm{lab}})\,\rho'_{LT}f_{LT}^{\prime \,21+}(X)& = &-\sqrt{\frac{2}{3}}\Big[
                 S_{ed}^T(X;\frac{\pi}{2},\frac{\pi}{2},\frac{\pi}{4})
                -S_{ed}^T(X;\frac{\pi}{2},\frac{\pi}{2},\frac{3\pi}{4})\Big]\,,\\
c(k_1^{\mathrm{lab}},\,k_2^{\mathrm{lab}})\,\rho'_{LT}f_{LT}^{\prime \,21-}(X)& = &\sqrt{\frac{3}{2}}\Big[
                 S_{ed}^T(X;0,\frac{\pi}{4},\theta_d^0)
                -S_{ed}^T(X;\pi,\frac{\pi}{4},\theta_d^0)\Big]\,,\\
c(k_1^{\mathrm{lab}},\,k_2^{\mathrm{lab}})\,\rho'_{LT}f_{LT}^{\prime \,22+}(X)& = &\sqrt{\frac{2}{3}}\Big[2
                 S_{ed}^T(X;\frac{\pi}{2},\frac{\pi}{4},\frac{\pi}{2})
                +S_{ed}^T(X;\frac{\pi}{2},\tilde\phi,0)\Big]\,,\\
c(k_1^{\mathrm{lab}},\,k_2^{\mathrm{lab}})\,\rho'_{LT}f_{LT}^{\prime \,22-}(X)& = &\sqrt{\frac{3}{2}}\Big[
                 S_{ed}^T(X;0,\frac{\pi}{2},\theta_d^0)
                -S_{ed}^T(X;\pi,\frac{\pi}{2},\theta_d^0)\Big]\,.
\end{eqnarray}

In this last set of equations as well as for what follows
we have made a particular choice of angles to allow the separations.
This choice is not unique and other choices will lead to
different coefficients in the linear combinations.
The terms $\rho_Lf_L^{10}(X)+\rho_Tf_T^{10}(X)$  and 
$\rho_Lf_L^{11}(X)+\rho_Tf_T^{11}(X)$ in $A_d^V(X)$ can 
also be determined from two 
asymmetry measurements, i.e., 
\begin{eqnarray}
c(k_1^{\mathrm{lab}},\,k_2^{\mathrm{lab}})\,\rho_Lf_L^{10}(X)+\rho_Tf_T^{10}(X)&=& \frac{1}{2}\Big[
                 S_d^V(X;\frac{\pi}{4},\tilde\phi,0)
                +S_d^V(X;\frac{3\pi}{4},\tilde\phi,0)\Big]\,,\\
c(k_1^{\mathrm{lab}},\,k_2^{\mathrm{lab}})\,\rho_Lf_L^{11}(X)+\rho_Tf_T^{11}(X)&=&- \frac{1}{\sqrt{2}}\Big[
                 S_d^V(X;\frac{\pi}{4},0,\frac{\pi}{2})
                +S_d^V(X;\frac{3\pi}{4},0,\frac{\pi}{2})\Big]\,.
\end{eqnarray}
In order to separate 
the longitudinal from the transverse part one needs in addition a
{\sc Rosenbluth} analysis.

Increasing the number of asymmetry measurements to three allows 
one to determine only two further structure functions, namely
\begin{eqnarray}
c(k_1^{\mathrm{lab}},\,k_2^{\mathrm{lab}})\,\rho_{TT}f_{TT}^{11+}(X)& = &-\frac{1}{\sqrt{2}}\Big[
                 S_d^V(X;\frac{\pi}{4},0,\frac{\pi}{2})
                +S_d^V(X;\frac{3\pi}{4},0,\frac{\pi}{2})
                -2 S_d^V(X;\frac{\pi}{2},0,\frac{\pi}{2})\Big]\,,\\
c(k_1^{\mathrm{lab}},\,k_2^{\mathrm{lab}})\,\rho_{TT}f_{TT}^{10+}(X)& = &\frac{1}{2}\Big[
                 S_d^V(X;\frac{\pi}{4},\tilde \phi,0)
                +S_d^V(X;\frac{3\pi}{4},\tilde \phi,0)\Big]
                -S_d^V(X;\frac{\pi}{2},\tilde \phi,0)\,.
\end{eqnarray}
Three settings also allow a determination of the combination      
\begin{equation}
c(k_1^{\mathrm{lab}},\,k_2^{\mathrm{lab}})\,\rho_Lf_L^{22}(X)+\rho_Tf_T^{22}(X)=\sqrt{\frac{2}{3}}
\Big[S_d^T(X;\frac{\pi}{4},\frac{\pi}{4},\frac{\pi}{2})
+S_d^T(X;\frac{3\pi}{4},\frac{\pi}{4},\frac{\pi}{2})
+\frac{1}{\sqrt{2}}S_d^T(X;\frac{\pi}{2},\tilde\phi,0)\Big]\ .\label{b24}
\end{equation}
All fourteen vector structure functions and sixteen of 
the tensor ones are then determined.
For the remaining functions 
$f_{TT}^{21+}(X)$, $f_{TT}^{21-}(X)$, 
$f_{TT}^{22+}(X)$, $f_{TT}^{22-}(X)$ and
($\rho_Lf_L^{21}(X)+\rho_Tf_T^{21}(X)$)
one needs four measurements in order to 
determine them. The last combination and the one in (\ref{b24}) 
requires an additional {\sc Rosenbluth} separation. A survey on the number of 
asymmetry measurements needed for the separation of a given structure 
function is given in Tab.~\ref{tab2}. 

\begin{table}
\caption{Number of asymmetry measurements for a structure function of an 
observable $X$. The symbol ($R$) indicates the need of an additional 
{\sc Rosenbluth} $L$-$T$-separation.}
\begin{ruledtabular}
\begin{tabular}{ccccccccc}
&&&&$X \in A$  & & & & \\ 
\colrule
$IM$& $L,\,T$ & $LT+$ & $LT-$ & $TT+$ &  $TT-$ & $T'$ & $LT'+$ & $LT'-$ \\ 
\colrule
10 & $-$  & $-$ & 1  & $-$ &  2 &  1 & $-$& 2   \\
11 & 2(R)& 2  & 1  &  3 & 2  &  1 & 1 & 2   \\
20 & 2(R)& 2  & $-$ &  3 & $-$ &  $-$& 1 & $-$  \\
21 & 4(R)& 2  & 2  &  4 & 4  &  2 & 2 & 2   \\
22 & 4(R)& 2  & 2  &  4 & 4  &  2 & 2 & 2   \\
\colrule
&&&&$X \in B$  & & & & \\ 
\colrule
$IM$& $L,\,T$ & $LT+$ & $LT-$ & $TT+$ &  $TT-$ & $T'$ & $LT'+$ & $LT'-$ \\ 
\colrule
10 & 2(R)& 2  & $-$ &  3 & $-$ & $-$ & 1 & $-$  \\
11 & 2(R)& 2  & 1  &  3 & 2  &  1 & 1 & 2   \\
20 & $-$ & $-$  & 1  & $-$ & 2  &  1 & $-$& 2   \\
21 & 4(R)& 2  & 2  &  4 & 4  &  2 & 2 & 2   \\
22 & 3(R)& 2  & 2  &  4 & 4  &  2 & 2 & 2   \\
\end{tabular}
\end{ruledtabular}
\label{tab2}
\end{table}


\setcounter{equation}{0}
\section{Explicit Expressions for the Matrix Representation of 
${\cal U}^{\lambda' \lambda I M}_{X}$}\label{MatRep}

For the matrix representation of the ${\cal U}^{\lambda' \lambda I M}_{X}$ 
in (\ref{CIMX}) we switch to the spherical representation according to
(\ref{cartsph})
\begin{equation}
\widetilde C^{I M \lambda' \lambda}_{j'j}(X) = 
\sum_{\tau'\nu'\tau\nu}s_{\alpha'}^{\tau'\nu'}s_{\alpha}^{\tau\nu}\,
C^{m_1'm_2'\lambda_d'}_{m_1 m_2 \lambda_d}(\tau'\nu'\tau\nu IM)\,.
\end{equation}
The explicit forms of the 
$C^{m_1'm_2'\lambda_d'}_{m_1 m_2 \lambda_d}(\tau'\nu'\tau\nu IM)$ for the 
helicity, hybrid and standard bases are: \\
(i) Helicity basis with labeling $(\lambda^{(\prime)}_p,
\lambda^{(\prime)}_n, \lambda^{(\prime)}, \lambda^{(\prime)}_d)$  
\beqa
C^{\lambda_p' \lambda_n' \lambda_d'}_{\lambda_p \lambda_n \lambda_d}
(\tau'\nu'\tau\nu IM)&=& 
  2\sqrt{3}\,(-)^{\lambda'_p+\lambda'_n -\lambda_d+\tau'+\tau} 
  \hat\tau'\hat\tau\hat I
  \left(\begin{matrix}
1 & 1 & I\cr \lambda_d' & -\lambda_d & M\cr\end{matrix}\right)
\left(\begin{matrix}
\frac{1}{2} & \frac{1}{2} & \tau' \cr 
  \lambda_p' & -\lambda_p & \nu'\cr\end{matrix}\right)
  \left(\begin{matrix}
\frac{1}{2} & \frac{1}{2} & \tau \cr
       \lambda_n' & -\lambda_n & \nu\cr\end{matrix}\right)\,.
\eeqa
(ii) Hybrid basis with the labeling
$(\tilde\lambda^{(\prime)}_p,
\tilde\lambda^{(\prime)}_n, \lambda^{(\prime)}, \tilde\lambda^{(\prime)}_d)$, 
where for proton, neutron and deuteron the spin projections refer to the 
transverse $y$-axis, 
\beqa
C^{\tilde\lambda_p' \tilde\lambda_n' \tilde\lambda_d'}
_{\tilde\lambda_p \tilde\lambda_n \tilde\lambda_d}
(\tau'\nu'\tau\nu IM)&=&2\sqrt{3}\, 
  \omega^{IM}_{1\tilde\lambda_d \tilde\lambda^{\prime}_d}\,
  \Big(\omega^{\tau' \nu'}
  _{\frac{1}{2}\tilde\lambda^{\prime}_p \tilde\lambda_p}\Big)^*
  \Big(\omega^{\tau \nu}
  _{\frac{1}{2}\tilde\lambda^{\prime}_n \tilde\lambda_n}\Big)^*\,,
\eeqa
where 
\beqa
\omega^{JM}_{j m' m} &=&(-)^{j-m'}\hat J \sum_{M'}
  \left(\begin{matrix}
j & J & j\cr -m' & M' & m\cr\end{matrix} 
\right)i^{M'-M}d^J_{M'M}(\frac{\pi}{2})\,.
\eeqa
(iii) Standard basis with index labeling
$(s^{(\prime)},m^{(\prime)}_s, \lambda^{(\prime)}, 
m^{(\prime)}_d)$ 
\beqa
C^{s' m_s' \lambda_d'}_{s m_s \lambda_d}(\tau'\nu'\tau\nu IM)&=& 
  {2\sqrt{3}\,}(-)^{1 -\lambda_d +s'-m'_s+\tau'+\tau}\hat s \hat s' \hat\tau'\hat\tau\,
  \hat I\left(\begin{matrix}
1 & 1 & 1\cr \lambda_d' & -\lambda_d & M\cr\end{matrix} \right)
\nonumber\\&& 
  \sum_{S\sigma}{\hat S}^2
    \left(\begin{matrix}
\tau' & \tau & S\cr \nu' & \nu & -\sigma\cr \end{matrix}\right)
    \left(\begin{matrix}
s' & s & S\cr m'_s & -m_s & -\sigma\cr \end{matrix}\right)
\left\{\begin{matrix}
\frac{1}{2} & \frac{1}{2} & \tau' \cr
       \frac{1}{2} & \frac{1}{2} & \tau \cr
       s' & s & S\cr\end{matrix}\right\}\,.
\eeqa

\setcounter{equation}{0}
\section{Explicit expressions of the one-body current matrix in 
plane wave {\sc Born} approximation in the standard representation}\label{Born}

In the plane wave {\sc Born} approximation (PWBA), 
the final $np$-scattering state is 
replaced by a pure plane wave. The deuteron wave function has the form
\begin{equation}
\langle \vec r\;|1\lambda_d\rangle =\langle r\theta\phi|1\lambda_d\rangle 
=\sum_{l=0,2}\frac{u_l(r)}{r}\langle 
\theta\phi|(l1)1\lambda_d\rangle \,,
\end{equation}
and the final state plane wave
\begin{equation}
\langle \vec r\;|\vec k sm_s\rangle =\frac{e^{i\vec k\cdot\vec r}}
{(2\pi)^{3/2}}|sm_s\rangle \,.
\end{equation}
The quantization axis is chosen along $\vec k$, the relative $np$ momentum 
in the final state.

Taking the current in the {\sc Dirac-Pauli} form, one obtains for the 
one-body current matrix element in the standard representation for the 
nonrelativistic charge density~\cite{BeA92}
\begin{eqnarray}
\langle sm_s| \rho^{NR} (\vec q,\vec P)|1\lambda_d\rangle & = &
 \delta_{s1}\sum_{j=1,2}\sum_{lm_l} C_{lm_l1m_s}^{1\lambda_d}
F_{1j} \langle \vec k_j|lm_l\rangle \,,\label{rnr}
\end{eqnarray}
where $C_{lm_l1m_s}^{1\lambda_d}$ denotes a {\sc Clebsch-Gordan} 
coefficient $\langle lm_l 1m_s\,|1\lambda_d\rangle$. 
The leading order relativistic contributions arise from the $p/M$-expansion 
of the {\sc Dirac} current and from the wave function boost. The 
corresponding charge density operators are denoted by $\rho^{R}$ and 
$\rho^{B}$, respectively,
\begin{eqnarray}
\langle sm_s| \rho^{R} (\vec q,\vec P)|1\lambda_d\rangle &=&
-\frac{1}{8M^2}\sum_{j=1,2}\sum_{lm_lm_s'} C_{lm_l1m_s'}^{1\lambda_d}
(2G_{Mj}-F_{1j})
\Big\{\delta_{s1}\delta_{m_sm_s'}q^2 +S_{j,k\times q}^{sm_sm_s'}
\Big\}\langle \vec k_j|lm_l\rangle \,,\\
\langle sm_s| \rho^{B} (\vec q,\vec P)|1\lambda_d\rangle &=&
\frac{1}{16M^2}\sum_{j=1,2}\sum_{lm_lm_s'} C_{lm_l1m_s'}^{1\lambda_d}F_{1j}
\Big\{(-q^2\delta_{s1}\delta_{m_sm_s'} 
+S_{1,k\times q}^{sm_sm_s'}
+S_{2,k\times q}^{sm_sm_s'})\langle \vec k_j|lm_l\rangle
\nonumber\\& &\hspace*{2cm} 
-(s_jq^2 -\Pi_{q})\delta_{s1}\delta_{m_sm_s'}
\langle \vec k_j|i\vec q\cdot \vec r\,|lm_l\rangle \Big\} \,.\label{rb}
\end{eqnarray}
Similarly, one has for the nonrelativistic transverse current ($\lambda=\pm 1$)
\begin{eqnarray}
\langle sm_s| J_\lambda^{NR} (\vec q,\vec P)|1\lambda_d\rangle &=&
\frac{e^{i\lambda \phi}}{2M}
\sum_{j=1,2}\sum_{lm_lm_s'} C_{lm_l1m_s'}^{1\lambda_d}
\Big\{F_{1j} \delta_{s1}\delta_{m_sm_s'} \Pi_{j,\lambda}
+G_{Mj}S_{j, q, \lambda}^{sm_sm_s'}
\Big\}\langle \vec k_j|lm_l\rangle \,.\label{jnr}
\end{eqnarray}
For the leading order relativistic current contributions we again distinguish 
$J_\lambda^{R}$ from expanding the {\sc Dirac} current and 
two contributions from the boost, $J_{c,\lambda}^{B}$ and 
$J_{s,\lambda}^{B}$, with respect to the 
nonrelativistic convection and spin current, 
\begin{eqnarray}
\langle sm_s| J_\lambda^{R} (\vec q,\vec P)|1\lambda_d\rangle &=&
-\frac{e^{i\lambda \phi}}{16M^3}
\sum_{j=1,2}\sum_{lm_lm_s'} C_{lm_l1m_s'}^{1\lambda_d}
\Big\{ \Pi_{j,\lambda} \Big[(F_{1j}k_j^2 +G_{Mj}q^2)
\delta_{s1}\delta_{m_sm_s'} +(G_{Mj}-F_{1j})S_{j,k\times q}^{sm_sm_s'}
\Big]\nonumber\\
& &
\hspace*{2cm}
+G_{Mj} \Big[(k_j^2+ q^2) S_{j, q, \lambda}^{sm_sm_s'}+ \Pi_{j,q}
S_{j,\pi,\lambda}^{sm_sm_s'}
\Big]\Big\}
\langle \vec k_j|lm_l\rangle \,,\\
\langle sm_s| J_{c,\lambda}^{B} (\vec q,\vec P)|1\lambda_d\rangle &=&
\frac{e^{i\lambda \phi}}{32M^3}
\sum_{j=1,2}\sum_{lm_lm_s'} C_{lm_l1m_s'}^{1\lambda_d}F_{1j}\Pi_{j,\lambda}
\Big\{(-q^2 \delta_{s1}\delta_{m_sm_s'} +S_{1,k\times q}^{sm_sm_s'}
+S_{2,k\times q}^{sm_sm_s'}) \langle \vec k_j|lm_l\rangle 
\nonumber\\
& &
\hspace*{2cm} 
-(s_jq^2 -\Pi_{q})\delta_{s1}\delta_{m_sm_s'}
\langle \vec k_j|i\vec q\cdot \vec r\,|lm_l\rangle \Big\} \,,\label{jcb}\\
\langle sm_s| J_{s,\lambda}^{B} (\vec q,\vec P)|1\lambda_d\rangle &=&
\frac{e^{i\lambda \phi}}{32M^3}
\sum_{j=1,2}\sum_{lm_lm_s'} C_{lm_l1m_s'}^{1\lambda_d}G_{Mj}
\Big\{-(s_jq^2 -\Pi_{q})S_{j,q,\lambda}^{sm_sm_s'}
\langle \vec k_j|i\vec q\cdot \vec r\,|lm_l\rangle\nonumber\\
&&\hspace*{-2cm}+ 
\Big[-((q^2-s_j\Pi_q) S_{j,q,\lambda}
^{sm_sm_s'} +q^2S_{j,k,\lambda}^{sm_sm_s'})
+q^2\Pi_{j,\lambda}\delta_{s1}\delta_{m_sm_s'}
+2\lambda k q^2 \sum_{j`\neq j}s_{j'}S_{j,j'}^{sm_sm_s'}
\Big]\langle \vec k_j|lm_l\rangle 
\Big\} \,,\label{jsb}
\end{eqnarray}
where $\vec k_j=\vec k-s_j\vec q/2$ with $s_j=\pm 1$ for $j=1$ and 2, 
respectively. The {\sc Fourier} transform of the orbital components of the 
deuteron wave function is given by
\begin{eqnarray}
\langle \vec p\;|l\,m_l\rangle&=&\int d^3r e^{-i\vec p\cdot\vec r}
\langle\vec r\;|l\,m_l\rangle\nonumber\\
&=&(-)^{l/2}4\pi\, Y_{lm_l}(\hat p)\int_0^\infty dr r u_l(r)j_l(pr)\,,
\end{eqnarray}
and
\begin{equation}
\langle \vec p\;|i\vec q\cdot \vec r\,|lm_l\rangle =
-\vec q\cdot \nabla_p\langle \vec p\;|l\,m_l\rangle\,.
\end{equation}
For convenience the various spin matrix elements have been denoted by
\begin{eqnarray}
S_{1j}^{s m_sm_s'} & =  &\langle (\frac{1}{2}\frac{1}{2})s m_s|
\sigma_{j,m_s-m_s'}|(\frac{1}{2}\frac{1}{2})1 m_s'\rangle
=(-1)^{s^{}-m_{s}}
\left( \begin{array}{ccc} s & 1 & 1\\ 
-m_{s} & m_{s}-m_{s}'& m_{s}' 
\end{array} \right)
\langle s \Vert \sigma_{j}^{[1]} \Vert 1 \rangle \,,\\
S_{j,k\times q}^{sm_sm_s'}&=&e^{-i\lambda \phi}
\langle (\frac{1}{2}\frac{1}{2})s m_s|
2s_ji\vec \sigma_{j}\cdot(\vec k\times \vec q\,)
|(\frac{1}{2}\frac{1}{2})1 m_s'\rangle
=2s_j k q (m_s-m_s')S_{1j}^{s m_s m_s'}
d^1_{0, m_s-m_s'}(\theta)\,,\\
S_{j, q, \lambda}^{sm_sm_s'}&=&e^{-i\lambda \phi}
\langle (\frac{1}{2}\frac{1}{2})s m_s|
i(\vec \sigma_{j}\times \vec q\,)_\lambda
|(\frac{1}{2}\frac{1}{2})1 m_s'\rangle
=\lambda q S_{1j}^{s m_s m_s'}
d^1_{\lambda , m_s-m_s'}(\theta)\,,\\
S_{j,k , \lambda}^{sm_sm_s'}&=&e^{-i\lambda \phi}
\langle (\frac{1}{2}\frac{1}{2})s m_s|
2s_ji(\vec \sigma_{j}\times \vec k)_\lambda
|(\frac{1}{2}\frac{1}{2})1 m_s'\rangle
=2 s_j k (m_s-m_s')S_{1j}^{s m_s m_s'}
d^1_{\lambda , m_s-m_s'}(\theta)\,,\\
S_{j,\pi , \lambda}^{sm_sm_s'}&=&e^{-i\lambda \phi}
\langle (\frac{1}{2}\frac{1}{2})s m_s|
i(\vec \sigma_{j}\times (2s_j\vec k -\vec q\,))_\lambda
|(\frac{1}{2}\frac{1}{2})1 m_s'\rangle
=S_{j,k , \lambda}^{sm_sm_s'}
-S_{j,q , \lambda}^{sm_sm_s'}\,,\\
S_{j,j'}^{s m_sm_s'} & = & (-)^{s- m_s}\sum_{s'' m_s''}(m_{s}''-m_{s}')
\Big( \begin{array}{ccc} s & 1 & s''\\ -m_{s} & m_{s}-m_{s}''
& m_{s}'' \end{array} \Big)
\nonumber\\& &\hspace{2cm}
d_{\lambda,m_s-m_s''}^1(\theta)
d_{0,m_s''-m_s'}^1(\theta)S_{1j'}^{s''m_s''m_s'}
\langle s \Vert \sigma_{j}^{[1]} \Vert s'' \rangle \,.
\end{eqnarray}
Furthermore, 
\begin{eqnarray}
\Pi_{q}(\theta)&=& 2kq \cos{\theta}\,,\\
\Pi_{j,q}(\theta)&=&s_j\Pi_{q}-q^2\,,\\
\Pi_{j,\lambda}(\theta)&=&2s_j k d^1_{\lambda ,0 }(\theta)\,.
\end{eqnarray}
Taking the current in the {\sc Sachs} form, the expressions for the 
nonrelativistic and boost currents are simply obtained 
from (\ref{rnr}), (\ref{rb}), (\ref{jnr}), (\ref{jcb}) and (\ref{jsb}) by 
substituting $F_{1j}$ by $G_{Ej}$. For the other relativistic contributions
one finds 
\begin{eqnarray}
\langle sm_s| \rho^{R} (\vec q,\vec P)|1\lambda_d\rangle &=&
-\frac{1}{8M^2}\sum_{j=1,2}\sum_{lm_lm_s'} C_{lm_l1m_s'}^{1\lambda_d}
\Big\{G_{Ej}\delta_{s1}\delta_{m_sm_s'}q^2 
+(2G_{Mj}-G_{Ej})S_{j,k\times q}^{sm_sm_s'})
\Big\}\langle \vec k_j|lm_l\rangle \,,\\
\langle sm_s| J_\lambda^{R} (\vec q,\vec P)|1\lambda_d\rangle &=&
-\frac{e^{i\lambda \phi}}{16M^3}
\sum_{j=1,2}\sum_{lm_lm_s'} C_{lm_l1m_s'}^{1\lambda_d}\Big\{ G_{Mj}
\Big[ (k_j^2+q^2) S_{j, q, \lambda}^{sm_sm_s'}+
\Pi_{j,q}S_{j,\pi,\lambda}^{sm_sm_s'}
\Big]\nonumber\\
& &
+\Pi_{j,\lambda} \Big[(G_{Ej}(2q^2+k_j^2 )-G_{Mj}q^2)
\delta_{s1}\delta_{m_sm_s'}-(G_{Ej}-G_{Mj})S_{j,k\times q}^{sm_sm_s'}\Big]
\Big\}\langle \vec k_j|lm_l\rangle \,.
\end{eqnarray}



\begin{thebibliography}{99}

\bibitem{LeT91} 
W. Leidemann, E.L. Tomusiak, H. Arenh\"ovel,
Phys.\ Rev.\ C {\bf 43}, 1022 (1991).

\bibitem{ArL92} 
H. Arenh\"ovel, W. Leidemann, E.L. Tomusiak,
Phys.\ Rev.\ C {\bf 46}, 455 (1992).

\bibitem{ArL93} 
H. Arenh\"ovel, W. Leidemann, E.L. Tomusiak,
Few-Body Syst.\ {\bf 15}, 109 (1993).

\bibitem{ArL95} 
H. Arenh\"ovel, W. Leidemann, E.L. Tomusiak,
Phys.\ Rev.\ C {\bf 52}, 1232 (1995).

\bibitem{ArL98} 
H. Arenh\"ovel, W. Leidemann, E.L. Tomusiak,
Nucl.\ Phys.\ {\bf 641}, 517 (1998).

\bibitem{ArL00} 
H. Arenh\"ovel, W. Leidemann, E.L. Tomusiak,
Few-Body Syst.\ {\bf 28}, 147 (2000).

\bibitem{Are88} 
H.\ Arenh\"ovel, Few-Body Syst.\ {\bf 4}, 55 (1988).

\bibitem{ArS90} 
H.\ Arenh\"ovel and K.-M.\ Schmitt, Few-Body Syst.\ {\bf 8}, 77 (1990).

\bibitem{DmG89} 
V.\ Dmitrasinovic and F.\ Gross, Phys.\ Rev.\ C {\bf 40}, 2479 (1989)

\bibitem{ArL02} 
H. Arenh\"ovel, W. Leidemann, E.L. Tomusiak, 
Eur. Phys. J. A {\bf 14}, 491 (2002).

\bibitem{Hal68} 
F.R. Halpern, Special Relativity and Quantum Mechanics. 
Englewood Cliffs, Prentice-Hall 1968

\bibitem{MaS70} 
M.A. Martin and T.D. Spearman, Elementary Particle Theory. 
Amsterdam, North-Holland 1970

\bibitem{Gie85} 
D.R. Giebink, Phys.\ Rev.\ C {\bf 32}, 502 (1985)

\bibitem{DoS86}
T.W. Donnelly and A.S. Raskin, Ann. Phys. (N.Y.) {\bf 169}, 247 (1986).

\bibitem{Ros57}
E.M. Rose, {\it Elementary Theory of Angular Momentum}, Wiley New York 1957.

\bibitem{Rob74} 
B.A. Robson, The Theory of Polarization Phenomena. 
                   Oxford, Clarendon Press 1974  

\bibitem{Par64}
F. Partovi, Ann. Phys. (N.Y.) {\bf 27}, 79 (1964).

\bibitem{ArS91} 
H. Arenh\"ovel and M. Sanzone, 
                    Few-Body Syst.\ Suppl.\ {\bf 3}, 1 (1991). 
\bibitem{Kaw58}
M. Kawaguchi, Phys. Rev. {\bf 111}, 1314 (1958).

\bibitem{CaM82}
A. Cambi and B. Mosconi, Phys. Rev. C {\bf 26}, 2358 (1982).

\bibitem{RaD89}
A.S. Raskin and T.W. Donnelly, Ann. Phys. (N.Y.) {\bf 191}, 78 (1989).

\bibitem{BlB52} 
J.M. Blatt and L.C. Biedenharn, Phys. Rev. {\bf 86}, 399 (1952).

\bibitem{FaA79}
W. Fabian and H. Arenh\"ovel, Nucl. Phys. A {\bf 314}, 253 (1979).

\bibitem{ArD71}
H. Arenh\"ovel, M. Danos, and H.T. Williams, Nucl. Phys. A {\bf 314}

\bibitem{LeA87}
W. Leidemann and H. Arenh\"ovel, Nucl. Phys. A {\bf 465}, 573 (1987).

\bibitem{RiG97}
F. Ritz, H. G\"oller, T. Wilbois, and H. Arenh\"ovel, 
Phys. Rev. C {\bf 55}, 2214 (1997).

\bibitem{MaH87} 
R. Machleidt, K. Holinde, and Ch. Elster, 
Phys.\ Rep.\ {\bf 149}, 1 (1987). 

\bibitem{NaR78} 
M.M. Nagels, T.A. Rijken, and J.J. de Swart,
Phys.\ Rev.\ D {\bf 17}, 768 (1978).

\bibitem{LaL80} 
M. Lacombe, B. Loiseau, J.M. Richard, R. Vinh Mau, 
J. C\^ot\'e, P. Pir\`es, and R. de Tourreil, 
Phys.\ Rev.\ C {\bf 21}, 861 (1980). 

\bibitem{WiS84} 
R.W. Wiringa, R. Smith, and T.L. Ainsworth, 
Phys.\ Rev.\ C {\bf 29}, 1207 (1984). 

\bibitem{WiS95} 
R.W. Wiringa, V.G.J. Stoks, and R. Schiavilla,
Phys.\ Rev.\ C {\bf 51}, 38 (1995). 

\bibitem{BeA92} 
G. Beck and H. Arenh\"ovel, Few-Body Syst. {\bf 13}, 165 (1992); 
{\bf 15}, 37 (1993) (E).

\bibitem{GaK71}
S. Galster {\it et al.}, Nucl. Phys. B {\bf 32}, 221 (1971).

\bibitem{BlZ74}
S. Blatnik, N. Zovko, Acta Phys. Austriaca {\bf 39}, 62 (1974).

\bibitem{HoP76}
G. H\"ohler {\it et al.}, Nucl. Phys. B {\bf 114}, 505 (1976).

\bibitem{GaK92}
M.F. Gari, W. Kr\"umpelmann, Phys. Lett. B {\bf 274}, 159 (1992).

\bibitem{HaM96}
H.W. Hammer, U.-G. Mei{\ss}ner, and D. Drechsel, 
Phys. Lett. B {\bf 385}, 343 (1996).

\bibitem{Are87}
H. Arenh\"ovel, Phys. Lett. B {\bf 199}, 13 (1987).

\bibitem{ToA88}
E.L. Tomusiak and H. Arenh\"ovel, Phys. Lett. B {\bf 206}, 187 (1988).

\bibitem{ArL88} 
H. Arenh\"ovel, W. Leidemann, E.L. Tomusiak,
Z. Phys.\ A {\bf 331}, 123 (1988); A {\bf 334}, 363 (1989) [Erratum].

\bibitem{ScA01}
M. Schwamb and H. Arenh\"ovel, Nucl. Phys. A {\bf 696}, 556 (2001).

\bibitem{Ris85}
D.O. Riska, Phys. Scrip. {\bf 31}, 471 (1985).

\bibitem{BuL85}
A. Buchmann, W. Leidemann, and H. Arenh\"ovel, 
Nucl. Phys. A {\bf 443}, 726 (1985).

\bibitem{ArS01}
H. Arenh\"ovel and M. Schwamb, Eur. Phys. J. A {\bf 12}, 207 (2001).

\bibitem{Are81}
H. Arenh\"ovel, Z. Phys. A {\bf 302}, 25 (1981).

\bibitem{GaH81}
M. Gari and H. Hebach, Phys. Rep. {\bf 72}, 1 (1981).

\bibitem{WeA78}
H.J. Weber and H. Arenh\"ovel, Phys. Rep. {\bf 36}, 279 (1978).

\bibitem{GrS82}
A.M. Green and M. Sainio, J. Phys. G {\bf 8}, 1337 (1982).

\bibitem{Lei80}
W. Leidemann, diploma thesis, Mainz 1980.

\bibitem{Fri77}
J.L. Friar, Ann. Phys. (NY) {\bf 104}, 180 (1977).

\bibitem{TrA89}
E. Truhlik and J. Adam, Jr., Nucl. Phys. A {\bf 492}, 529 (1989).

\bibitem{GoA92}
H. G\"oller and H. Arenh\"ovel, Few-Body Syst. {\bf 13}, 117 (1992).

\bibitem{TaN92}
K. Tamura, T. Niwa, T. Sato, and H. Ohtsubo, 
Nucl. Phys. A {\bf 536}, 597 (1992).

\bibitem{AdA97}
J. Adam, Jr. and H. Arenh\"ovel, Nucl. Phys. A {\bf 614}, 289 (1997).

\bibitem{WiB93}
T. Wilbois, G. Beck, and H. Arenh\"ovel, Few-Body Syst. {\bf 15}, 39 (1993).

\bibitem{Aue85} 
S.\ Auffret {\it et al.},
Phys.\ Rev.\ Lett.\ {\bf 55}, 1362 (1985).

\bibitem{Boe90}
R.G.\ Arnold {\it et al.},
Phys.\ Rev.\ C {\bf 42}, R1 (1990).

\bibitem{SiW79}
G.G. Simon {\it et al.}, Nucl. Phys. A {\bf 324}, 277 (1979).

\bibitem{QuB88}
B.P. Quinn {\it et al.}, Phys. Rev. C {\bf 37}, 1609 (1988).

\bibitem{BoB92}
B. Boden {\it et al.}, Nucl. Phys. A {\bf 549}, 471 (1992).

\bibitem{BuA95}
H.J. Bulten {\it et al.}, Phys. Rev. Lett. {\bf 74}, 4775 (1995).

\bibitem{PeA97}
A. Pellegrino {\it et al.}, Phys. Rev. Lett. {\bf 78}, 4011 (1997).

\bibitem{BlB98}
K.I. Blomqvist {\it et al.}, Phys. Lett. B {\bf 424}, 33 (1998).

\bibitem{KaH98}
W.-J. Kasdorp {\it et al.}, Few-Body Syst. {\bf 25}, 115 (1998).

\bibitem{DoA99}
S.M. Dolfini {\it et al.}, Phys. Rev. C {\bf 60}, 064622 (1999).

\bibitem{Ost99}
M. Ostrick {\it et al.}, Phys. Rev. Lett. {\bf 83}, 276 (1999).

\bibitem{NiA01}
D.M. Nikolenko {\it et al.}, Nucl. Phys. A {\bf 684}, 525c (2001).

\bibitem{ZhA01}
H. Zhu {\it et al.}, Phys. Rev. Lett. {\bf 87}, 081801 (2001).

\bibitem{ZhC01}
Z.-L. Zhou {\it et al.}, Phys. Rev. Lett. {\bf 87}, 172301 (2001).

\bibitem{vNR02}
P. von Neuman-Cosel {\it et al.},
Phys. Rev. Lett. {\bf 88}, 202304 (2002).

\bibitem{UlA02}
P.E. Ulmer {\it et al.}, Phys. Rev. Lett. {\bf 89}, 062301 (2002).

\bibitem{Mul03}
 U. M\"uller {\it et al.}, Proc. Int. School on Nuclear Physics, Erice, 
Italy, Prog. Part. Nucl. Phys. {\bf 50}, 483 (2003).

\bibitem{NiA03}
D.M. Nikolenko {\it et al.}, Phys. Rev. Lett. {\bf 90}, 072501 (2003).

\bibitem{MaS03}
R. Madey {\it et al.}, Eur. Phys. J. A {\bf 17}, 323 (2003).

\bibitem{MaS03a}
R. Madey {\it et al.}, Phys. Rev. Lett. {\bf 91}, 122002 (2003).

\end{thebibliography}
\end {document}